\documentclass[aps,prd,twocolumn,groupedaddress,asmsymb,amsmath,amssymb,longbibliography]{revtex4-1}

\usepackage{graphicx}
\usepackage{calligra}
\usepackage{upgreek}
\usepackage{color}
\usepackage{inputenc,caption}
\usepackage{stackrel}
\usepackage[usenames,dvipsnames]{xcolor}
\usepackage{subcaption}

\numberwithin{equation}{section}

\DeclareMathAlphabet{\mathpzc}{OT1}{pzc}{m}{it}

\newcommand{\be}{\begin{equation}}
\newcommand{\ee}{\end{equation}}
\newcommand{\bea}{\begin{eqnarray}}
\newcommand{\eea}{\end{eqnarray}}
\newcommand{\lb}{\label}

\newcommand{\bv}{{\bf v}}

\newcommand{\bx}{{\bf x}}

\newcommand{\br}{{\bf r}}

\newcommand{\grad}{{\mbox{\boldmath $\nabla$}}}

\newcommand{\bzed}{{\mbox{\boldmath $0$}}}

\newcommand{\calD}{{\mathcal D}}

\newcommand{\hit}{\!\mbox{$\mathpzc{h}$}}
\newcommand{\sign}{{\rm sign}} 

\begin{document}


\title{A Renormalization Group Approach to Spontaneous Stochasticity}


\author{Gregory L. Eyink${\,\!}^{1,2}$ and Dmytro Bandak${\,\!}^3$}
\affiliation{${\,\!}^1$Department of Applied Mathematics \& Statistics, The Johns Hopkins University, Baltimore, MD, USA, 21218}
\affiliation{${\,\!}^2$Department of Physics \& Astronomy, The Johns Hopkins University, Baltimore, MD, USA, 21218}
\affiliation{${\,\!}^3$Department of Physics, University of Illinois at Urbana-Champaign, Urbana, IL, USA, 61801} 


\date{\today}

\begin{abstract}
We develop a theoretical approach to ``spontaneous stochasticity'' in classical dynamical systems that are nearly 
singular and weakly perturbed by noise. This phenomenon is associated to a breakdown in uniqueness of solutions for 
fixed initial data and underlies many fundamental effects of turbulence (unpredictability, anomalous dissipation, 
enhanced mixing).  Based upon analogy with statistical-mechanical critical points at zero temperature, we 
elaborate a renormalization group (RG) theory that determines the universal statistics obtained for sufficiently 
long times after the precise initial data are ``forgotten''.  We apply our RG method to solve exactly the ``minimal model'' 
of spontaneous stochasticity given by a 1D singular ODE. Generalizing prior results for the infinite-Reynolds limit 
of our model, we obtain the RG fixed points that characterize the spontaneous statistics in the near-singular, 
weak-noise limit, determine the exact domain of attraction of each fixed point, and derive the universal approach 
to the fixed points as a singular large-deviations scaling, distinct from that obtained by the standard saddle-point 
approximation to stochastic path-integrals in the zero-noise limit. We present also numerical simulation results 
that verify our analytical predictions, propose possible experimental realizations of the ``minimal model'', and 
discuss more generally current empirical  evidence for ubiquitous spontaneous stochasticity in Nature. Our RG 
method can be applied to more complex, realistic systems and some future applications are briefly outlined. 
\end{abstract}

\pacs{?????}

\maketitle


\section{Introduction}\label{sec:I}

Classical dynamical systems (ODE's and PDE's) 
\be \dot{\bx}=\bv(\bx,t), \quad \bx(0)=\bx_0 \lb{ODE} \ee 
with smooth vector fields $\bv$ have unique solutions of the Cauchy initial-value problem 
for any fixed initial data $\bx_0$. \\ However, understanding has slowly consolidated that unicity is 
effectively broken when the field $\bv$ becomes nearly singular and then small random perturbations of the deterministic 
dynamics can persist in the joint limit of vanishing noise and increasing singularity. In that case, deterministic 
trajectories become non-unique and solutions remain stochastic. This phenomenon occurs 
for Lagrangian fluid particles in high Reynolds-number turbulence \cite{gawedzki2001turbulent,
kupiainen2003nondeterministic}, where it has been termed {\it spontaneous stochasticity}
\cite{chaves2003lagrangian}. Although the effect was implicit in the work of L. F. Richardson 
on two-particle dispersion \cite{richardson1926atmospheric}, it was only realized much later that classical 
determinism of Lagrangian trajectories breaks down and provides the mechanism for anomalous dissipation of scalars advected 
(passively or actively) by a turbulent flow  \cite{bernard1998slow,eijnden2000generalized,e2003note,drivas2017lagrangian}. 
Similar Lagrangian spontaneous stochasticity of magnetic field-lines in turbulent astrophysical plasmas 
breaks the flux-freezing constraint at high electrical conductivity, necessary to explain fast magnetic reconnection
\cite{lazarian1999reconnection,eyink2011fast,eyink2013flux}. Entropy cascade in nearly collisionless plasmas 
likewise requires spontaneous stochasticity of Hamiltonian charged-particle motions in the self-consistent 
electromagnetic fields of the limiting Vlasov kinetic equation \cite{eyink2018cascades,bardos2019onsager}. 
Even before these Lagrangian manifestations were clearly understood, it was pointed out by E. N. Lorenz in a 
pioneering work \cite{lorenz1969predictability} that Eulerian solutions of fluid equations can exhibit similar 
indeterminacy for multiscale turbulent flow.  This Eulerian spontaneous stochasticity must be directly connected 
with non-uniqueness of dissipative weak solutions to the Cauchy problem for ideal fluid equations obtained 
in the high Reynolds limit or the ``Nash non-rigidity" phenomenon 
\cite{delellis2010admissibility,delellis2017high,daneri2020non}. 
There are profound scientific implications for predictability of weather and climate \cite{palmer2014real,palmer2019stochastic}, 
but more generally for practical indeterminism of diverse problems in geophysics, astrophysics, and cosmology
\cite{mailybaev2016spontaneously,biferale2018rayleigh,thalabard2020butterfly}.    

Despite the significant implications, there are few general theoretical tools available to 
predict when spontaneous stochasticity will occur and to determine what the observed statistics will be.  
We develop here a novel approach based on the renormalization group method from statistical physics 
and field theory \cite{wilson1974renormalization,wilson1975renormalization,goldenfeld2018lectures}. 
Although the method can be applied much more generally, we focus 
on the ``minimal model'' that exhibits the basic phenomenon, the one-dimensional ODE  
\be \dot{x}=v_*(x):=A|x|^h\sign(x), \quad x(0)=0 \lb{vstar-eq} \ee 
with H\"older exponent/roughness exponent $h$. Here $v_*(x)$ may be regarded as a caricature of a turbulent 
velocity field in the limit $Re\to\infty$ which, according to Onsager's theorem \cite{onsager1949statistical,
eyink2006onsager}, must develop singularities 
$h\leq 1/3$ in order to account for anomalous dissipation of kinetic energy. Although the model velocity $v_*(x)$
is singular at only a single point ($x=0$) when $0<h<1,$ it provides the standard textbook example of 
breakdown of uniqueness (e.g. see \cite{hartman1982ordinary}, p.2). The non-unique solutions, labelled 
by $\alpha=(\pm,\tau),$ have the form of delayed power-laws, with $\delta=1-h$: 
\be x^\alpha(t)=\left\{\begin{array}{ll}
                                0 & 0\leq t\leq \tau,  \cr 
                                \pm (A\,\delta\, (t-\tau))^{1/\delta} & t>\tau, 
                                \end{array} \right. \lb{nonuniq} \ee
where $\pm$ is the direction and $\tau\in [0,\infty]$ is a waiting time at the origin. The set of non-unique 
solutions is here uncountably infinite and, in fact, this is generally true for any bounded, continuous 
$\bv$ by Kneser's theorem (\cite{hartman1982ordinary}, Theorem 4.1). 
The above simple one-dimensional ODE and various generalizations have been extensively 
studied in the probability theory literature when supplemented with a white-noise random perturbation:
\be  \dot{x}=v_*(x) + \sqrt{2D} \,\eta(t), \quad x(0)=0. \lb{vstar-D-eq} \ee 
See \cite{bafico1982small,gradinaru2001singular,attanasio2009zero,flandoli2013topics}. 
This initial-value problem is stochastically well-posed and has a unique (random) solution. In the limit 
of vanishing noise $D\to 0,$ it has been shown that the solution remains stochastic, with equal probabilities 
of $1/2$ for the two {\it extremal solutions} with no initial delay ($\tau=0$) to occur:
\be x^\pm(t) :=\pm (A\,\delta\,t)^{1/\delta},  \lb{extrm} \ee
and all other solutions of the limiting deterministic ODE have 
zero probability to occur \cite{bafico1982small,attanasio2009zero}. The precise rate of vanishing of the probability 
to observe non-extremal solutions has also been derived as a ``singular large deviations'' 
\cite{gradinaru2001singular}, distinct from Freidlin-Wentzell predictions for smooth dynamical systems 
\cite{freidlin2012random}. 

We wish to reconsider this model from the perspective of statistical physics. To render the model
in principle experimentally realizable, it is necessary to impose an IR cut-off length $L$ and also a   
short-distance (UV) regularization scale $\ell,$ for example,
\be v(x) = \frac{A\,x}{(x^2+\ell^2)^{\delta/2}}, \quad |x|<L \lb{vreg} \ee 
so that ``rough'' power-law scaling $v(x)\sim A|x|^h$ occurs only in the range $\ell\ll |x|\ll L.$ In a physical
turbulent fluid flow, the UV cut-off $\ell$ is provided by viscosity. The precise behavior for $|x|>L$ is 
immaterial but one of our main findings is that the details of the cutoff at $|x|<\ell$ influence the 
asymptotic behavior in the limit $L/\ell\gg 1,$ and render generally inapplicable many of the specific
results obtained in the probability theory literature for the idealized mathematical problem with $\ell=0.$  
The statistics of the time-histories $x(t)$ with added white-noise, as in \eqref{vstar-D-eq}, can be 
expressed in terms of path-integrals with the Onsager-Machlup action \cite{onsager1953fluctuations,
wio2013path} 
\be S[\xi]=\frac{1}{4D} \int_{t_0}^{t} ds\, |\dot{\xi}(s)-v(\xi(s))|^2 \lb{OM} \ee
e.g. for the transition probability density function:    
\be
P(x,t|x_0,0) =  \int_{\xi(0)=x_0} \calD \xi\, \delta(x-\xi(t)) e^{-S[\xi]}. 
\lb{OM-PI} \ee 
The stochastic dynamics thus corresponds to a (0+1)-dimensional Euclidean field theory, or a $1$-dimensional
equilibrium statistical system with the Onsager-Machlup action $S[\xi]$ as ``Hamiltonian''.  The joint limit 
$L/\ell\to \infty$ and $D\to 0$ is then quite closely analogous to a zero-temperature phase-transition 
in an equilibrium system such as the 1D Ising model, which appears in the joint thermodynamic limit 
$L/a\to\infty$ and low-temperature limit $T\to 0$ \cite{nelson1975soluble}. 
In this analogy, the non-unique solutions \eqref{nonuniq}
of the initial-value problem play the role of non-unique ground states of the Hamiltonian and the 
physics problem is to determine which ``ground-states'' appear in the limit and their precise probabilities. 

We shall apply a version of the renormalization group (RG) method to answer this question. 
To understand why RG is applicable to this problem, there are two key points. First, the 
physical mechanism of spontaneous stochasticity is always the {\it forgetting of the 
initial data} at sufficiently long times. This mechanism can be well-illustrated by the singular, deterministic 
ODE (\ref{vstar-eq}) whose exact solution for general initial data $x_0\neq 0$ is easily found 
for $h<1$ to be 
\be x(t) =\pm (|x_0|^\delta + A\,\delta\,t)^{1/\delta}, \quad \pm=\sign(x_0) \lb{forget} \ee 
When $t\gg |x_0|^\delta/A\delta,$ then the general solution becomes indistinguishable from the 
extremal solutions (\ref{extrm}) and all memory of the initial data is lost except for the overall $\pm$ 
sign. At long times one expects a universal statistical state to be achieved independent of the 
precise initial data and their statistics. There is thus a close analogy with critical phenomena in 
equilibrium spin systems, in which a universal statistical behavior is expected at length scales 
much greater than the lattice spacing $a,$ independent of the microscopic Hamiltonian
within some broad universality class, which is determined by an IR stable RG fixed point and its 
domain of attraction \cite{wilson1974renormalization,wilson1975renormalization,
goldenfeld2018lectures}. The second key fact that makes RG relevant is that the 
singular dynamics (\ref{vstar-eq}) possesses an exact {\it scale symmetry} according to which 
the transformation
\be x_b(t) := b^{-1/\delta} x(bt) \lb{scal-sym} \ee 
for any real number $b>0$ maps a solution $x(t)$ to another solution $x_b(t).$ This symmetry is 
broken by the regularization of type \eqref{vreg} at length-scales $<\ell$ and also by the Langevin 
noise proportional to $\sqrt{D}$ in the stochastically perturbed dynamics \eqref{vstar-D-eq}. However,
the symmetry is expected to be restored, at least statistically, in the joint limit $\ell\to 0$ and 
$D\to 0.$ Here we may note that the extremal solutions \eqref{extrm} are scale-invariant,
$x_b^\pm(t)=x^\pm(t)$ for all $b>0,$ and are thus perfectly ``ordered'' states, analogous to the homogeneous 
ground states of the 1D Ising model with all spins up or down.  The existence of such a scale-symmetry 
(or set of scale-symmetries) of the limiting singular dynamics is not required for spontaneous stochasticity, 
but it is a typical feature of the ideal equations that govern high Reynolds-number fluids and plasmas
\cite{frisch1995turbulence,palmer2019stochastic}. 
Based upon these two ingredients, we shall construct an RG framework that permits a systematic analysis 
of SS in the 1D model problem. 

\begin{figure}[h!]\label{phase_diagram}
 \begin{center}
 \includegraphics[width=240pt]{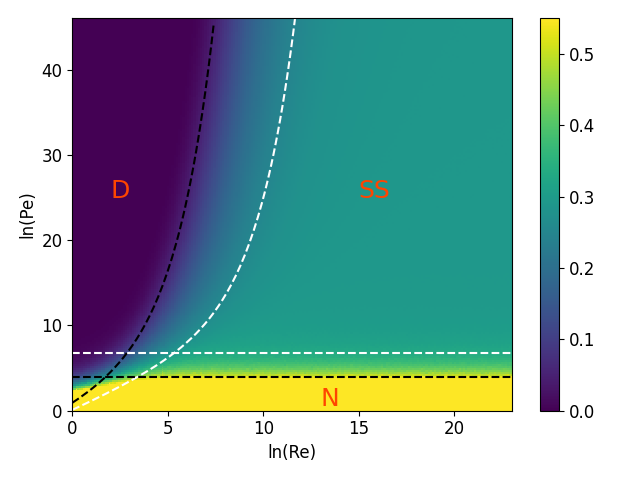}
 \end{center}
 \caption{Phase diagram for the position variance $\Upsilon$ at the inertial time $\bar{t}=1$ 
 as a function of the state variables $X=\ln(Re)$ and $Y=\ln(Pe)$. The white (black) solid lines plot
 $Y=X+c_{1(2)}\exp(X/2)$ and $Y=c_{3(4)}$ for $c_1=0.10$, $c_3=6.74$ ($c_2=0.94$,  $c_4=3.97$), the theoretical phase boundaries for large $X,$ $Y$ from the RG analysis.}\label{4.png}
 \end{figure}

The fundamental result we obtain is that there is an RG fixed point that governs the ``critical limit'' 
$\ell\to 0,$ $D\to 0$ of our model. Just as for closely analogous $T=0$ critical phenomena such as 
quantum critical points 
\cite{sachdev2000quantum,coleman2005quantum} 
or the order-disorder transition in the 1D Ising model
\cite{nelson1975soluble}, the singular critical point in spontaneous stochasticity can never be strictly 
achieved but its drastic effects are felt for finite values of $\ell$ and $D$ over a very wide region. 
This is illustrated by the phase diagram in $X-Y$-parameter plane, Figure 1, with $X=\ln(Re)$ and $Y=\ln(Pe),$ where $Re$ is the 
``Reynolds number'' and $Pe$ the ``P\'eclet number'', dimensionless versions of $1/\ell$ and $1/D,$
respectively. (See section \ref{dimension}). As order parameter for spontaneous stochasticity we take 
the variance of a similarly non-dimensionalized position $\bar{x}(\bar{t})$ at time $\bar{t}=1$:\\

\vspace{-20pt}
\be \Upsilon(X,Y) :=\langle \bar{x}^2(\bar{t})\rangle_{X,Y} -\langle \bar{x}(\bar{t})\rangle^2_{X,Y} \lb{order} \ee 
This variance is suitable as order parameter because deterministic behavior is implied by $\Upsilon=0$ while spontaneous 
stochasticity corresponds to a value $\Upsilon=\Upsilon_*>0$ determined by the RG fixed point and independent 
of both $X$ and $Y.$ The plot in Figure 1 clearly distinguishes three phases, a ``deterministic phase'' D
for extremely large Y where $\Upsilon\simeq 0,$ a ``noise-driven phase'' $N$ at small $Y$ where $\Upsilon(Y)$  
strongly increases as $Y$ decreases, and an extensive ``spontaneously stochastic phase'' (SS) at 
sufficiently large values of $X,$ $Y$ where $\Upsilon=\Upsilon_*.$ The values of $\Upsilon$ are obtained 
from a numerical simulation of the model (section \ref{numerics}) but the 
``phase boundaries'' shown as solid lines are analytic results from an RG analysis of the approach 
to the fixed point (section \ref{approach}). Although the boundaries between the ``phases'' are somewhat 
arbitrary, they are qualitatively well-defined and match well our analytical predictions.  In any case,  it is clear 
from Figure 1 that the critical behavior at $X=Y=\infty$ is attained already for a broad range of finite $X,$ $Y$ values. 

Before we present the details of our RG analysis, we must emphasize that spontaneous stochasticity 
is qualitatively distinct from the so-called ``butterfly effect''  or  ``deterministic chaos''
of differentiable dynamics \cite{lorenz1963deterministic,ChaosBook}. This 
familiar behavior corresponds to Lyapunov exponential growth of errors, 
which is observed in our 1D toy for $h=1.$ In that case, the solution of equation 
\eqref{vstar-eq} for general initial data is 
\be  x(t) = x_0 e^{At} \quad (h=1). \lb{recall} \ee 
Regarding $x_0\neq 0$ as a small perturbation away from $0,$ it is obvious that for $h=1$
the initial error is exponentially magnified over time but never ``forgotten.'' In particular, 
for any fixed time $t$ one observes that $x(t)\to 0$ as $x_0\to 0,$ whereas from \eqref{forget}
for $h<1$ one observes at any fixed time $t$ that $x(t)\to x^\pm(t)\neq 0$ as $x_0\to 0.$ This 
fundamental distinction is what characterizes spontaneous stochasticity. Even though the 1D toy 
problem has a smooth velocity when regularized as in \eqref{vreg}, we shall see that the 
exponential error growth \eqref{recall} either does not occur at all or else is an extremely short-lived
transient that rapidly crosses over to power-law growth as in \eqref{extrm}.  As we shall argue 
below, this is the generic situation in high Reynolds-number turbulent flows and in similar problems 
with near-singular dynamics in geophysics and astrophysics. 

\section{Review of Current Evidence for Spontaneous Stochasticity}\lb{evidence} 

Spontaneous stochasticity is conjectured or even demonstrated to underlie some of the most basic turbulence 
phenomena and should be ubiquitous in turbulent flows. A simple Lagrangian
manifestation of spontaneous stochasticity is the universal Richardson $t^3$-growth in mean-square width 
of plumes of smoke or of other aerosols advected by a turbulent flow \cite{bernard1998slow}, which many observations
suggest to be independent of the details of the source and of the mass diffusivity of the aerosol
(\cite{monin2013statistical}, section 24.3). More recently, it has been argued \cite{thalabard2020butterfly}
that Eulerian spontaneous stochasticity explains how universal statistics are attained in a finite time
for turbulent mixing layers during their phase of non-stationary, self-similar ``equilibrium''  spread 
\cite{brown2012turbulent,suryanarayanan2014free}. In addition to many hints and indirect pieces 
of evidence for spontaneous stochasticity, there are a few controlled, precision studies which provide 
direct corroboration of the phenomenon. It is useful to review these here in order to provide context and
motivation for our subsequent analytical investigations of our model problem.
  

We first consider Lagrangian fluid particles in turbulent flows, where spontaneous stochasticity is 
necessary to explain enhanced turbulent mixing. In fact, anomalous 
scalar dissipation for $Re\gg 1,$ $Pe\gg 1$ strictly independent of molecular transport coefficients $D,$ $\nu$
can occur only if Lagrangian particle trajectories are spontaneously stochastic (at least away from 
solid walls) \cite{bernard1998slow,eijnden2000generalized,drivas2017lagrangian}. Therefore, the accumulated 
evidence from experiments and simulations for a scalar dissipative anomaly \cite{donzis2005scalar}
provides indirect support for spontaneous stochasticity. It is impossible from empirical observations, 
however, to ever conclusively rule out a very slow decrease of scalar dissipation for increasing $Re,$ $Pe.$ 
It is therefore desirable to have direct evidence for spontaneous stochasticity. The most basic 
observable effect on displacements $\br(t)=\bx'(t)-\bx(t)$ of pairs of advected particles  
at positions $\bx(t),$ $\bx'(t)$ would be the ``forgetting'' of the initial displacement $\br(0),$ analogous 
to the ``forgetting'' of $x_0$ in \eqref{forget} for our model problem in the limit $t\gg |x_0|^\delta/A\delta.$

Unfortunately, this phenomenon has proved difficult to verify in laboratory experiments. The papers 
\cite{bourgoin2006role,ouellette2006experimental} describe an experiment in a turbulent flow at 
$Re\simeq 4.4\times 10^4,$ seeded with neutrally bouyant polystyrene particles which were 
illuminated by two crossed laser beams and tracked with high-speed cameras. Those experiments 
observed a ``ballistic'' separation of particles, with r.m.s. separation $r'(t)\propto v'(r(0))t,$ exhibiting 
a strong dependence upon $r(0)$ through $v'(r(0))=\sqrt{\langle |\bv(\br(0))-\bv(\bzed)|^2\rangle}.$ 
The analogous dependence is seen in our model problem at short times by Taylor-expanding \eqref{forget} 
for $t\ll |x_0|^\delta/A\delta$ to obtain $x(t)-x_0\simeq v_*(x_0)t.$ Since the time $t_0=r(0)/v'(r(0))$ required 
to ``forget'' $\br(0)$ in a turbulent flow grows $\propto (r^2(0)/\bar{\varepsilon})^{1/3}$ (assuming the Kolmogorov
``2/3-law'' \cite{kolmogorov1941local} with $\bar{\varepsilon}$ the mean energy dissipation per mass) a plausible 
explanation of the observations of \cite{bourgoin2006role,ouellette2006experimental} is that initial separations 
$r(0)$ were insufficiently small to observe ``forgetting''. In fact, difficulties in tracking particles in densely 
seeded flows prevented \cite{bourgoin2006role,ouellette2006experimental} from accessing separations 
smaller than about $43$ Kolmogorov dissipation lengths. 

Numerical simulations are not subject to such limitations on $r(0).$ The study 
\cite{bitane2013geometry} computed two forced isotropic turbulent flows at Reynolds numbers 
$Re\simeq 1.4\times 10^4$ and $3.5\times 10^4$ and investigated deterministic Lagrangian particles
for initial separations ranging from $2-24$ Kolmogorov lengths. Figure 2 of \cite{bitane2013geometry} 
showed a very clear ``forgetting'' effect, with mean-square displacement $\langle r^2(t)\rangle$
transitioning from a ``ballistic'' regime $\propto (v'(r(0))^2 t^2$ for $t\lesssim t_0$  instead to a universal regime 
$\propto \bar{\varepsilon} t^3$ 
as predicted  by Richardson \cite{richardson1926atmospheric}, independent of $r(0)$ for $t\gg t_0$
A similar effect had been seen even earlier for stochastic Lagrangian particles satisfying a Langevin 
equation such as \eqref{model} but with velocity taken from a numerical simulation of forced, isotropic 
turbulence at $Re\simeq 1.25\times 10^4$ \cite{eyink2011fast}. This study released all realizations of the 
simulated Brownian particles at the same point and averaged over that release location. For two different 
values of Schmidt number, $Sc=1$ and $Sc=0.1$ \footnote{Paper \cite{eyink2011fast} used the terminology
``(magnetic) Prandtl number'' rather than ``Schmidt number'', because the application considered there was
the turbulent kinematic dynamo}, Figure 1 of \cite{eyink2011fast} showed that 
$\langle r^2(t)\rangle$ crossed from a diffusive scaling $\propto Dt$ for $t\ll t_D\sim (D^3/\bar{\varepsilon})^{1/4}$
over to a universal Richardson law $\propto \bar{\varepsilon} t^3$ for $t\gg t_D,$ independent of the magnitude 
of $Sc.$ The noise level  measured by $Sc$ is thus ``forgotten'' at sufficiently long times. 
Using the same simulation database, paper \cite{drivas2017lagrangian} extended the previous study by 
considering stochastic Lagrangian particles without averaging over the release location $\bx_0$ and 
computed the probability distributions $P(\bx,t)$ for three values $Sc=0.1,$ $1,$ $10.$ It was verified 
for $t$ equal to about one large-eddy turnover time that these distributions were nearly the same 
for all three $Sc$ values. 

Empirical evidence has emerged also for Eulerian spontaneous stochasticity of the type first proposed 
by Lorenz \cite{lorenz1969predictability} for multiscale turbulent systems. This effect must 
occur constantly in laboratory turbulence, as obvious from the apparent non-reproducibility 
of the specific details of such flows. It would be very challenging, however, to perform controlled 
laboratory studies because of the difficulty in creating an ensemble of initial conditions whose 
members differ in only fine details. Numerical experiments, on the other hand, can be readily 
performed and the most important scientific implications are in fact for the unpredictability and 
extreme ill-conditionedness of simulations for such singular turbulent flows. 
Numerical studies of predictability of three-dimensional isotropic turbulence \cite{boffetta2017chaos,berera2018chaotic} 
have considered the mean-square ``pair-separation'' $\langle \|\bv(t)-\bv'(t)\|^2\rangle$ in energy-norm
$\|\cdot\|$ for two Navier-Stokes solutions $\bv(t),$ $\bv'(t)$ with initial data slightly different at small scales. 
These works verify the prediction \cite{leith1972predictability} that that mean-square separation 
should grow $\propto \bar{\varepsilon} t$ for long enough times, but unfortunately the crucial ``forgetting'' of the 
initial separation $\|\bv(0)-\bv'(0)\|$ was not investigated. 

More direct evidence for Eulerian spontaneous stochasticity was found in \cite{palmer2014real}
which studied a solution of the surface quasi-geostrophic (SQG) model for turbulent flow over a 
fractal topography. It should be noted here that weak solutions of the inviscid SQG equations 
expected to describe the infinite-$Re$ limit are known to be non-unique for exactly prescribed initial data 
\cite{buckmaster2019nonuniqueness}. Figure 5 of \cite{palmer2014real} presented data on the 
r.m.s. separation $\sqrt{\langle \|\psi(t)-\psi'(t)\|^2\rangle}$ of the stream-functions $\psi(t)$ for the 
base flow  and $\psi'(t)$ for a flow with initial condition perturbed by three sets of ``error fields'' of 
decreasing magnitude and length-scale. This separation was found to be almost unchanged as 
the initial errors decreased, consistent with Eulerian spontaneous stochasticity, or 
what \cite{palmer2014real} termed ``the real butterfly effect''.  Even more compelling evidence 
comes from the paper \cite{thalabard2020butterfly} which studied the development of a 
turbulent mixing layer from a Kelvin-Helmholtz unstable singular vortex-sheet, where again 
solutions of the limiting inviscid Euler equations are known to be non-unique 
\cite{szekelyhidi2011weak,mengual2020dissipative}. 
Perturbing the initial vortex-sheet with random ``errors'' of decreasing magnitude and 
simultaneously decreasing the viscosity $\nu$, the crucial ``forgetting'' of initial  error $\|\bv(0)-\bv'(0)\|$
was observed in Figure 1 of \cite{thalabard2020butterfly} for the mean-square separation 
$\langle \|\bv(t)-\bv'(t)\|^2\rangle$ after a sufficient time-interval. Furthermore, after this initial 
short transient, the spontaneously stochastic ensemble of solutions entered a self-similar
growth phase with universal statistical properties independent of regularization and noise level,
consistent with earlier observations \cite{suryanarayanan2014free}.  

We may mention finally that quantum spontaneous stochasticity in the semiclassical limit $\hbar/m\to 0$
has been derived for a Hamiltonian version of our 1D model: 
\be m\ddot{x} = A |x|^h{\rm sign}(x), \quad 0<h<1 \lb{2ndord-eq} \ee
when that is quantized in the canonical manner. See \cite{athanassoulis2012strong,eyink2015quantum}. 
The second paper \cite{eyink2015quantum} also verified their analytical results for a regularized 
potential by numerical simulation of the Schr\"odinger equation and discussed some 
possible realizations by laboratory experiments. However, it would be extremely 
challenging in an experiment to verify quantum fluctuations as the origin of the stochasticity,   
precisely because the limiting spontaneous statistics are universal and do not depend upon 
the precise source of the random perturbations. Temperatures near absolute zero would 
be required to suppress thermal fluctuations and environmental noise sources 
would have to be very carefully controlled. 
  
\section{Renormalization Group Theory} 

The experimental and numerical evidence presented in the previous section provide some
solid confirmation that spontaneous stochasticity is generally present in high-Reynolds-number 
turbulence and that it underlies such basic phenomena as anomalous dissipation, enhanced 
mixing, unpredictability, and universal statistics in non-stationary, evolving flows. Presently, 
however, there are first-principles analytical derivations of spontaneous stochasticity only 
in some model problems, such as Lagrangian particle histories in the Kraichnan
model \cite{bernard1998slow,eijnden2000generalized,lejan2002integration,lejan2004flows}
and in Burgers shock solutions \cite{eyink2015spontaneous}, or semi-classical particle 
trajectories in some singular 1D quantum models \cite{athanassoulis2012strong,eyink2015quantum}.  
The mathematical techniques employed in these works are special to the models considered and
give no general physical intuition. We here develop an RG approach based upon a close analogy 
with zero-temperature critical phenomena, which provides not only physical insight into  
the nature of spontaneous stochasticity but also a powerful tool with which to derive the 
universal statistics, both analytically and numerically.   

\subsection{Dimensional Analysis of the Model}\lb{dimension} 

We first perform a careful dimensional analysis of the family of models we consider, 
which may all be expressed by the 1D  Langevin equation 
\be \frac{dx}{dt} = v(x) + \sqrt{2 D}\, \eta(t), \quad x(0)=x_0  \lb{model} \ee
where $\eta(t)$ is white-noise with mean zero and covariance 
\be \langle \eta(t)\eta(t')\rangle = \delta(t-t'), \lb{noise} \ee 
but with different choices of the drift velocity $v(x).$ In all cases, $v(x)$ is a ``quasi-singular'' function satisfying 
\be v(x) = v_*(x):=A\,\sign(x)|x|^h, \quad  |x|\geq \ell  \lb{outside} \ee 
with some roughness exponent $0<h<1,$ while it is regularized for $|x|<\ell.$ In most of our analysis 
we assume that $v'(x)\geq 0$ and a simple, concrete regularization satisfying that monotonicity
condition is the linear interpolation 
\be v(x) = A\, x \ell^{h-1}, \quad |x|\leq \ell. \lb{linear} \ee
Any such regularization must have exactly one point $|x_{eq}|<\ell$ where $v(x_{eq})=0$ and 
for most natural regularizations, such as the linear interpolation, $x_{eq}=0.$ Although it is not 
necessary to our arguments, we assume that below for convenience of presentation. It is also 
convenient to impose the strict equality \eqref{outside}, although it suffices in fact to assume 
only that 
\be \lim_{\lambda\to 0} \lambda^{-h} v(\lambda x)=v_*(x) \ee 
uniformly on compact sets, as in the example (\ref{vreg}). To simplify the mathematics we shall
analyze the model with no explicit IR cutoff $L$ and simply restrict attention to times less than 
$T=L^\delta/A\delta,$ before the extremal solutions reach $|x|=L.$ In a realizable experiment, 
$L$ would be set by the size of the apparatus but here we let $x$ range over the real line.
The details of the velocity field (or even its meaningfulness) for $|x|>L$ do not matter for the effects 
we study and the only requirement is that $L\gg \ell.$


These toy models may be interpreted physically in terms of a Brownian tracer particle
in a 1D ``turbulent'' flow, where $x$ represents particle position, $t$ is time, $D$ is molecular diffusivity,
$v(x)$ is fluid velocity, $\ell$ is viscous dissipation length, and $L$ is a large-scale outer length. The exponent
$0<h<1$ gives the velocity H\"older exponent in the ``inertial range''  of the flow for $\ell<|x|<L,$ with $h=1/3$
the Kolmogorov-Onsager value. The physical dimensions of the various quantities appearing are 
\be \begin{array}{ll} 
       [x]=length & [t]=time \cr
       [D]=\frac{(length)^2}{time} & [A]=\frac{(length)^\delta}{time} \cr 
       [\ell]= length & [L]=length
       \end{array} \ee   
In defining dimensionless parameter groups it is useful to adopt the fluid-mechanical interpretation.
We thus introduce a ``kinematic viscosity'' $\nu$ with dimensions $(length)^2/time$ via
\be  \nu:= A \ell^{1+h} \ee
and an ``integral-scale velocity'' $V=AL^h,$ from which we construct the dimensionless parameters
\be Re:= \frac{VL}{\nu} = \left(\frac{L}{\ell}\right)^{1+h}, \quad  Pe:= \frac{VL}{D}= \frac{A L^{1+h}}{D}, \ee 
which we refer to as ``Reynolds number'' $Re$ and ``P\'eclet number'' $Pe.$ The first number is a dimensionless
measure of the singularity of the velocity, while the second number is a dimensionless measure of the weakness
of noise. We may also introduce their ratio 
\be Sc:=\frac{\nu}{D} = \frac{Pe}{Re} = \frac{A\ell^{1+h}}{D} \ee
the ``Schmidt number'' $Sc,$ which measures the importance of noise-weakness relative to velocity-singularity. 
In a physical fluid both molecular transport coefficients $\nu,\,D$ are thermodynamic functions only of temperature and 
pressure, and the second coefficient depends also upon molecular properties of the solute and solvent, such as their 
molar masses. Thus, in fluid turbulence experiments, $Re$ and $Pe$ are usually made large by increasing $V$ 
or especially $L,$ while $Sc$ remains nearly constant. Of course, in the context of our mathematical toy model, 
an arbitrary dependence of $Sc$ upon $Re$ may be assumed. 

Dimensionless variables can be introduced in a macroscopic or ``inertial-range'' description as 
\be \bar{x} = x/L,   \quad \bar{t} = t/(L/V),  \quad \bar{v}=v/V, \ee 
and the resulting non-dimensionalized equations are 
\be \frac{d\bar{x}}{d\bar{t}} = \bar{v}_{Re}(\bar{x}) + \sqrt{\frac{2}{Pe}}\,\eta(\bar{t}), \quad \bar{x}(0)=\bar{x}_0. 
\lb{langevin-int} \ee
The velocity field in these variables becomes Reynolds-dependent, or 
explicitly for the linear model:  
\be \bar{v}_{Re}(\bar{x}) = \left\{ \begin{array}{ll}  
                          \bar{x} Re^{\frac{1-h}{1+h}} &  |\bar{x}|\leq\bar{\ell}:=Re^{-\frac{1}{1+h}},  \cr
                           {\rm sign}(\bar{x})|\bar{x}|^h    &   \bar{\ell}\leq |\bar{x}|\leq 1 
                          \end{array} \right. \ee 
In the limit $Re\to\infty,$ $\bar{v}_{Re}(\bar{x})\to v_*(\bar{x}):=\sign(\bar{x})|\bar{x}|^h.$ 
The problem to be solved in these variables is to consider the limit $Re,\, Pe\to\infty$ with 
random initial data selected from a distribution 
\be \bar{P}^{Re}_0(\bar{x}_{0}) := \frac{1}{\bar{\ell}} P_0(\bar{x}_{0}/\bar{\ell}) \to \delta(\bar{x}_{0}) \quad 
\mbox{as $Re\to\infty$} \ee 
and with all other quantities held fixed.  

Another non-dimensionalization corresponds to a microscopic or ``dissipation-range'' description  
\be  \hat{x}=x/\ell\to x,   \quad \hat{t}=t/(\ell/\upsilon_\nu)\to t,  \quad \hat{v}=v/\upsilon_\nu\to v, \ee
where $\upsilon_\nu := A\ell^h$ is the ``dissipation-range velocity''. We take these variables as the 
default, without superscript, expressed in terms of which the dynamics becomes 
\be \frac{dx}{dt} = v(x) + \sqrt{\frac{2}{Sc}}\, \eta(t), \quad x(0)=x_{0}
\lb{langevin-dis} \ee
In this picture the velocity field is $Re$-independent except for the extent of its power-law 
``inertial range'', as indicated explicitly for the linear regularization: 
\be v(x) = \left\{ \begin{array}{ll}  
                          x &  |x|\leq 1 \cr
                          {\rm sign}(x)|x|^h    &   1 \leq |x|\leq L:=Re^{1/(1+h)}
                         \end{array} \lb{dis-vel} \right. \ee 
The two sets of dimensionless variables are related by 
\be x=  Re^{\frac{1}{1+h}}\,\bar{x}, \quad t = Re^{\frac{1-h}{1+h}}\, \bar{t} , 
\lb{relate} \ee
implying that  
\be \bar{x}_{Re}(\bar{t}) = Re^{\frac{-1}{1+h}} x\left(Re^{\frac{1-h}{1+h}} \bar{t}\right). \lb{int-dis} \ee
The limit $Re\to\infty$ with inertial-range variables $\bar{x},$ $\bar{t}$ fixed thus becomes 
in dissipation-range variables a  {\it long-time, large-distance limit} $x\to\infty,$ $t\to\infty$
for some specified probability distribution $P_0(x_0)$ of initial data.  
It should be noted here that the extremal solutions \eqref{extrm} are invariant under the 
change of variables \eqref{int-dis}, so that $\bar{x}^\pm(\bar{t})=x^\pm(t).$

\subsection{The RG Flow and Fixed Points}

To investigate the long-time, large-distance behavior of the solutions $x(t)$ of the dissipation-range
stochastic dynamics \eqref{langevin-dis} we shall appeal to the analogy with critical phenomena
discussed in the Introduction. Recall that in near-critical equilibrium lattice systems with spacing $a$ there 
is scaling behavior at length-scales $r$ in the range $a\ll r\ll \xi=$ the correlation length, 
which is universal within a broad class of microscopic Hamiltonians ${\mathcal H}$. This universal behavior 
is described by effective Hamiltonians ${\mathcal H}_b$ of Kadanoff block-spin variables over $b$ lattice sites,
obtained by integrating out the microscopic degrees of freedom between scales $a$ and $ba.$ When rescaled 
from lattice constant $ba$ back to $a,$ these effective Hamiltonians approach an RG fixed point ${\mathcal H}_*$
for $b\gg 1$ \cite{wilson1974renormalization,wilson1975renormalization,goldenfeld2018lectures}. In classical 
dynamics with spontaneous stochasticity there is expected to be a similar universality of the long-time statistics, which 
should be independent of the specific initial data and stochastic perturbations of the dynamics, within broad universality 
classes. This issue can be studied with path-integrals such as \eqref{OM-PI}. Because the early-time 
statistics and dynamics are supposed to be irrelevant, one can integrate out the pre-histories before 
some arbitrary time-window $t_w,$ altering the ``bare'' Onsager-Machlup action $S[\xi]$ to some 
new window-dependent effective action $\Gamma_{t_w}[\xi].$  Varying the window size as $bt_w$ and then rescaling 
back to $t_w$ yields a $b$-dependent effective-action $\Gamma_{t_w,b}[\xi].$ This effective action 
can be expected to flow for $b\gg 1$ to a scale-invariant fixed-point $\Gamma_{t_w,*}[\xi]$ which will describe 
the spontaneous statistics within some broad universality class. 

Carrying out this program for general classes of stochastic perturbations will require methods of functional 
renormalization group \cite{delamotte2012introduction,canet2011general}. However, for the specific white-in-time 
perturbations in the Langevin equation \eqref{langevin-dis} the stochastic dynamics is Markovian and the 
single-time probability density function $P(x,t)$ satisfies a closed partial differential equation, the Fokker-Planck equation
\cite{risken2012fokker} with Markov operator $\hat{L}$: 
\be \partial_t P = -\frac{\partial}{\partial x} (v(x) P) + \frac{1}{Sc} \frac{\partial^2}{\partial x^2} P:=\hat{L}P.  \lb{FPeq} \ee
This permits a more elementary approach, because ``integrating out'' the early times $t<t_w$ corresponds 
simply to solving this equation with initial distribution $P_0(x)$ over the time interval $0<t<t_w$. 
By varying the time-window as $bt_w,$ the goal in this simplified RG approach is to find new effective 
initial data and effective stochastic dynamics for $b\gg 1.$ Analogous to ``block-spins'' on  a rescaled lattice 
with fixed spacing $a,$ we may define the random ``effective initial data'' at fixed time $t=t_w$ as 
\be x_w^{(b)}:= b^{-1/\delta} x(bt_w). \lb{effect-init} \ee
The multiplicative field-renormalization' $b^{-1/\delta}$ is here motivated by the 
scaling symmetry \eqref{scal-sym} of the singular deterministic dynamics and we 
shall see that it guarantees that fixed point distributions of the ``block-spin" variables 
$x_w^{(b)}$ exist as $b\to\infty.$ Denoting the probability density of $x_w^{(b)}$ 
as $Q_b(x_w,t_w),$ it follows directly from the definition (\ref{effect-init}) that 
\be Q_b(x_w,t_w) = b^{1/\delta} P\left(b^{1/\delta} x_w,b t_w\right), 
\quad b\geq 1 \lb{Q-def} \ee
where $P(x,t)$ is the solution at time $t$ of the Fokker-Planck equation \eqref{FPeq}. 
This formulation of the RG method is very similar to that used previously to study self-similar 
solutions, fronts and blow-ups of nonlinear PDE's \cite{goldenfeld2018lectures,goldenfeld1990anomalous,
goldenfeld1991asymptotics,bricmont1995renormalizing}. Using \eqref{Q-def} and the formula \eqref{int-dis}, 
one can express the probability  
density of solution $\bar{x}_{Re}(\bar{t})$ of the inertial-range stochastic dynamics \eqref{langevin-int} as 
\begin{eqnarray}
 \bar{P}_{Re}(\bar{x},\bar{t}) &=& \left(\frac{t_w}{\bar{t}}\right)^{1/\delta} 
Q_b\left(\left(\frac{t_w}{\bar{t}}\right)^{1/\delta}\bar{x},\, t_w\right), \cr
&& \hspace{50pt} b=Re^{\frac{1-h}{1+h}} \frac{\bar{t}}{t_w} 
\lb{large-Re-b} \end{eqnarray} 
This key ``bridging relation'' directly connects the large-$Re$ limiting behavior of the inertial-range 
dynamics to the large-$b$ asymptotics of the RG flow. 

The RG flow equation for $Q_b(x_w,t_w)$ is easily derived from \eqref{Q-def} 
for $b\geq 1$ to be 
\be
b\partial_b Q_b(x_w,t_w) =-\frac{\partial}{\partial x_w} 
\left( w_b(x_w)Q_b\right) 
+ \frac{t_w}{Sc \cdot b^{\frac{1+h}{1-h}}} \frac{\partial^2 Q_b}{\partial x_w^2}, 
\lb{Qb-flow} 
\ee 
with an {\it RG-modified drift} 
\be w_b(x_w):=t_w v_b(x_w) - \frac{1}{\delta} x_w \lb{RG-drift} \ee 
for 
\be v_b(x):=b^{-h/\delta} v(b^{1/\delta}x). \ee
This equation must be solved with the initial data $Q_1(x_w,t_w)=P(x_w,t_w).$ Although this equation 
is non-autonomous in $b,$ it can be straightforwardly rewritten in autonomous form as 
\begin{eqnarray} 
b\partial_b Q_b &=& -\frac{\partial}{\partial x_w} \big(
w_b Q_b\big)
+ \varepsilon_b \frac{\partial^2 Q_b}{\partial x_w^2}, \cr 
b\partial_b v_b &=&  \frac{1}{\delta}\left(-h+x_w\frac{\partial}{\partial x_w} \right)v_b \cr  
b\partial_b \varepsilon_b &=& -\frac{1+h}{1-h}\varepsilon_b 
\lb{RG-auto} \end{eqnarray} 
with additional initial data $v_1(x_w)=v(x_w),\, \varepsilon_1=t_w/Sc.$ This flow equation
makes clear that the effective dynamics $v_b$ and effective noise $\varepsilon_b$ change under renormalization,
as well as the effective initial data $Q_b.$

Note that the $(v_b,\varepsilon_b)$ sector of the RG flow is decoupled from the $Q_b$ sector. 
With our assumptions on the regularized velocity $v(x)$
\be \lim_{b\to\infty} (v_b(x),\varepsilon_b)= (v_*(x),0). \ee 
The limiting behavior of $Q_b$ is also easy to infer from its RG flow equation. Since the effective 
noise vanishes as $b\to\infty,$ the probability $Q_b$ in the limit is conserved along the characteristics 
of the limiting hyperbolic PDE. In fact, the RG flow equation for $Q_b$ is equivalent to 
a Langevin equation for the random realization $x_w(\tau):=x_w^{(b)}$  evolving under 
the ``RG-time'' $\tau=\ln b \geq 0$ by
\be \frac{dx_w}{d\tau} = w(x_w,\tau) 
+ \sqrt{2\varepsilon(\tau)}\, \eta(\tau), \quad x_w(0)=x_w. \lb{langevin-dis-RG} \ee
It is thus clear that as $\tau\to\infty$ the probability must be concentrated on the 
stable equilibria $x_w^{eq}$ of the modified flow, satisfying $\lim_{\tau\to\infty} w(x_w^{eq},\tau)=0,$ 
and in our model there are exactly three such points
\be x_w^o=0, \  \ x_w^+=(\delta t_w)^{1/\delta},\  \ x_w^+=-(\delta t_w)^{1/\delta}. \ee 
The origin is always strongly linearly unstable for sufficiently large RG time $\tau$ since
\be  w'(0,\tau)= e^\tau v'(0)-\frac{1}{\delta} \ee 
and under our assumptions $v'(0)>0.$ Hence, any stable RG fixed point $Q_*$
will satisfy $Q_*(0)=0.$ One might worry that an initial $Q_1(x_w)$ with a part 
$\propto \delta(x_w)$ will have that part conserved in time under the RG flow. However, recall 
that $Q_1(x_w)=P(x_w,t_w)$ and, even if the initial distribution $P_0(x_0)$ contained 
a piece $\propto \delta(x_0)$, it is diffused away by integrating over any small window-time $t_w>0.$ 
We give a careful analytic proof that $Q_*(0)=0$ in the following section \ref{approach},
where we consider the rate of approach to the RG fixed points. The other two equilibrium points 
$x_w^\pm$ are the stable attractors of the RG drift and, in fact,
\be w'(x_w^\pm,\tau)=-1 \ee 
independent of $\tau.$ All of the probability $Q_b$ thus flows as $b=e^\tau\to\infty$ into 
the two equilibria $x_w^\pm$. Their large-$\tau$ stability is due, obviously, to the  additional 
linear term $-x_w/\delta$ that appears in the RG drift \eqref{RG-drift}. 

The conclusion of this analysis is that the stable RG fixed points must have the form 
\be Q_*^{(p)}(x_w)= p \delta(x_w-(\delta t_w)^{1/\delta})     
+ (1-p) \delta(x_w+(\delta t_w)^{1/\delta}),  \lb{fxpt-p}   \ee   
for some $p$ in the range $0\leq p\leq 1$ or, equivalently, in terms of random realizations 
\be \lim_{\tau\to\infty} x_w(\tau) = x_w^\pm \quad \mbox{with probabilities $p$, $1-p$.} 
\lb{fxpt-p2} \ee 
In section \ref{domains} we shall completely characterize the domains of attractions 
of these RG fixed points, which are non-empty for each $p$ and govern the long-time statistics 
of suitable initial distributions $P_0(x_0),$ velocity regularizations $v(x)$ and Schmidt 
numbers $Sc$. This result is trivial for $Sc=\infty,$ since in that case the 
integrals 
\begin{eqnarray}
p&=&\int_0^\infty dx_w \, Q_b(x_w,t_w)=\int_0^\infty dx_0 \, P_0(x_0), \cr
1-p&=&\int_{-\infty}^0 dx_w \, Q_b(x_w,t_w)=\int_{-\infty}^0 dx_0 \, P_0(x_0)\cr
&& 
\lb{LR-prob} \end{eqnarray}  
are invariants of the RG flow. 
Such continuous lines of RG fixed points occur in some equilibrium statistical mechanics 
models, such as the eight-vertex model, and are associated there with a breakdown 
of universality, in the sense that critical scaling exponents become continuous functions    
of the interaction parameters in the microscopic Hamiltonian \cite{kadanoff1971some,
van1975singularities}. There is a comparable breakdown of universality in the present 
case since the probabilities $p,$ $1-p$ asymptotically assigned to $x_w^\pm$ depend 
upon fine details of  $P_0(x_0),$ $v(x)$ and $Sc$ (see section \ref{domains}). This 
non-universality is associated with the lack of strong ``chaotic'' properties for 
the limiting deterministic dynamics, where every orbit has zero Lyapunov exponent except 
for the unstable fixed point at $x=0$ which has infinite Lyapunov exponent.  There is 
robust universality only for the case of symmetric initial distributions satisfying $P_0(-x_0)=P_0(x_0)$
and symmetric regularized velocities satisfying $v(-x)=-v(x).$ These symmetry properties are 
preserved by the stochastic dynamics \eqref{langevin-dis} and by the RG flow \eqref{RG-auto} 
so that, within this class, there remains only the symmetric fixed point  
\be Q_*(x_w,t_w) = \frac{1}{2} \delta(x_w-(\delta t_w)^{1/\delta})     
+ \frac{1}{2} \delta(x_w+(\delta t_w)^{1/\delta}) \lb{fxpt}  \ee
with equal probabilities $p=1-p=1/2$. 

The convergence of the RG flow to a stable fixed point $Q_*^{(p)}$ yields the limiting statistics 
of time-histories. Inverting the definition (\ref{effect-init}) of the ``effective initial data'' for $b=t/t_w$ gives 
\be x(t) = (t/t_w)^{1/\delta} x_w^{(t/t_w)}  \lb{invert} \ee
and thus \eqref{fxpt-p2} implies that as $t\to\infty$
\be x(t) \sim x^\pm(t)=\pm (\delta t)^{1/\delta}  \quad \mbox{with probabilities $p$,$1-p$.} \ee 
We can then obtain also the $Re\to\infty$ limit of $\bar{x}_{Re}(\bar{t})$ in the inertial-range
formulation of the problem by means of the scaling relation (\ref{int-dis}), which yields as $Re\to\infty$
\be \bar{x}_{Re}(\bar{t}) \sim \bar{x}^\pm(\bar{t}) =\pm (\delta \bar{t})^{1/\delta}  \quad 
\mbox{with probabilities $p$,$1-p$.} \ee 
In particular, all of these conclusions hold with $p=1/2$ if $P_0(x_0)=\delta(x_0)$, i.e. if $x(0)=0$ with probability 1.
It follows that these models exhibit spontaneous stochasticity of a very simple type, with exactly 
two random ``ground states'' selected from the continuum of solutions \eqref{nonuniq} of the 
singular, deterministic ODE. Just as for the idealized model \eqref{vstar-D-eq} previously 
considered in the probability theory literature 
\cite{bafico1982small,gradinaru2001singular,attanasio2009zero,flandoli2013topics},
which in our language has vanishing Schmidt number $Sc=0,$ the solutions selected 
in the zero-noise limit are the two extremal solutions. It is worth mentioning, however, 
that this result is regularization-dependent and relies upon our assumption that 
$v'(x)\geq 0.$ If instead the regularization of the velocity makes the origin very strongly 
stable, then it is possible that the equilibrium $x_w^o=0$ of the RG drift persists 
with positive probability in the limit $\tau\to\infty$. (We thank A. Mailybaev for this observation.) 
Because it is instructive, we show in Appendix \ref{stable0} explicitly how the origin 
may be stabilized by such an ``unnatural'' regularization. In that case, the spontaneously 
stochastic limit can include with positive probability the identically zero solution $x^\infty(t)\equiv 0$  
(infinite delay) in addition to the extremal solutions $x^\pm(t)$ with zero delay. 

\subsection{Approach to the Fixed Point and Singular Large Deviations}\lb{approach} 

The preceding discussion has identified the limiting statistical distribution of solutions of the singular 
ODE \eqref{vstar-eq} obtained in the ``critical limit''  $Re\to\infty$, associated to an RG fixed point.
A general question of interest is corrections to critical scaling which may be obtained from the RG flow near 
the fixed point, e.g. the low-temperature behavior of the 1D Ising model near its analogous $T=0$ 
critical point \cite{nelson1975soluble}. The limiting statistics in the 1D models we consider have great 
simplicity, because all histories other than the two extremal solutions  having vanishing probabilities 
in the limit and become rare events. Thus, the finite $Re$-corrections that describe these rare 
fluctuations have the form of probabilistic large-deviations results. This issue was first studied 
for the idealized $Sc=0$ version of the model in \cite{gradinaru2001singular}, which derived 
``singular large deviations" estimates that differ in scaling from those in the Freidlin-Wentzell theory. 
 We here derive similar results for our more general models using the RG approach.  

\subsubsection{Deductions from the RG Flow}

The relation \eqref{Q-def} connects large-$b$ asymptotics of $Q_b(x_w,t_w)$ with 
large-$x,$ large-$t$ asymptotics of the Fokker-Planck solution $P(x,t)$, where exponential decay 
is generally expected. We shall rigorously justify this exponential decay analytically in 
the following section, but here we simply make the ansatz of exponential decay
\be Q_b(x_w,t_w) \sim \exp\left[- b\, \phi(x_w,t_w) \right], \quad b\gg 1.  
\lb{approach2} \ee
at leading-order and then derive consequences from the RG flow \eqref{Qb-flow}. 
We expect that $\phi(x_w,t_w)>0$ for $|x_w|\neq (\delta t_w)^{1/\delta}$ and $\phi(x_w,t_w)=0$ for  
$|x_w|=(\delta t_w)^{1/\delta},$ to be consistent with the limiting fixed-point behavior 
\eqref{fxpt-p},\eqref{fxpt-p2}. It is also plausible that $\phi(x_w,t_w)$ is monotonically decreasing in $|x_w|$
for $|x_w|<(\delta t_w)^{1/\delta}$ and increasing in $|x_w|$ for $|x_w|>(\delta t_w)^{1/\delta}$. 
If the ansatz (\ref{approach2}) is substituted into the RG flow equation (\ref{Qb-flow}), it yields
\begin{eqnarray}
-b\phi(x_w,t_w) &=& -\left(t_w v_b'(x_w)-\frac{1}{\delta}\right) \cr 
                   \qquad &&  + b\left(t_w v_b(x_w)-\frac{x_w}{\delta}\right)\phi'(x_w,t_w) \cr
                   \qquad && + \frac{t_w}{Sc\cdot b^{\frac{1+h}{1-h}}}\left[ b^2(\phi'(x_w,t_w))^2-b \phi''(x_w,t_w)\right]. \cr
                   &&
\lb{phi-bal}                    
\end{eqnarray}    
Since $0<h<1$ implies that $2-\frac{1+h}{1-h} = \frac{1-3h}{1-h} <1$, the dominant balance in (\ref{phi-bal}) 
as $b\to\infty$ is between the terms proportional to $b$ so that $\phi(x_w,t_w)$ must satisfy 
\be  \left(t_w v_*(x_w)-\frac{x_w}{\delta}\right)\phi'(x_w,t_w)=-\phi(x_w,t_w). \lb{phi-eq} \ee 
This equation is obviously consistent with the conjecture that 
$\phi(x_w,t_w)$ is positive and decreasing in $|x_w|$ for $|x_w|<(\delta t_w)^{1/\delta}.$ Furthermore,
straightforward integration of the ODE (\ref{phi-eq}) starting from $x_w=0$ yields
\be \ln \phi(x_w,t_w) =\ln\left(1-\frac{|x_w|^\delta}{t_w\delta}\right) + C \ee
with $C=\ln \phi(0,t_w)$ an unknown constant, or, equivalently, 
\be   \phi(x_w,t_w) =\epsilon_0 \left(t_w-\frac{|x_w|^\delta}{\delta}\right) \lb{phi} \ee 
with $\epsilon_0=\phi(0,t_w)/t_w>0.$ This functional form is quite general, independent of $Sc,$ 
the velocity regularization $v(x)$ (so long as $Q_*(0,t_w)=0$), and the initial distribution $P_0(x_0).$ 
We shall see in section \ref{expansatz} that the value of prefactor $\epsilon_0$ does depend upon 
some of these details.  

These large-$b$ asymptotics are directly transformed to large-$Re$ asymptotics in 
the inertial-range description by means of the ``bridging relation'' \eqref{large-Re-b}. 
Substituting the ansatz \eqref{approach2}, one obtains a local large-deviations result 
\be \bar{P}_{Re}(\bar{x},\bar{t}) \sim 
\exp\left[-Re^{\frac{1-h}{1+h}} \cdot \bar{\Phi}(\bar{x},\bar{t}) \right], \quad |\bar{x}|<(\delta \bar{t})^{1//\delta},  \lb{Grad-int} \ee
for $Re\gg 1,$ with the specific {\it action function}  
\begin{eqnarray}
 \bar{\Phi}(\bar{x},\bar{t}) &:=& \left(\frac{\bar{t}}{t_w}\right)  \phi\left( \left(\frac{t_w}{\bar{t}}\right)^{1/\delta} \bar{x} ,t_w\right) \cr
 &=& \epsilon_0 \left(\bar{t}-\frac{|\bar{x}|^\delta}{\delta}\right), 
\lb{Phi} 
\end{eqnarray} 
where the second line uses \eqref{phi}. It should be noted that the equation (\ref{phi-eq}) determining $\phi(x_w,t_w)$ 
is a ``quasi-steady'' Hamilton-Jacobi equation:
\be H(x_w,\phi'(x_w,t_w)) = -\phi(x_w,t_w), \lb{quasi-HJ-sing} \ee
with a zero-noise Freidlin-Wentzell Hamiltonian \cite{freidlin2012random} for the RG drift \eqref{RG-drift}: 
\be H(x_w,p_w) =  w_*(x_w)p_w=\left(t_w v_*(x_w)-\frac{x_w}{\delta}\right)p_w.  \lb{FWH-RG-zero} \ee 
By combining (\ref{quasi-HJ-sing}) and the first line of (\ref{Phi}) it is straightforward to see that the action 
$\bar{\Phi}$ likewise satisfies a time-dependent Hamilton-Jacobi equation 
\be  \frac{\partial\bar{\Phi}(\bar{x},\bar{t})}{\partial \bar{t}} = 
-\bar{H}\left(\bar{x},  \frac{\partial\bar{\Phi}(\bar{x},\bar{t})}{\partial \bar{x}}   \right) 
\lb{HJeq-sing} \ee
with zero-noise Freidlin-Wentzell Hamiltonian 
\be \bar{H}(\bar{x},\bar{p}) = v_*(\bar{x})\bar{p} \lb{FW-sing} \ee
for the singular velocity $v_*(x)$ in the limit equation \eqref{vstar-eq}. 

The $Re$-scaling in \eqref{Grad-int} is ``anomalous" with respect to standard Freidlin-Wentzell 
asymptotics  \cite{freidlin2012random} for the weak-noise limits of Langevin equations, 
like the inertial-range model \eqref{langevin-int}.  The standard predictions would follow 
from the corresponding path-integral
\be \bar{P}(\bar{x},\bar{t}|\bar{x}_0,0) =  \int_{\bar{\xi}(0)=\bar{x}_0} \calD \bar{\xi}\, 
\delta(\bar{x}-\bar{\xi}(\bar{t})) e^{-Re\cdot \bar{S}[\bar{\xi}]}. 
\lb{OM-PI-int} \ee 
with the inertial-range Onsager-Machlup action 
\be \bar{S}_{Re}[\bar{\xi}]=\frac{Sc}{4} \int_{\bar{t}_0}^{\bar{t}} d\bar{s}\, |\dot{\bar{\xi}}(\bar{s})-\bar{v}_{Re}(\bar{\xi}(\bar{s}))|^2 \lb{OM-int} \ee
if naively evaluated by the Laplace method for $Re\gg 1.$ This would lead one to expect instead that  
\be \bar{P}_{Re}(\bar{x},\bar{t}) \sim 
\exp\left[-Re \cdot \bar{\Psi}(\bar{x},\bar{t}) \right] 
\lb{Grad-ext} \ee
for $Re\gg1,$ with a different $Re$-scaling than in \eqref{Grad-int} and with the single-time action function 
\be \bar{\Psi}(\bar{x},\bar{t})=\inf_{\bar{\xi}: {\scriptsize \begin{array}{l}
                                                                    \bar{\xi}(\bar{t})=\bar{x} \cr
                                                                    \bar{\xi}(0)=0
                                                                    \end{array}}} \bar{S}[\bar{\xi}]. \ee 
Equivalently, $\bar{\Psi}$ would naively be expected to be the solution of the time-dependent 
Hamilton-Jacobi equation
\be  \frac{\partial\bar{\Psi}(\bar{x},\bar{t})}{\partial \bar{t}} = 
-\bar{H}\left(\bar{x},  \frac{\partial\bar{\Psi}(\bar{x},\bar{t})}{\partial \bar{x}}   \right) 
\lb{HJeq-stand} \ee
with inertial-range Freidlin-Wentzell Hamiltonian 
\be \bar{H}(\bar{x},\bar{p}) = v_*(\bar{x})\bar{p} +  \frac{1}{Sc} \bar{p}^2. 
\lb{FW-stand} \ee
which is Legendre dual to the Onsager-Machlup Lagrangian, whereas the correct Hamiltonian 
in \eqref{FW-sing} corresponds to vanishing noise. The Laplace method to evaluate 
\eqref{OM-PI-int} breaks down because $\bar{v}_{Re}(\bar{x})$ in the action \eqref{OM-int} 
is $Re$-dependent and tends to the singular velocity $v_*(\bar{x})$ as $Re\to\infty,$ so that 
the limiting action has non-unique ``ground-states" $\xi_*$ satisfying $S_*[\xi_*]=0.$ Because 
$\frac{1-h}{1+h}<1$ the probabilities of rare events decay more slowly with $Re$ than would 
be naively expected \footnote{The exponent $\frac{1-h}{1+h}\neq 1$ is not, however, an ``anomalous dimension''
in the sense of critical phenomenon and quantum-field theory. As discussed in section \ref{dimension},
the factor $Re^{\frac{1-h}{1+h}}$ appears in the ``bridging relation'' \eqref{large-Re-b} because of 
simple dimensional analysis.}. 

The non-standard large-deviations prediction \eqref{Grad-int} can, however, be valid only for 
$|\bar{x}|<(\delta \bar{t})^{1//\delta},$ since the explicit expression in the second line of 
\eqref{Phi} shows that $\bar{\Phi}(\bar{x},\bar{t})<0$ for $|\bar{x}|>(\delta \bar{t})^{1//\delta}.$
Likewise, the basic exponential-decay ansatz (\ref{approach2}) cannot be valid for $|x_w|>(\delta t_w)^{1/\delta}$ because 
it leads to the conclusion that $\phi(x_w,t_w)<0.$ The assumption of a stretched-exponential 
decay with $b\to b^\sigma$ leads to similar inconsistency for $\sigma<\frac{1+h}{1-h},$ but
the ansatz with $\sigma=\frac{1+h}{1-h},$ or explicitly 
\be Q_b(x_w,t_w) \sim \exp\left[- b^{\frac{1+h}{1-h}} \psi(x_w,t_w) \right],  \quad |x_w|>(\delta t_w)^{1/\delta}, 
\lb{approach3} \ee   
for $b\gg 1,$ leads to a consistent picture.   
Indeed, if the ansatz (\ref{approach3}) is substituted into the RG flow equation (\ref{Qb-flow}), it yields
\begin{eqnarray}
&& -\left(\frac{1+h}{1-h}\right) b^{\frac{1+h}{1-h}}\psi(x_w,t_w) 
= -\left(t_w v_b'(x_w)-\frac{1}{\delta}\right) \cr
&&\qquad + b^{\frac{1+h}{1-h}}\left(t_w v_b(x_w)-\frac{x_w}{\delta}\right)\psi'(x_w,t_w) \cr
&& \qquad + \frac{t_w}{Sc}b^{\frac{1+h}{1-h}}\left[ (\psi'(x_w,t_w))^2-  \frac{1}{b^{\frac{1+h}{1-h}}}\psi''(x_w,t_w)\right]. \cr
&& 
\lb{psi-bal}                    
\end{eqnarray}    
The new dominant balance in (\ref{psi-bal}) as $b\to\infty$ is between the terms proportional to 
$b^{\frac{1+h}{1-h}}$ so that $\psi(x_w,t_w)$ must satisfy the equation
\begin{eqnarray}
&& \left(t_w v_*(x_w)-\frac{x_w}{\delta}\right)\psi'(x_w,t_w)  +  \frac{t_w}{Sc} (\psi'(x_w,t_w))^2 \cr
&& \qquad \qquad =-\left(\frac{1+h}{1-h}\right)\psi(x_w,t_w). \lb{psi-eq} \end{eqnarray} 
Note that this equation is consistent with the requirement that $\psi(x_w,t_w)$ is positive and increasing in $|x_w|$ 
for $|x_w|>(\delta t_w)^{1/\delta}.$ 

Furthermore, note that the equation (\ref{psi-eq}) determining the action function $\psi$ is a ``quasi-steady'' Hamilton-Jacobi equation
\be H(x_w,\psi'(x_w,t_w)) = -\left(\frac{1+h}{1-h}\right)\psi(x_w,t_w). \lb{quasi-HJ} \ee
with the standard Freidlin-Wentzell Hamiltonian appropriate to the RG Langevin model \eqref{langevin-dis-RG}: 
\be H(x_w,p_w) =  \left(t_w v_*(x_w)-\frac{x_w}{\delta}\right)p_w +  \frac{t_w}{Sc} p^2_w. \lb{FWH-RG} \ee 
In contrast to \eqref{FWH-RG-zero}, the noise contributes now a quadratic term $\propto p^2_w.$ 
The intuition is that effects of singularity are so strong for $|x_w|<(\delta t_w)^{1/\delta}$ that 
external noise does not influence the probabilities of rare events, but the dynamics is smooth 
for $|x_w|>(\delta t_w)^{1/\delta}$ and external noise reasserts its role in producing rare fluctuations. 
In fact, the modified ansatz \eqref{approach3} in conjunction with the ``bridging relation'' 
\eqref{large-Re-b} implies a standard large-deviations result of the form \eqref{Grad-ext}
with the expected $Re$-scaling and with the specific action function 
\be \bar{\Psi}(\bar{x},\bar{t}) = \left(\frac{\bar{t}}{t_w}\right)^{\frac{1+h}{1-h}} 
\psi\left( \left(\frac{t_w}{\bar{t}}\right)^{1/\delta} \bar{x} \right). 
\lb{Psi} \ee
It is furthermore straightforward using (\ref{quasi-HJ}) to show that the action $\bar{\Psi}$ in 
\eqref{Psi} satisfies the time-dependent Hamilton-Jacobi equation \eqref{HJeq-stand} with the 
inertial-range Freidlin-Wentzell Hamiltonian \eqref{FW-stand}. We therefore conclude that standard 
Freidlin-Wentzell large-deviations theory must hold for the space-time region 
$|\bar{x}|>(\delta \bar{t})^{1//\delta}$ exterior to the extremal solutions. This conclusion is entirely 
plausible, because optimal histories that produce the rare events in that region must depart the 
origin even faster than do the extremal solutions, through the agency of external noise. 

Our results from analysis of the RG flow equations are completely consistent with the 
``singular large-deviations'' theory of \cite{gradinaru2001singular} for the idealized special 
case with $Sc=0.$ That paper rigorously obtained large-deviations results corresponding 
to our anomalous/singular estimate \eqref{Grad-int} in the interior region and our standard/regular 
estimate \eqref{Grad-ext} in the exterior region. The analysis of \cite{gradinaru2001singular} showed 
further that the prefactor $\epsilon_0$ in the large-deviations rate function $\bar{\Phi}$ 
in \eqref{Phi} is the ground-state energy eigenvalue of a certain Schr\"odinger operator.  In 
the following section we show for our more physical models with $Sc>0$ that the constant prefactors $\epsilon_0$ are
likewise ground-state energies of suitable Schr\"odinger operators, using a more elementary 
argument than that employed in \cite{gradinaru2001singular} and recovering their results in the 
limit $Sc\to 0.$

\subsubsection{Derivation of the Exponential-Decay Ansatz}\lb{expansatz} 

We here derive carefully the exponential-decay ansatz \eqref{approach2} for the distribution $Q_b(x_w,t_w)$ of 
effective initial data in the 
interior region $|x_w|<(\delta t_w)^{1/\delta}$, which is linked by \eqref{Q-def} to the decay of 
$P\left(b^{1/\delta} x_w,b t_w\right)$ for $b\to\infty.$ We make a direct analysis of the long-time, 
large-distance limit of the Fokker-Planck solution by using a standard method of orthogonal eigenfunction 
expansions. See \cite{risken2012fokker}, section 5.4. As discussed there, the transformation
\be P(x,t)= e^{Sc \,V(x)/2} \varPsi(x,t)  \ee
for any anti-derivative $V(x)$ of the velocity $v(x)$ transforms the Fokker-Planck equation 
\eqref{FPeq} into 
the imaginary-time Schr\"odinger equation
\be \partial_t\varPsi=-\hat{H}\varPsi:= -\left[-\frac{1}{Sc} \frac{\partial^2}{\partial x^2}+U(x)\right]\varPsi 
\lb{Scheq}\ee
with potential
\be U(x) = \frac{Sc}{4}(v(x))^2 + \frac{1}{2}v'(x). \lb{Udef} \ee
Note that $U(x)\geq 0$ when $v'(x)\geq 0$ and that outside the regularization region 
\be U(x)=  \frac{Sc}{4}|x|^{2h} + \frac{h}{2|x|^\delta}, \quad |x|>1. \lb{Ustar} \ee
Setting $Sc=1,$ the potential \eqref{Udef} thus coincides with the singular potential $U_*(x)$ of 
\cite{gradinaru2001singular} for $|x|>1$ (up to a trivial change of notations \footnote{The equation (2) of \cite{gradinaru2001singular}
differs from our inertial-range Langevin equation \eqref{langevin-int} by some simple factors of 2. Setting $x\to x/2^{1+h},$ 
$t\to t/2^{\frac{1-h}{1+h}}$ in (2) of \cite{gradinaru2001singular} gives our \eqref{langevin-int}, with their 
$\varepsilon$ related to our P\'eclet number by $\varepsilon^2=1/Pe.$ By this change of variables, all 
results of \cite{gradinaru2001singular} are transformed into the $Sc\to 0$ limit of ours.})
but differs from their potential in the regularized region $|x|<1.$ 

Because $U(x)\geq 0,$ the eigenvalues $\epsilon_n,$ $n=0,1,2,...$ of the 1-particle Hamiltonian 
$\hat{H}$ are all positive, with corresponding eigenfunctions  
\be \hat{H}\varPsi_n=\epsilon_n \varPsi_n \ee
related to the eigenfunctions of $\hat{L}$ and $\hat{L}^*$ 
\be \hat{L}\varPhi_n=-\epsilon_n \varPhi_n, \quad \hat{L}^*\bar{\varPhi}_n=-\epsilon_n \bar{\varPhi}_n
\ee
by the formulas 
\be \varPhi_n(x)= e^{+Sc \,V(x)/2} \varPsi_n(x), \quad \bar{\varPhi}_n(x)= e^{-Sc \,V(x)/2} \varPsi_n(x).  
\lb{Phi-Psi} \ee
The transition probability density can be expanded in these eigenfunctions as 
\be
 P(x,t|x_0,0) = \sum_{n=0}^\infty e^{-\epsilon_n t} \varPhi_n(x) \bar{\varPhi}_n(x_0) 
\lb{expand}
\ee  
See \cite{risken2012fokker}, section 5.4, for all of these results. 

From the expansion (\ref{expand}) one can see that the solution $P(x,t)$ of the Fokker-Planck equation 
(\ref{FPeq}) for any initial data $P_0(x_0)$ is given by the series
\be
 P(x,t) =\sum_{n=0}^\infty e^{-\epsilon_n t} \varPhi_n(x) \langle \bar{\varPhi}_n,P_0\rangle \lb{P-expand} \ee 
 which converges uniformly on compact sets of $x$ for all $t>0.$ This result suffices to give large-time
 asymptotics of $P(x,t)$ for fixed $x,$ but to study $Q_b(x_w,t_w)$ for $b\to\infty$ we need as well a result 
 on the large-distance asymptotics of the eigenfunctions $\Phi_n(x),$ as follows:
 \be \Phi_n(x)\sim \exp\left(+\epsilon_n \frac{|x|^\delta}{\delta}\right), \quad |x|\to \infty. \ee  
 We briefly sketch here the proof. Since the derivation is independent of $n,$ we suppress that index 
 in the following. Starting with the eigenvalue equation 
\be -\frac{1}{Sc} \frac{\partial^2\varPsi}{\partial x^2}+U(x)\varPsi=\epsilon\varPsi \ee
we substitute 
\be \varPsi(x) =\exp(S(x)) \ee 
to obtain the equation 
\be -\frac{1}{Sc} \left[(S'(x))^2+S''(x)\right]+U(x)=\epsilon. \lb{S-lam} \ee
For $|x|\to \infty$ the dominant balance is between the pair of terms involving $S'(x)$
and $U(x)\sim \frac{Sc}{4}|x|^{2h},$ yielding the asymptotic result 
\be S'(x) \sim \pm \frac{Sc}{2}|x|^h \  \ \Longrightarrow \ \ S(x)\sim \pm \frac{Sc}{2(1+h)}|x|^{1+h}. \ee  
Only the negative sign is consistent with square-integrability $\int dx \, |\varPsi(x)|^2<\infty,$
so we conclude that 
\be \varPsi(x) \sim \exp\left(-\frac{Sc}{2(1+h)}|x|^{1+h}\right), \quad |x|\to\infty.  \lb{Psi-asympt} \ee
However, in order to determine the asymptotics of $\varPhi(x)$ from that of $\varPsi(x)$ via (\ref{Phi-Psi}) 
the leading-order asymptotic formula in  (\ref{Psi-asympt}) is  inadequate,  since 
\be \frac{Sc}{2}V(x)\sim \frac{Sc}{2(1+h)}\,|x|^{1+h} \ee 
and the leading-order term is exactly cancelled. Thus, we must determine a 
correction:  
\be S(x)= -\frac{Sc}{2(1+h)}|x|^{1+h} + S_1(x). \ee
Substituting into (\ref{S-lam}) gives  
\be |x|^hS_1'(x) -\frac{1}{Sc}[(S_1(x))^2+S''(x)] + U_1(x) = \epsilon \ee
with $U_1(x) =h/2|x|^\delta$ from (\ref{Ustar}).  The dominant balance here 
is between the term linear in $S_1'(x)$ and the constant $\epsilon,$ so that
\be S_1'(x) \sim \epsilon |x|^{-h} \ \ \Longrightarrow \ \ S_1(x)\sim \epsilon \frac{|x|^\delta}{\delta}, \quad |x|\to\infty,  \ee
yielding as claimed that 
\be  \varPhi(x)\sim \exp\left(\epsilon \frac{|x|^\delta}{\delta}\right), \quad |x|\to\infty. \ee 

We thus obtain from the eigenfunction expansion \eqref{P-expand} an 
exact series representation for the effective distribution and also an asymptotic formula for 
large-$b$: 
\begin{eqnarray} 
Q_b(x_w,t_w) &=&  b^{1/\delta} \sum_{n=0}^\infty e^{-\epsilon_n bt_w} 
\varPhi_n(b^{1/\delta} x_w) \langle \bar{\varPhi}_n,P_0\rangle \cr 
&\sim& b^{1/\delta} \sum_{n=0}^\infty 
\exp\left[-b \epsilon_n\left(t_{w}-\frac{|x_{w}|^\delta}{\delta}\right)\right] 
\langle \bar{\varPhi}_n,P_0 \rangle, \cr
&& \qquad \qquad \qquad \qquad b\to\infty. 
\lb{Qb-series} \end{eqnarray}  
The series converges for all $b$ uniformly on compact subsets of the interior region 
$|x_w|<(\delta t_w)^{1/\delta}$ and the first term in the series \eqref{Qb-series} dominates 
as $b\to\infty$ whenever $\langle \bar{\varPhi}_0,P_0 \rangle\neq 0.$  Recall that the ground 
state wavefunction $\varPsi_0$ of the 1-particle Hamiltonian has no nodes and can always be 
chosen positive, and it has two more derivatives than the potential $U(x)$ defined in \eqref{Udef} 
from the regularized velocity \cite{gulisashvili1996exact}. Furthermore, from \eqref{Psi-asympt}, $\varPsi_0$ decays 
rapidly for $|x|\to\infty.$ The corresponding eigenfunction $\bar{\varPhi}_0$ of $\hat{L}^*$
inherits all of these properties of $\varPsi_0$ via the formula \eqref{Phi-Psi} and, therefore,
$0<\langle \bar{\varPhi}_0,P_0 \rangle<\infty$ for all possible choices of $P_0.$ We conclude 
that as $b\to\infty$
\be
Q_b(x_w,t_w) 
\sim b^{1/\delta}\exp\left[-b\epsilon_0\left(t_{w}-\frac{|x_{w}|^\delta}{\delta}\right)\right] 
\langle \bar{\varPhi}_0,P_0 \rangle, \lb{Qb-largeb} \ee 
in the region $|x_{w}|<(\delta t_{w})^{1/\delta},$ thereby establishing the exponential ansatz \eqref{approach2}
and also identifying $\epsilon_0$ as the ground-state energy of the 1-particle Hamiltonian defined in 
\eqref{Scheq}-\eqref{Udef}\footnote{An attentive reader will have noticed that this same argument 
derives directly the singular large-deviations estimate \eqref{Grad-int}, without any use of the RG
flow equations. In fact, the relation \eqref{int-dis} between inertial and dissipation range variables 
implies the relation of their PDF's 
$\bar{P}_{Re}(\bar{x},\bar{t}) = Re^{\frac{1}{1+h}} P( Re^{\frac{1}{1+h}}\bar{x},Re^{\frac{1-h}{1+h}}\bar{t})$
and this result may be use to estimate $\bar{P}_{Re}(\bar{x},\bar{t})$ as $Re\to\infty$ in the same 
manner as \eqref{Q-def} was used to derive an estimate for $Q_b(x_w,t_w)$ as $b\to\infty.$}

\subsubsection{Discussion of the Results}\lb{discussion} 

The conclusion of our argument is that the approach to fixed points $Q_*^{(p)}(x_w,t_w)$ 
is the same for all $p$, independent of $P_0$ and with identical functional form for all $Sc$ 
and for all regularized velocities $v(x)$ satisfying $v'(x)\geq 0.$ However, the prefactor $\epsilon_0$
depends upon $Sc$ and $v(x)$ through the potential $U(x)$ in \eqref{Udef}. It is quite reasonable 
that the rate of approach to the limiting statistics as $Re\to\infty$ should be regularization-dependent 
in the interior region $|\bar{x}|<(\delta\bar{t})^{1/\delta},$ since the histories that contribute 
to such rare events must have lingered in the regularization region $|\bar{x}|<\bar{\ell},$ with a macroscopic 
time-delay near the origin 
\be \bar{\tau}:=\bar{t}-\frac{|\bar{x}|^\delta}{\delta} \ee 
before departing along the corresponding non-extremal solution trajectory defined in \eqref{nonuniq}. 
This argument provides a physical interpretation of the action function \eqref{Phi} for singular large-deviations 
as $\bar{\Phi}=\epsilon_0\bar{\tau},$ with rate function directly proportional to the time-delay $\bar{\tau}$. 
The random noise plays no direct role in such large-deviations, because these rare events are 
mediated by the non-unique deterministic solutions.  

The prior analysis in \cite{gradinaru2001singular} for the idealized $Sc=0$ model made also 
implicit regularizations at $x=0$ \footnote{For example, the Tanaka formula that was used in 
\cite{gradinaru2001singular} for the It$\bar{\rm o}$-differential $d|B(t)|$ of the absolute value 
of a Wiener process $B(t)$ is derived by a regularization such as $|x|\to \frac{x^2}{2\ell}+\frac{\ell}{2}$
for $|x|<\ell$ with $\ell\to 0$}. It is therefore useful to discuss in more detail the relation of our 
work with that in \cite{gradinaru2001singular} and to discuss possible regularization-dependence
of their results. Since they took $\ell=0$ from the outset, the ``dissipation range'' non-dimensionalization 
that we adopted in section \ref{dimension} is unavailable. The corresponding non-dimensionalization
for the $Sc=0$ model is in ``diffusive range'' variables based on the diffusive length-scale $\ell_D=(D/A)^{\frac{1}{1+h}}$
and time-scale $t_D=\ell_D^2/D,$ and the only dimensionless group is the ``P\'eclet number'' $Pe.$
The diffusive-range non-dimensionalization of the $Sc=0$ model yields a Langevin equation
of the same form as our dissipation-range equation \eqref{langevin-dis} with $Sc\mapsto 1,$ while the inertial-range 
non-dimensionalization remains the same as in our section \ref{dimension} and yields the Langevin equation 
\eqref{langevin-int} with $\bar{v}_{Re}(\bar{x})\mapsto \bar{v}_*(\bar{x}).$ The macroscopic weak-noise limit is now 
$Pe\to\infty$ and, in our language, \cite{gradinaru2001singular} obtain large-deviations results 
in the parameter $Pe.$ To formally derive their results as a special case of ours, we must transform
our formulas in ``dissipation-range'' $x$ instead to ``diffusion-range'' $\hat{x}$ using $\hat{x}=Sc^{\frac{1}{1+h}}x,$
since $\ell/\ell_D=Sc^{\frac{1}{1+h}},$ and then take $Sc\to 0.$ With this transformation, one easily obtains 
from our 1-particle Hamiltonian 
\begin{eqnarray}
&&-\frac{1}{Sc} \frac{\partial^2}{\partial x^2}+\frac{Sc}{4}(v(x))^2 + \frac{1}{2}v'(x) \cr
&& \sim Sc^{\frac{1-h}{1+h}} \left[-\frac{\partial^2}{\partial \hat{x}^2}+\frac{1}{4}|\hat{x}|^{2h} + \frac{h}{2|\hat{x}|^\delta} \right] 
\lb{smallSc-Ham} 
\end{eqnarray} 
in the idealized limit $Sc\to 0.$ Note that the Hamiltonian in the square bracket coincides with that in 
\cite{gradinaru2001singular} (see footnote [46]). We thus obtain the relationship of our ground-state eigenvalue 
$\epsilon_0=\epsilon_0(Sc)$ with the corresponding eigenvalue of 
\cite{gradinaru2001singular}, $\hat{\epsilon}_0$, as 
\be \epsilon_0(Sc) \sim Sc^{\frac{1-h}{1+h}} \hat{\epsilon}_0\ee 
for $Sc\to 0.$ As long as $Sc\ll 1$ while $Pe=Re\,Sc\gg 1$ then our singular large-deviation estimates in 
$Re$ yield the large-deviations results of \cite{gradinaru2001singular} in $Pe$ to good accuracy and show 
how to realize those predictions in a physical manner. We shall not attempt here to make this $Sc\to 0$ 
limit rigorous, but our formal argument does suggest that the results of \cite{gradinaru2001singular} 
should hold for all natural regularizations $v(x)$ satisfying $v'(x)\geq 0$ in the limit $\ell\to 0.$

The asymptotics of the opposite case $Sc\gg 1$ are a bit more subtle. This is a singular limit in which 
the kinetic term of the 1-particle Hamiltonian \eqref{Scheq} formally vanishes and one expects a sort of 
``boundary-layer''  of thickness $1/\sqrt{Sc}$. Making the change to the stretched variable $y=\sqrt{Sc}\,x$ in the 
dissipation-range Langevin model \eqref{langevin-dis} yields the modified equation 
\begin{eqnarray} 
\dot{y} &=& \sqrt{Sc}\ v\left(\frac{y}{\sqrt{Sc}}\right) + \sqrt{2} \eta(t) \cr
             &\doteq & \gamma y + \sqrt{2}\, \eta(t), \qquad Sc\gg 1 \lb{Sc-large} 
\end{eqnarray}        
for the variable $y$ fixed and $\gamma:=v'(0)>0.$ In the limit $Sc\to \infty,$ one thus obtains the 
Langevin model with unstable linear drift corresponding to an inverted parabolic potential $-\frac{\gamma}{2}y^2$
(see \cite{risken2012fokker}, \S 5.5.2). This has the form of our starting model (\ref{vstar-D-eq}) for 
$A\mapsto \gamma,$ $h\mapsto 1,$ $D\mapsto 1.$  Although we have shown that spontaneous stochasticity 
appears for any finite value of $Sc,$ however large, this $Sc=\infty $ limiting equation has smooth exponent 
$h=1$ and thus no spontaneous stochasticity. In fact, the transition probability is given explicitly for our 
model (\ref{vstar-D-eq}) with $A\mapsto \gamma,$ $h\mapsto 1$ by \cite{risken2012fokker}, Eq.(5.70) as: 
 \be P(x,t|x_0,0)=\sqrt{\frac{\gamma}{2\pi D(e^{2\gamma t} -1)}}
\exp\left[-\frac{\gamma (x-e^{\gamma t}x_0)^2}{2D(e^{2\gamma t}-1)}\right],  \lb{invert-parab} \ee
a Gaussian unimodal distribution centered at the unique deterministic solution $x=e^{\gamma t} x_0$ 
for all times $t,$ with position width $\ \sim (D/\gamma)^{1/2} e^{\gamma t}$ also growing exponentially.\\
Applied to the linear model \eqref{Sc-large}, the result \eqref{invert-parab} implies that for any initial 
distribution $P_0(y_0)$ with position spread $\sim 1$ 
\be P(y,t) \simeq \sqrt{\frac{\gamma}{2\pi e^{2\gamma t}}} \exp\left[ -\frac{\gamma y^2}{2 e^{2\gamma t}}\right],
\quad t\gg 1.  \ee
There is no contradiction with our claim of spontaneous stochasticity for $Sc\gg 1,$ because the approximation 
\eqref{Sc-large} breaks down as soon as positions $|y|\gg \sqrt{Sc}$ are attained with appreciable 
probability and this occurs within a time of order $t\sim (1/2\gamma) \ln(Sc).$
 
Finally, we make a methodological remark about our RG analysis. In traditional RG calculations 
for spin systems, the asymptotic approach to a fixed point follows from the linearized RG flow 
near the fixed point \cite{wilson1974renormalization,wilson1975renormalization,goldenfeld2018lectures}. 
Similar linear analysis does not suffice to derive the corresponding results 
for our problem, such as the singular large-deviations \eqref{Grad-int}. In fact, the linearization of the 
autonomous RG flow \eqref{RG-auto} for small departures $(\tilde{Q}_b,\tilde{v}_b,\tilde{\varepsilon}_b)$ 
from any fixed point $(Q_*,v_*,0)$ has the form 
\begin{eqnarray} 
b\partial_b \tilde{Q}_b&=& -\partial_{x_w}\cdot \left( t_w  \tilde{v}_b(x_w) Q_*\right) \cr
&& -\partial_{x_w}\cdot \left[\left( t_w v_*(x_w) - \frac{1}{\delta} x_w\right)\tilde{Q}_b\right] \cr
&& + \tilde{\varepsilon}_b \triangle_{x_w}Q_*\cr
b\partial_b \tilde{v}_b&=& -\frac{h}{\delta} \, \tilde{v}_b(x_w)+ \frac{1}{\delta} (x_w\cdot\partial_{x_w})
\tilde{v}_b(x_w), \cr
b\partial_b \tilde{\varepsilon}_b &= &  -\frac{1+h}{1-h} \tilde{\varepsilon}_b 
\lb{lin-RG-flow} \end{eqnarray} 
The crucial observation is that the first term vanishes in the linearized flow equation for the 
perturbation $\tilde{Q}_b$, or  
\be \partial_{x_w}\cdot \left( t_w  \tilde{v}_b(x_w) Q_*\right) \equiv 0 \ee 
because $\tilde{v}_b(x_w)\equiv 0$ for $|x_w|>b^{-1/\delta}$ and $Q_*(x_w)$ 
is supported entirely in that region (at least when $x_w^o=0$ is made unstable by the regularization). 
Thus, the linearized RG equations for $(\tilde{Q}_b,\tilde{\varepsilon}_b)$ and $\tilde{v}_b$ are decoupled, 
and, in particular, the rate at which $\tilde{Q}_b\to 0$ as $b\to\infty$ must be independent of $\tilde{v}_b:=
v_b-v_*$ according to the linearized equations. We have seen, however, that the asymptotic approach 
of $Q_b$ to any fixed point $Q_*$ (and even which fixed point is attained) depends upon the precise form 
of $v_b$ in the regularization region where $\tilde{v}_b\neq 0.$ The linearized RG flow near each fixed 
point is thus insufficient to describe the approach to that fixed point. This is a fundamental difference from 
RG analyses of equilibrium critical systems, where the RG flow is generally well-described by an autonomous 
ODE in a finite-dimensional space of ``relevant variables''. For our problem, however, the RG flow is 
intrinsically in a space of functions $(Q_b(x_w),v_b(x_w),\varepsilon_b)$ and governed by the set of 
PDE's \eqref{RG-auto}. In that case, even the question whether $(Q_b,v_b,\varepsilon_b)$ is ``close'' 
to the fixed point $(Q_*,v_*,0)$ becomes subtle, because there are many inequivalent norms, and 
a global analysis of the RG flow equations is required to study the asymptotic approach. 

\subsection{Domains of Attraction}\lb{domains} 

Each of the RG fixed points $Q_*^{(p)}$ for $p\in [0,1]$ has its own domain of initial distributions 
$P_0$ that it attracts, which depends sensitively upon both velocity regularization $v(x)$
and Schmidt number $Sc.$ Fortunately, for this simple 1D model, we can completely 
characterize the domains of attraction {\it a priori}, as we show now.

Because the fixed points assign non-vanishing
probabilities to only the two states $x_w^-=-x_w^+,$ it is clearly sufficient to consider the probabilities 
to be right or left of the origin \footnote{There is nothing special about the origin for our argument
and we could consider equally well the probabilities of the two complementary events $x(t)>a$ 
and $x(t)<a,$ for any fixed real constant $a$. However, it is natural to choose $a=0.$}, which are defined as 
\begin{eqnarray}     
p_+(t|x,t_0) &=& P(x(t)>0|x(t_0)=x)\cr
                    &=&\int_0^\infty dx'\ p(x',t|x,t_0), 
\end{eqnarray}   
and                   
\begin{eqnarray}       
p_-(t|x,t_0) &=& P(x(t)<0|x(t_0)=x)\cr
                  &=&\int_{-\infty}^0 dx' \ p(x',t|x,t_0). 
\end{eqnarray}  
and to determine their long-time asymptotics for $t\gg t_0$. \\ Using the the fact that the transition 
probability $p(x',t'|x,t)$ satisfies the backward Kolmogorov equation in its initial space-time arguments 
(see \cite{risken2012fokker}, section 4.7)
\begin{eqnarray}
0&=&(\partial_t+\hat{L}_x^*)p(x',t'|x,t) \cr
  &=& \left(\partial_t + v(x)\frac{\partial}{\partial x} +\frac{1}{Sc} \frac{\partial^2}{\partial x^2}\right) p(x',t'|x,t) =0 \cr
 &&
\end{eqnarray} 
and using $p(x',t|x,t_0)=p(x',0|x,t_0-t)$ by time-homogeneity, it then follows that 
\be \partial_t p_\pm(t|,x,t_0)= 
\left(v(x)\frac{\partial}{\partial x} + \frac{1}{Sc} \frac{\partial^2}{\partial x^2}\right) p_\pm(t|x,t_0). \ee 
Since the adjoint Fokker-Planck operator $\hat{L}^*$ has all of its spectrum in the left
half of the complex plane, the long-time limits exist and are independent of $t_0$
\be   p_\pm(x)=\lim_{t\to \infty} p_\pm(t-t_0|x,0) = \lim_{t\to\infty} p_\pm(t|x,t_0),  \ee
satisfying the stationary backward equation 
\be  \hat{L}^*p_\pm(x)= \left(v(x)\frac{\partial}{\partial x} + \frac{1}{Sc} \frac{\partial^2}{\partial x^2}\right) 
p_\pm(x)=0\ \ee
subject to the boundary conditions $p_\pm(\mp \infty)=0.$  

The latter equations can be explicitly integrated by quadratures to give  
\be                 p_+(x)=  (1/N) \int_{-\infty}^x dy \ \exp(-Sc\  V(y))  \lb{p-plus} \ee
and 
\be                 p_-(x)=  (1/N) \int_x^{+\infty} dy \ \exp(-Sc\  V(y) )  \lb{p-minus}  \ee
where $V(x)$ as before is any anti-derivative of $v(x).$ For the velocity regularizations satisfying $v'(x)\geq 0$ 
that we consider, $p_0=\lim_{t\to \infty} P(x(t)=0)=0,$ so that the normalization $N$ is fixed by $p_+(x) + p_-(x)= 1$ 
to be 
\be  N =   \int_{-\infty}^{+\infty} dy\ \exp(-Sc\ V(y)). \lb{Nnorm}  \ee
Note that $p_\pm(x)$ are non-trivial {\it zero modes} of the backward Kolmogorov operator $\hat{L}^*,$
in addition to the trivial constant zero mode \footnote{These zero modes belong to the Banach space 
${\mathcal L}^\infty$ of bounded, measurable functions, where the Markov semigroup $e^{t\hat{L}^*}$ 
naturally acts.  None of these modes appears in the eigenfunction expansion (\ref{expand}), because 
the corresponding wavefunctions $\varPsi$ via the relation (\ref{Phi-Psi}) are not square-integrable.}. 
For any solution $P(x,t)$ of the Fokker-Planck equation, therefore, the integrals 
\be           p_\pm = \int dx \ p_\pm (x) P(x,t)      \lb{integrals}   \ee
are exact constants of the motion and give the asymptotic probabilities to be right or left 
of the origin. If $Sc\to\infty$ then $p_\pm(x)$ converge to the Heaviside step functions 
$\theta(\pm x),$ consistent with the fact 
that the initial values $p_+(0)=\int_0^\infty dx\ P_0(x),$  $p_-(0)=\int_{-\infty}^0 dx\ P_0(x)$ are 
conserved in time for $Sc=\infty$. 

The argument made above for the Fokker-Planck equation fails to carry over directly to the RG Fokker-Planck 
flow equation (\ref{Qb-flow}), because the latter is non-autonomous and also the integrals such 
as \eqref{Nnorm} diverge for the RG-drift \eqref{RG-drift}. Similar zero modes and associated 
conservation laws do not apparently exist for the RG flow.  It follows directly from definition (\ref{Q-def}) 
of the distribution of ``block-spins'' that
\begin{eqnarray} 
&&  \int_0^\infty dx_w \ Q_b(x_w,t_w) =  \int_0^\infty dx \ P(x,t) \cr 
&& =\int_{-\infty}^\infty dx\ p_+(t|x_0,0) P_0(x_0), \quad b=t/t_w 
\end{eqnarray}  
but these integrals all change with $b=t/t_w.$ On the other hand, after taking the joint limit $t,\,b\to\infty$
the above relation implies that
\be p=\int_0^\infty dx_w \ Q_*^{(p)}(x_w,t_w) =\int_{-\infty}^\infty dx\ p_+(x_0) P_0(x_0). \ee
Therefore, the fixed-point $Q_*^{(p)}$ that attracts any initial distribution $P_0$ is fully 
characterized by the constants of motion \eqref{integrals}, with $p=p_+,$ $1-p=p_-.$ 

The existence of these exact integrals of motion help to justify our RG strategy for this degenerate model, 
with its continuous line of fixed points. The RG method 
is most successful when statistics are {\it universal}, e.g. when the details of the microscopic
Hamiltonian are irrelevant to the long-distance critical scaling and simplified coarse-grained 
models belong to the same universality class. Our strategy of considering ``effective initial data''
after some arbitrary time window $t_w>t_0$ is premised on the idea that the early-time statistics 
and dynamics are similarly irrelevant and may be ignored. Although universality breaks 
down in this specfic 1D model, fortunately the true initial distribution $P_0(x_0,t_0)$ and the ``effective 
initial distribution''  $Q_1(x_w,t_w)=P(x_w,t_w)$ have the same integrals of motion \eqref{integrals}. 
One can thus infer that 
\begin{eqnarray} 
\int_0^\infty dx_w \ Q_*^{(p)}(x_w,t_w) &= & \int_{-\infty}^\infty dx\ p_+(x_w) P(x_w,t_w)\cr
&=& \int_{-\infty}^\infty dx\ p_+(x_w) Q_1(x_w,t_w) \cr
&& \end{eqnarray} 
even though the latter integral is not an invariant quantity under the RG flow. We are therefore justified 
in ignoring early times $t<t_w,$ since the universality class of the initial distribution $P_0$ is fully encoded
in $Q_1.$

It is only for the completely symmetric case, with $P_0(-x_0)=P_0(x_0)$ and $v(-x)=-v(x),$ that there 
is robust universality in this simple 1D problem, of the same sort encountered in most short-range, equilibrium
critical spin systems. For any symmetric velocity $v(x)$
\be p_-(-x)=p_+(x) \lb{p-sym} \ee
from \eqref{p-plus},\eqref{p-minus}, so that any symmetric initial distribution $P_0(x_0)$ then has $p=1-p=1/2.$ 
The symmetric RG fixed-point $Q_*=Q_*^{(1/2)}$ in \eqref{fxpt} governs the long-time 
dynamics for all initial distributions and velocity regularizations in this symmetric class. It is worth 
remarking, on the other hand, that the domain of attraction of the symmetric fixed point is larger 
than the class of symmetric data and for any $Sc<\infty$ there are non-symmetric $P_0$ and $v$ 
that flow into $(Q_*,v_*)$. We construct in Appendix \ref{symdomain} explicit examples of non-symmetric 
initial distributions $P_0$ with invariants $p_+=p_-=1/2,$ which thus belong to the basin of attraction of $Q_*.$ 

\section{Empirical Study of the 1D Model}\lb{numerics} 

The rather exhaustive analytical investigation of our model problem \eqref{model} in the preceding section 
yields detailed predictions that can be verified by direct numerical simulations. We present such 
numerical results here,  because they render concrete our rich set of mathematical results and 
also motivate some additional discussion of the theory.  Furthermore, it is worthwhile to explore
physical systems that could realize our 1D model in a controlled laboratory setting, even if such 
experiments do not correspond directly to any naturally occurring circumstances. The scarcity of 
direct experimental evidence for spontaneous stochasticity is vexing, especially considering 
its presumed ubiquity in Nature. 

\subsection{Possible Experimental Realizations} 

\begin{figure}[h!]
  \begin{center}
 \includegraphics[width=240pt]{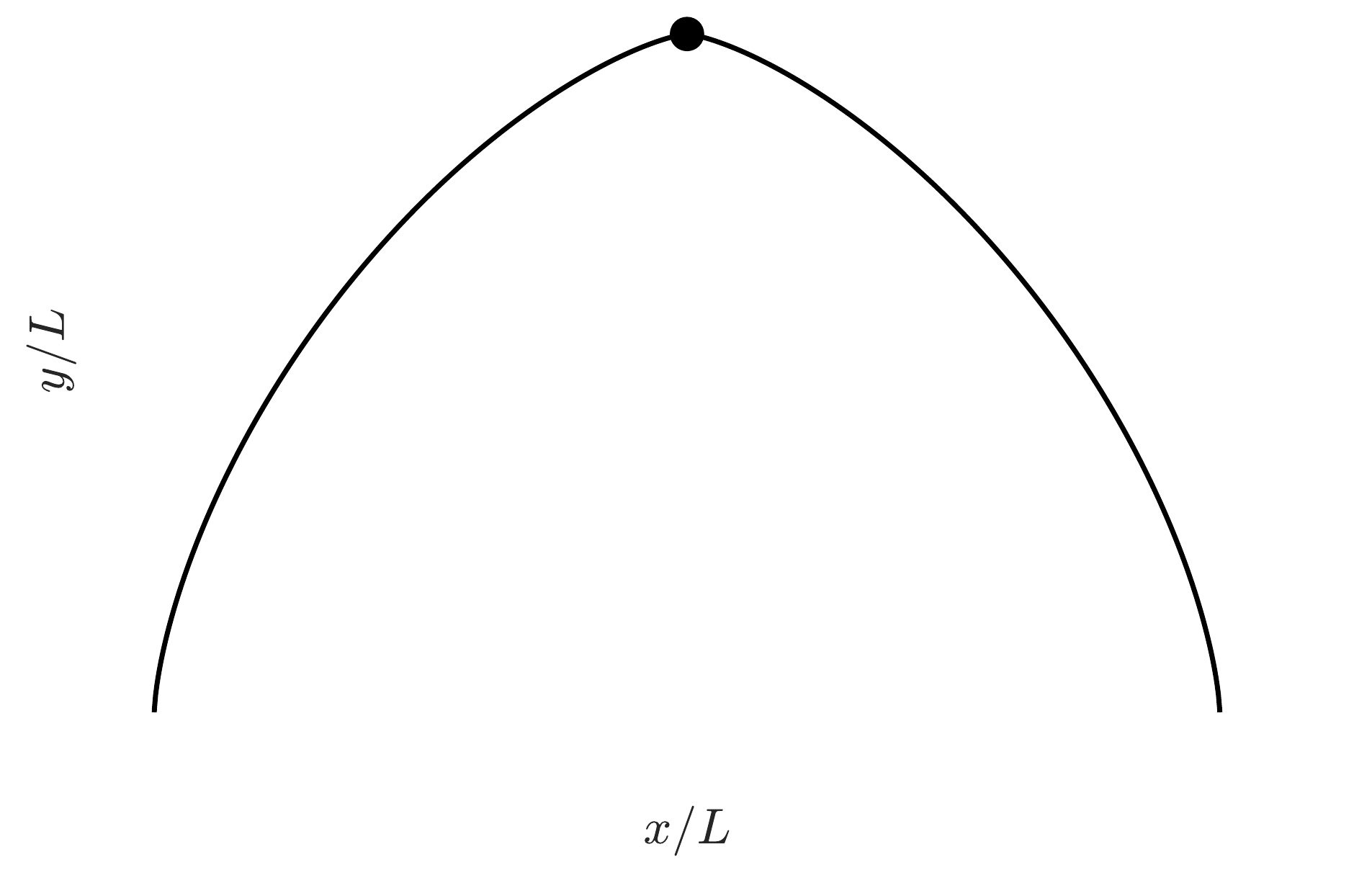}
 \end{center}
 \caption{Bead sliding with large friction on a curved wire with shape given by \eqref{fx-eq},
\eqref{bead-eq} for $h=1/3$.  The curve \\ in this plot is regularized only implicitly by the step-size\\
$\Delta s$ used in integrating the ODEs \eqref{fx-eq}. }\label{FigA} 
 \end{figure}

The simplest set-up to realize our 1D model \eqref{model} is a mechanical experiment using a 
bead sliding with strong friction on a bent thin wire, 
as pictured in Figure \ref{FigA}. The equation for arclength $s(t)$ as a function of time $t$ can be obtained 
by ignoring inertia $m\ddot{s}$ and balancing friction force $-\gamma \dot{s}$ and parallel component of 
gravitational force $-mg \, dy/ds.$ Note that the maximum velocity obtained when the wire is bent vertically 
is $v_*=mg/\gamma$ and any velocity field satisfying $|v(s)|\leq v_*$ can be achieved by choosing a wire 
in the shape $y=y(x)$ given implicitly by the equations 
\begin{eqnarray}
    y'(s)&=&\hat{v}(s) \cr
    x'(s)&=&\sqrt{1-\hat{v}^2(s)} 
\lb{fx-eq} \end{eqnarray}     
where $\hat{v}(s)=v(s)/v_*.$ Figure \ref{FigA} shows the shape required so that the equation 
\be \dot{s}= v_*\left|\frac{s}{L}\right|^h {\rm sign}(s),  \quad |s|<L \lb{bead-eq} \ee 
is satisfied for a wire of length $2L$ and $h=1/3.$  In any physical experiment this shape will be regularized 
somehow for $|s|<\ell,$ where the length scale $\ell$ will be no smaller than the radial diameter of the wire.  
A Langevin white-noise force $\eta(t)$ can be applied to the bead, for example by using commercially 
available white-noise voltage generators \cite{lin2011gaussian,schultz2018pocket} and a transducer device.  
In fact, one could add noise in many other ways (for example, randomly jiggling the wire) and,
as long as the noise is small, it should not matter for the spontaneous stochasticity phenomenon. 
A practical alternative
to experiments with a mechanical  apparatus would be an electric-circuit realization of the system 
\eqref{bead-eq}, based on the standard impedance analogy with mechanics. 

Large friction in a mechanical experiment (or large resistance in an electrical circuit) yields 
the first-order system \eqref{vstar-eq} only as a singular limit and, as well-known, there will 
generally be transient, early-time deviations (\cite{strogatz2018nonlinear}, section 3.5).  It may therefore
be desirable to realize instead the second-order Hamiltonian equation \eqref{2ndord-eq} with $A=mg/L^h$
by the opposite limit of negligible friction.  In this case, the equation for arclength $s(t)$ as a function of time $t$ 
is obtained by balancing inertia $m\ddot{s}$ and and parallel component of gravitational force $-mg \, dy/ds.$ 
Similar to before, the maximum acceleration obtained when the wire is bent vertically is $g$ and any 
acceleration field satisfying $|a(s)|\leq g$ can be achieved by choosing a wire in the shape $y=y(x)$ 
given by the equations \eqref{fx-eq} with $\hat{v}(s)\mapsto \hat{a}(s),$ where $\hat{a}(s):=a(s)/g.$ 
 Because the equations are identical in form, the wire shape pictured in Figure \ref{FigA} is also that 
 required to realize the second-order model \eqref{2ndord-eq} with $h=1/3,$ in the limit of low friction. 
 Similar mechanical models have, in fact, been invoked previously in the natural philosophy literature
 in discussions of causality and determinism for classical dynamics \cite{norton2003causation,vanpoucke2020assigning}.
 In particular, the equations \eqref{fx-eq} can be explicitly integrated for exponent $h=1/2,$ as 
 \begin{eqnarray}
 x(s) &=& \frac{2L}{3}\left[1-\left(1-\left|\frac{s}{L}\right|\right)^{3/2}\right]\sign(s),   \cr
 y(s) &=& -\frac{2L}{3}\left|\frac{s}{L}\right|^{3/2} 
 \end{eqnarray} 
with $|s|<L$ and the resulting curve has been called ``Norton's dome'' \cite{norton2003causation,vanpoucke2020assigning}.
However, these works by philosophers and also those by mathematical probabilists 
\cite{bafico1982small,gradinaru2001singular,attanasio2009zero,flandoli2013topics}
have not recognized the physical necessity in any real experiment of a regularization of 
the singularity in the dynamics \eqref{2ndord-eq} at some non-zero length-scale $\ell.$
In particular, uniqueness of solutions holds for any finite regularization and some external 
source of noise, such as a Langevin force $\sqrt{2D}\,\eta(t)$ added to \eqref{2ndord-eq}, 
is required to supply stochasticity, which persists in the joint limit $D\to 0,$ $\ell\to 0.$
Although we shall not consider here further this stochastic perturbation of the Hamiltonian
model \eqref{2ndord-eq}, spontaneous stochasticity does occur for this model in the joint limit of 
vanishing noise and regularization just as for the quantum version studied in \cite{eyink2015quantum}. 

\subsection{Numerical Simulations}

In lieu of laboratory experiments, we present here the results of numerical simulations of our model problem, 
for H\"older exponent $h=1/3.$ We solve the Langevin model using the simplest Euler-Maruyama 
discretization \cite{kloeden2013numerical}, employing either the dissipation-range formulation 
\eqref{langevin-dis} or the inertial-range formulation \eqref{langevin-int} for numerical integration, 
depending on the situation. Empirical averages were then constructed by averaging over $N$ independent samples.    
Full details on numerical methods and  convergence tests are described in Appendix \ref{num-meth}. 
We consider only the deterministic initial distribution $P_0(x_0)=\delta(x_0)$, which lies in the domain 
of attraction of the symmetric RG fixed-point \eqref{fxpt}.

\begin{figure*}[t]
  \centering
  \begin{subfigure}[b]{0.24\linewidth}
    \includegraphics[width=\linewidth]{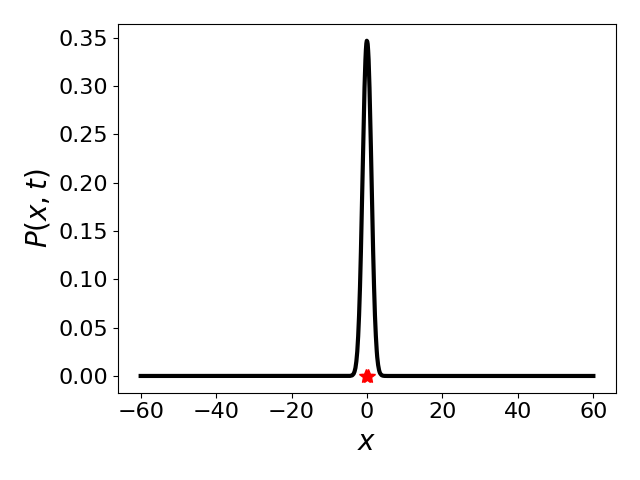}
    \caption{$t=0.5$}
  \end{subfigure}
  \begin{subfigure}[b]{0.24\linewidth}
    \includegraphics[width=\linewidth]{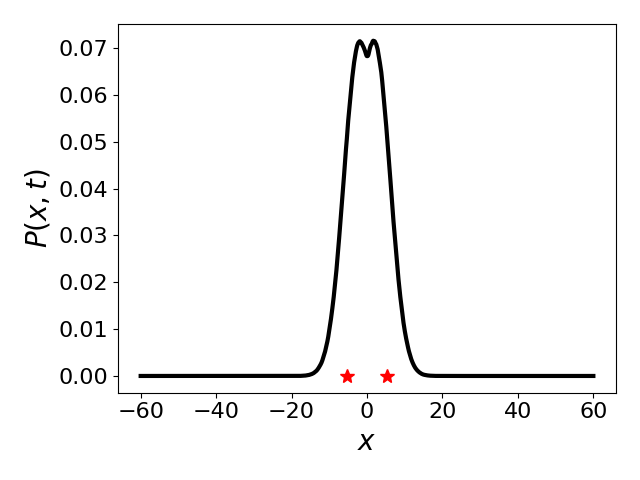}
    \caption{$t=3$}
  \end{subfigure}
    \begin{subfigure}[b]{0.24\linewidth}
    \includegraphics[width=\linewidth]{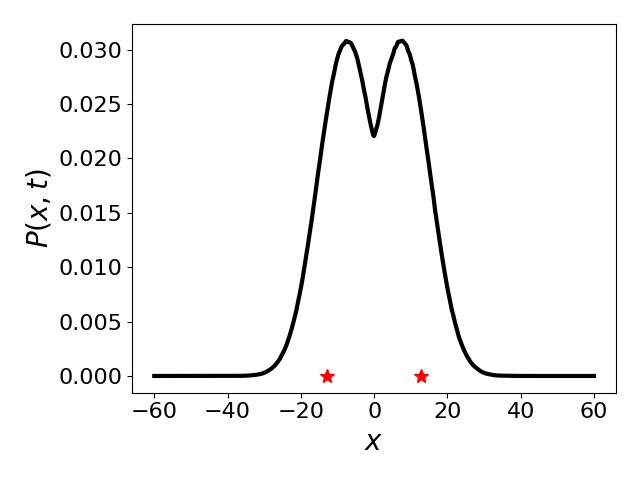}
    \caption{$t=5.5$}
  \end{subfigure}
    \begin{subfigure}[b]{0.24\linewidth}
    \includegraphics[width=\linewidth]{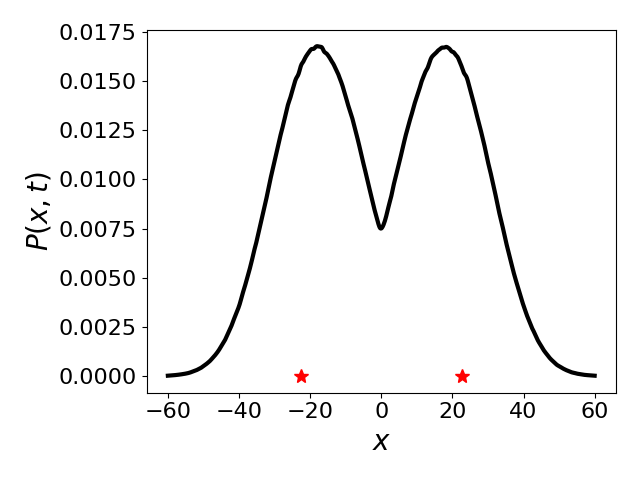}
    \caption{$t=8$}
  \end{subfigure}
   \caption{Probability density functions $P(x,t)$ for particle positions at four different times, all quantities in  dissipation-range scaling units. The asterisks indicate the positions of the two extremal solutions at the same four times.}
   \label{1.png} 
\end{figure*}

\begin{figure}[h!]
 \begin{center}
\includegraphics[width=240pt]{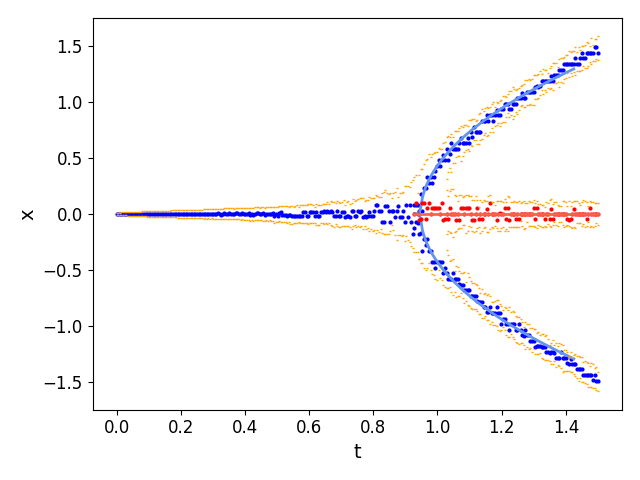}
 \end{center}
 \caption{Numerical bifurcation diagram, with local extrema of the empirical PDF $P(x,t)$ plotted versus time $t,$ local maximima in blue and local minima in red. The light orange markers indicate
 the region within which the values of the PDF vary by less than 0.1\% of its extremal value. The solid cyan line is a least mean square fit of a parabola to the outer prongs of the diagram, the good agreement being consistent with the normal form of a pitchfork bifurcation \cite{strogatz2018nonlinear}.}
 \label{6.png}
 \end{figure}

In Figure \ref{1.png} we plot the probability density functions $P(x,t)$ for particle positions in dissipation-range scaling, 
obtained for $Sc=1$ at times $t=0.5$, 3, 5.5, 8, starting at 
$x=0$ at time $t=0.$ The PDF's are close to Gaussian at early times,
but then flatten at time $t\doteq 0.95$ and thereafter become bimodal. The PDF's very near this splitting time are 
shown in Figure \ref{7.png} in Appendix \ref{num-meth}. The two peaks of the bimodal PDF's $P(x,t)$ approach
the extremal solutions $x_\pm(t)$ of the singular ODE \eqref{vstar-eq} as time advances. 
A plot of the local extrema (maxima and minima) of the PDF's in the $x$-$t-$plane,
Figure \ref{6.png}, shows the clear signature of a pitchfork bifurcation, as should be expected 
for the symmetry $x\leftrightarrow -x$
\cite{strogatz2018nonlinear}. In our simple model with just two extremal solutions selected 
in the zero-noise limit, this pitchfork bifurcation can be regarded as the ``onset'' 
of spontaneous stochasticity. 

 \begin{figure}[h!]
 \begin{center}
 \includegraphics[width=240pt]{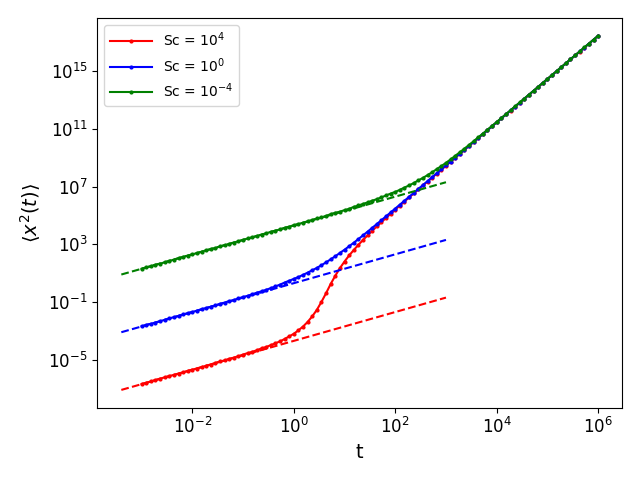}
 \end{center}
 \caption{Position variances $\langle x^2(t)\rangle$ plotted versus time $t$ in log-log, all quantities in dissipation-range scaled units, for three values of $Sc=10^{-4}$ (green), $Sc=1$ (blue), $Sc=10^{4}$ (red). The dashed lines plot the linear curves $(2/Sc)t$ with corresponding colors for each of the three 
 $Sc$ values. The final common line for all three $Sc$ values corresponds to a universal ``Richardson law'' $(2t/3)^3$}
 \label{5.png}
 \end{figure} 

\begin{figure*}[t]
  \centering
  \begin{subfigure}[b]{0.24\linewidth}
    \includegraphics[width=\linewidth]{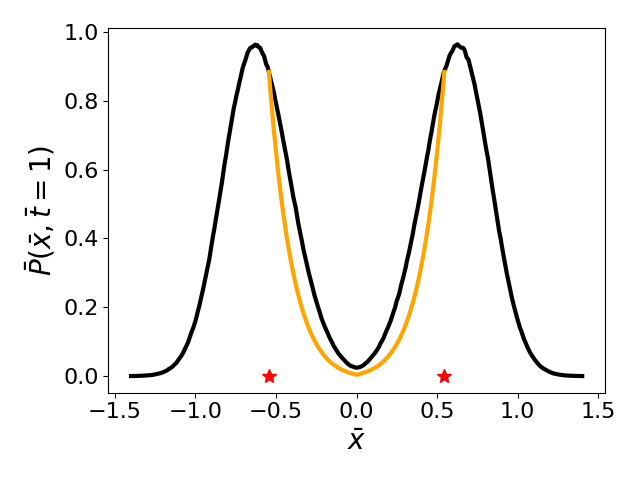}
    \caption{$Re = Pe =10^2 $}
  \end{subfigure}
  \begin{subfigure}[b]{0.24\linewidth}
    \includegraphics[width=\linewidth]{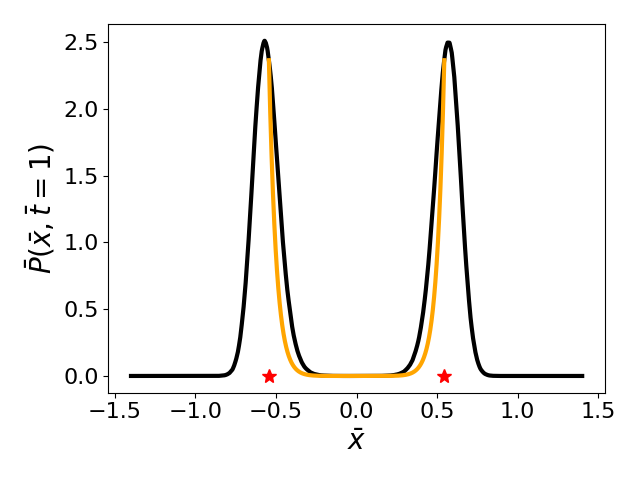}
    \caption{$Re = Pe =10^3 $}
  \end{subfigure}
    \begin{subfigure}[b]{0.24\linewidth}
    \includegraphics[width=\linewidth]{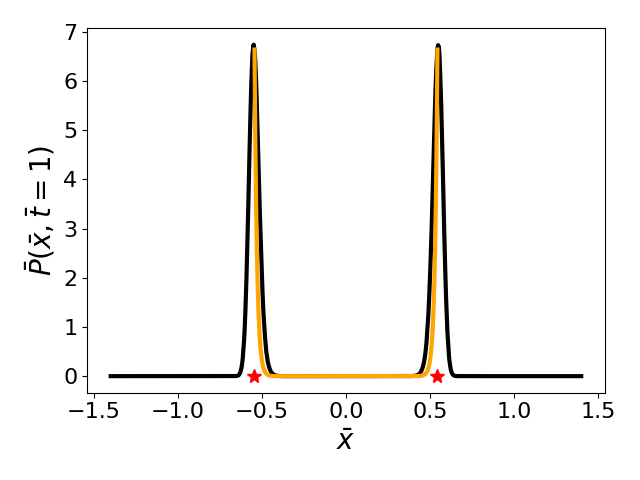}
    \caption{$Re = Pe =10^4 $}
  \end{subfigure}
    \begin{subfigure}[b]{0.24\linewidth}
    \includegraphics[width=\linewidth]{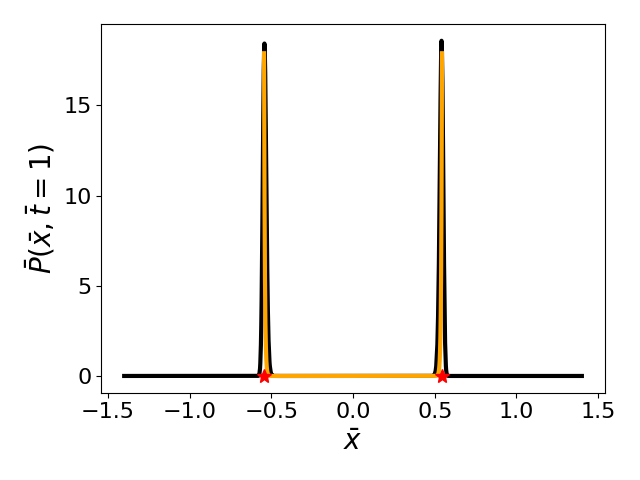}
    \caption{$Re = Pe =10^5 $}
  \end{subfigure}
   \caption{Probability density functions $\bar{P}(\bar{x},\bar{t})$ for particle positions at time $\bar{t} = 1$ for 
   four different values of $Re = Pe,$ all quantities in inertial-range scaling units. 
   The asterisks indicate the positions of the two extremal solutions.
   The orange line plots the singular large-deviations estimate given by (\ref{Grad-int}).}\label{3.png} 
\end{figure*}

Further insight can be obtained by plotting the variance $\langle x^2(t)\rangle=\int dx\, x^2\,P(x,t)$ versus time $t,$
in log-log coordinates.  For $Sc=1$ we obtain the data plotted with blue dots in Figure \ref{5.png},  
which shows as straight lines with distinct slopes the two power-laws 
\be \langle x^2(t)\rangle\sim \left\{\begin{array}{ll}
                                                       (2/Sc)t & t\lesssim t_c(Sc)   \cr 
                                                        (\delta t)^{2/\delta} & t\gtrsim t_c(Sc)
                                                      \end{array} \right. \ee   
for a crossover time $t_c(Sc)\sim (1/Sc)^{\frac{1-h}{1+h}},$ with $Sc\mapsto 1$. The same behavior is also observed 
for $Sc\ll 1$ as illustrated by the results for $Sc=10^{-4}$ plotted as green dots in Figure \ref{5.png}. The only effect of the 
increased noise is that the onset of spontaneous stochasticity is delayed. Consistent with our asymptotic analysis 
for $Sc\ll 1$ in section \ref{discussion}, this onset time is always $\hat{t}_c\sim 1$ when expressed in the diffusive scaling 
$\hat{t}=Sc^{\frac{1-h}{1+h}}t,$ as in \eqref{smallSc-Ham}. In the opposite limit $Sc\gg 1$ the linear model 
\eqref{Sc-large} predicts that the variance grows at early times instead as 
\be \langle x^2(t)\rangle\sim (1/\gamma Sc)(e^{2\gamma t}-1), \quad t\lesssim t_c(Sc)\sim \frac{1}{2\gamma}\ln Sc 
\lb{largeSc-var} \ee  
and then crosses over to the universal power-law $\sim (\delta t)^{2/\delta}$ at longer times. 
This behavior for $Sc\gg 1$ is verified with $\gamma=1$ by the data for $Sc=10^4$ plotted with red dots in Figure \ref{5.png}.
Note that \eqref{largeSc-var} yields also a linearly-growing variance $\sim (2/Sc)t$ at very early times $t\lesssim 1/2\gamma$ but 
an exponential growth $\sim (1/\gamma Sc) e^{2\gamma t}$ at the intermediate range of times 
$1/2\gamma\lesssim t\lesssim (1/2\gamma)\ln Sc.$ Both of these growth laws are a short 
transient behavior before the universal power-law scaling $\sim (\delta t)^{2/\delta}$ at long times,  
which is a key signature of spontaneous stochasticity.

With these results as backdrop and with the asymptotic analyses of section \ref{discussion} for 
$Sc\gg 1$ and $Sc\ll 1$ at hand, we now can explain the phase-boundaries plotted in Figure \ref{4.png} in the Introduction. 
We consider first the crossover between the ``deterministic phase'' D and the ``spontaneously stochastic phase'' SS. 
As is obvious from the ``bridging relation''  \eqref{large-Re-b},  inertial-range position variables 
$\bar{x}(\bar{t})$ will remain random in the limit $Re\to\infty,$ $Pe\to\infty$ only if the bifurcation 
in the dissipation-range PDF $P(x,t)$ occurs at a time less than or equal to $t\sim Re^{\frac{1-h}{1+h}}.$
(See also footnote [50].) By equating this time to $t_c\sim (1/2\gamma)\ln Sc$ for $Sc\gg 1,$ we obtain 
an approximate phase boundary   
\be \ln(Sc) \simeq c Re^{\frac{1-h}{1+h}} \ee 
for an order unity constant $c$ or, with $Pe=Sc\,Re,$
\be \ln(Pe) \simeq \ln(Re) + c_{1(2)} \exp(a \ln(Re)), \quad a= \frac{1-h}{1+h} \lb{D-SS} \ee 
These are the boundary lines plotted in Figure \ref{4.png} with the constant $c_{1(2)}$ chosen to provide 
a least-squares fit to the $\Upsilon-$isoline for the value  
$0.85\Upsilon_*$ (0.15$\Upsilon_*$), with $\Upsilon_*$ the 
fixed-point value. 
For the other crossover between the ``noise-driven phase'' N and the ``spontaneously stochastic phase'' SS, 
we note from the asymptotic analysis for $Sc\ll 1$ in section \ref{discussion} that spontaneous stochasticity 
will be observed as long as $Pe\gg 1,$  no matter how small $Sc$ may be. Thus, we obtain now as an 
approximate phase-boundary 
\be \ln(Pe) \simeq \ln(Pe_c):=c_{3(4)} \lb{N-SS} \ee 
where $Pe_c$ can be regarded as a ``transitional P\'eclet number''  for spontaneous stochasticity in the 
$Re\to\infty$ limit. This corresponds to the second set of boundary lines plotted in Figure \ref{4.png} with 
constants $c_{3(4)}$ selected for a best fit to the $\Upsilon$-isoline for the value $1.15\Upsilon_*$ 
($1.85\Upsilon_*).$

Finally, by simulating the model \eqref{langevin-int} in inertial-range units for $Sc=1,$ 
we have calculated the corresponding position PDF's $\bar{P}(\bar{x},\bar{t})$ at the fixed 
time $\bar{t}=1$ These PDF's are plotted in Figure \ref{3.png} for our default H\"older exponent $h=1/3$ 
at a sequence of increasing values of $Re=Pe.$ It is seen 
that the PDF's converge to a 
mixture of two delta-function spikes located at the extremal solutions $\bar{x}_\pm(\bar{t}),$ each with 
probability $1/2.$ This figure thus illustrates the most basic effect of spontaneous stochasticity, that the histories 
$\bar{x}(\bar{t})$ remain random for $Re,$ $Pe\to\infty.$ 
As a ``control experiment'' we have performed the same analysis 
for our model with $h=1,$ where instead $\bar{x}(\bar{t})$ converges
to the deterministic limit $\bar{x}^\infty(\bar{t})\equiv 0.$
See Appendix \ref{inverted}, Figure \ref{2.png}. 
Only for singular dynamics with $h<1$ can solutions be spontaneously stochastic. Also predicted by our theory 
is the rate of vanishing of probabilities to observe particles at positions $\bar{x}$ of non-extremal solutions in 
the limit $Re,$ $Pe\to\infty.$ To make a quantitative comparison, we have plotted in Figure \ref{3.png} over 
the interior interval $\bar{x}_-(\bar{t})<\bar{x}<\bar{x}_+(\bar{t})$ the singular large deviations estimate (\ref{Grad-int}). 
The constant prefactor of the rate function $\bar{\Phi}(\bar{x},\bar{t})$ in Eq.\eqref{Phi} was obtained numerically 
using the MATLISE software package \cite{ledoux2016matslise} to solve for the ground-state energy of the 
1-particle Hamiltonian \eqref{Scheq}, which gave $\epsilon_0\doteq 0.5545$ (see Appendix \ref{num-meth}). 
The theoretical prediction clearly matches the empirical PDF's better as $Re,$ $Pe$ increase.


\subsection{What is Special, What is General?}  

The 1D equation \eqref{vstar-eq} provides perhaps the simplest possible model of spontaneous stochasticity but 
it also suffers from a number of closely related special features that distinguish it from more generic cases
expected to occur in Nature. These special properties must be taken into account when using the results 
of the previous section to form expectations about observational signatures of spontaneous stochasticity 
in general. 
   
The first special feature of the model \eqref{vstar-eq} is that only a discrete subset of ``ground-states'' 
is selected from the continuum in the limit of increasing singularity and vanishing noise; in fact, precisely two. 
Most histories therefore have vanishing probability and the approach to the limit has the form of a large-deviations 
result, but with the novel feature that there is more than one minimum of the action and the usual ``law 
of large numbers" breaks down. Very similar behavior occurs for Lagrangian spontaneous stochasticity 
in shock solutions of Burgers equation, where only the two extremals are selected from all generalized characteristics
at the shock and the probabilities for all other histories vanish exponentially \cite{eyink2015spontaneous}. 
More generally, however, one expects that a continuous infinity of ``ground-state'' solutions will be selected, with 
a non-trivial limiting distribution over them. Numerical studies such as 
\cite{drivas2017lagrangian,palmer2014real,thalabard2020butterfly} exhibit no tendency for empirical distributions 
to be supported on only a finite number of solutions.  Nevertheless the limiting probability distributions are 
expected to be supported on (generalized) solutions of the limiting singular equations and there should 
quite generally be a large-deviations-type estimate on vanishing probability of non-solutions.

A second special feature of the model \eqref{vstar-eq} is that the dynamical vector field $v_*(x)$ has a singularity 
at just one point, $x=0,$ and the corresponding Cauchy problem has a non-unique solution only if the 
initial datum is chosen at that singular point. This feature is obviously related to the first, since a 
space-filling set of singularities would allow the solutions to ``branch'' at every point and the zero-noise 
limit could select an infinite ensemble of solutions. C.f. the example in \cite{hartman1982ordinary},
section II.5. There may be some problems of physical relevance where singular points are isolated in the 
dynamical state space, e.g. multi-body collisions in the classical $N$-body problem \cite{diacu1992singularities}. 
For typical situations in fluid turbulence, however, the limiting dynamics is singular at essentially every point 
and non-uniqueness should be generic. For example, the fluid velocity vector field that governs the motion of
Lagrangian particles for incompressible turbulence has as $Re\to \infty$ a space-filling set of H\"older singularities 
with exponent $\lim_{p\to 0} \zeta_p/p:=h_*\gtrsim 1/3$ (with $\zeta_p$ the velocity structure-function exponent) 
\cite{sreenivasan1996asymmetry,iyer2020scaling}. Likewise, Eulerian spontaneous stochasticity is expected 
to be generic in high-$Re$ incompressible fluid turbulence. As nicely reviewed in \cite{palmer2014real}, smooth 
(strong) solutions $\bv(t)$ of the incompressible Navier-Stokes equation are unique for fixed initial 
data $\bv_0$ and even a Leray (weak) solution $\bv'(t)$ with nearby initial condition $\bv_0'$ satisfies 
the inequality 
\be \|\bv(t)-\bv'(t)\|\leq \|\bv_0-\bv'_0\| \exp\left[\frac{c}{\nu^5} \int_0^t ds\  {\mathcal E}^2(s;\bv) \right]  \lb{NSuniq} \ee 
with ${\mathcal E}(t;\bv):=\nu\|\grad\bv(t)\|^2$ the volume-integrated energy dissipation per mass 
for solution $\bv.$ This inequality implies that $\bv'(t)\to \bv(t)$ as $\bv'_0\to \bv_0$ for fixed viscosity $\nu,$ 
but because of the turbulent ``dissipative anomaly'' the bound \eqref{NSuniq} is expected to diverge as $\nu\to 0$
and the smooth solution $\bv$ becomes more nearly singular. Although the inviscid $\nu\to 0$ limit is 
still very poorly understood mathematically, the arguments of \cite{lorenz1969predictability,leith1972predictability} 
on turbulent unpredictability suggest that limiting dissipative Euler solutions should be typically non-unique 
and exhibit the ``Nash non-rigidity'' phenomenon \cite{delellis2010admissibility,delellis2017high,daneri2020non}.  

Finally, a third special feature of the model \eqref{vstar-eq} is that it is ``almost non-chaotic''
since the non-fixed orbits must have zero Lyapunov exponent, as for any continuous-time 
dynamical system, and there are only two such orbits together tracing the entire right half-space
$\{x>0\}$ and left half-space $\{x<0\}.$ However, the spontaneous stochasticity phenomenon 
requires not only positive Lyapunov exponents as in standard deterministic chaos but in fact {\it infinite} 
Lyapunov exponents, since initial-data arbitrarily close to each other must separate to the same 
distance in finite time. In the model \eqref{vstar-eq} the unstable fixed point $x=0$ has  
the Lyapunov exponent $\bar{\lambda}=\frac{\partial \bar{v}}{\partial \bar{x}}(0)\sim Re^{\frac{1-h}{1+h}}$
generically for any natural regularization and thus all of the ``chaotic'' dynamics arises from 
$x=0$ where $\bar{\lambda}\to \infty$ as $Re\to \infty.$ On the other hand, fluid flows where spontaneous 
stochasticity is expected to arise as $Re\to\infty$ are known to possess much more robust forms of 
both Lagrangian and Eulerian chaos, even at moderate Reynolds numbers \cite{ChaosBook,bohr2005dynamical}.   
Here it is important to stress that spontaneous stochasticity requires not only positive Lyapunov exponents
but also increasingly singular dynamics in some limit, such as $Re\to\infty$ for fluids. The implications 
of spontaneous stochasticity are also much stronger. Whereas standard chaos theory for smooth 
dynamical systems implies the universality of infinite time-averages described by a natural invariant measure 
on a strange attractor in the weak-noise limit \cite{ChaosBook,eckmann1981roads}, spontaneous stochasticity 
produces universal statistics in a {\it finite} time for vanishingly small noise.  In order to exhibit such finite-time 
universality, spontaneously stochastic solutions do not have to lie on a chaotic attractor. In fact, 
a steady-state attractor does not even have to exist, as exemplified by such common flows 
as a decaying turbulent wake or a growing turbulent mixing layer.

\section{Future Directions}

We have developed in this paper a novel renormalization group (RG) approach to spontaneous stochasticity 
and applied it to the Langevin model \eqref{vstar-eq}, which is probably the simplest system conceivable 
which exhibits the basic phenomena. However, the general RG theory can be applied as well to much 
more complex and realistic systems with a limiting singular dynamics exhibiting spontaneous stochasticity. 
We here briefly indicate some directions for future RG work, first for Lagrangian particle histories and related 
time-histories governed by finite-dimensional ODE's, then for Eulerian space-time histories governed by 
infinite-dimensional ODE's/PDE's. 

\subsection{Lagrangian Spontaneous Stochasticity}

The simplest generalizations of the model \eqref{vstar-eq} are multi-dimensional extensions 
with point-singularities as considered in \cite{drivas2018life,drivas2020statistical}, where the singular radial 
dynamics is similar to that in the 1D model, while the angular dynamics in hyperspherical coordinates 
is smooth but otherwise arbitrary. As discussed in detail in \cite{drivas2018life,drivas2020statistical},
these mathematical model problems already exhibit a rich variety of behaviors including robust spontaneous 
stochasticity. These models have also the same fundamental scaling symmetry \eqref{scal-sym}
as the 1D model \eqref{vstar-eq} for a specified H\"older exponent $h.$ Thus, our RG theory carries over 
straightforwardly to this class of systems, although many new features now appear, such as continuous lines 
of RG limit cycles corresponding to spontaneous statistics with broken $O(2)$ symmetry. The class of models 
developed in \cite{drivas2018life,drivas2020statistical} provide important testbeds to understand better 
the role of chaotic dynamics in producing robust spontaneous stochasticity. As discussed already in 
section \ref{evidence}, chaotic dynamics and positive Lyapunov exponents are necessary to produce 
spontaneous stochasticity, but are not by themselves sufficient. The singularities in the dynamics play 
a crucial role in amplifying infinitesimal noise to macroscopic scales in finite times.  

 A step closer to realistic turbulence, our RG  method should provide a useful framework to study 
 Lagrangian particle histories also in the Kraichnan model of turbulent advection 
 \cite{bernard1998slow,eijnden2000generalized,gawedzki2001turbulent,kupiainen2003nondeterministic}.
Lagrangian spontaneous stochasticity was first discovered in this model and there are even rigorous 
demonstrations of its existence \cite{lejan2002integration,lejan2004flows}. However, little is known still 
in the Kraichnan model about (almost sure) properties of spontaneously stochastic particle-history 
distributions for fixed velocity realizations 
\footnote{It may be worth remarking in this context that there is no Eulerian spontaneous stochasticity of the 
scalar advection equations for the Kraichnan model. At least for incompressible (divergence-free) velocities,  
it has been shown that the scalar advection equations have strong stochastic solutions for fixed velocity
realizations \cite{lototskii2004passive}. This provides another example in addition to the Burgers equation
\cite{eyink2015spontaneous} where Lagrangian spontaneous stochasticity occurs without its Eulerian 
counterpart.}. 
The Kraichnan velocity is a fractional Brownian random field in space with specified H\"older exponent $h,$
so that it enjoys a statistical scaling symmetry analogous to \eqref{scal-sym} and a version of our RG 
method should be applicable. In addition to the Kraichnan model which is white-noise (delta-correlated) in time,
our RG method should apply also to particle advection by other self-similar Gaussian velocity fields 
with finite time-correlations \cite{chaves2003lagrangian}. Because all of these Gaussian advection models 
have velocity  fields that are everywhere singular in space and time, similar to real fluid turbulence, one 
expects that continuous branching will lead to spontaneously stochastic ensembles supported on an 
uncountably infinite set of particle histories. 

Real fluid turbulent flows and Lagrangian particles therein can be treated also by our RG theory. 
Although there are presently formidable difficulties to carrying out such a study analytically, the 
RG analysis can be implemented numerically following schemes similar to those used previously 
for front and self-similar solutions of deterministic PDE's \cite{chen1995numerical}. A very 
interesting feature of incompressible fluid turbulence is that the Euler fluid equations expected 
to describe the $Re\to\infty$ limit have infinitely-many scaling symmetries of the form 
\be \bv(\bx,t)\mapsto \bv_\lambda(\bx,t)=\lambda^{-h} \bv(\lambda \bx, \lambda^{1-h}t), \quad \lambda>0 \ee
for all real exponents $h,$ each of which gives an action of the one-dimensional group of 
dilatations, and these symmetries are expected to be statistically realized by multifractality of the  
velocity field \cite{frisch1985singularity,frisch1995turbulence}. In that picture, each spacetime event $(\bx,t)$ 
of the turbulent flow has its own ``local H\"older exponent'' $h=h(\bx,t)$ which is determined dynamically.
In the previous numerical study of spontaneously stochastic particle distributions in isotropic turbulence 
\cite{drivas2017lagrangian}, it was in fact observed that an intermediate-time nearly self-similar regime 
was attained for individual events, without  any need to average over $(\bx,t).$ This regime can be further 
explored with a numerical RG study, where now the field-renormalization exponent analogous to $1/\delta$
in \eqref{effect-init} must be determined self-consistently for each $(\bx,t)$ \cite{chen1995numerical}. 
The distribution of ``effective initial data'' that leads to self-similar growth backward in time is presumably 
universal for each $h$-exponent, since those singularities are expected to be created by dominant 
self-similar instanton solutions \cite{falkovich1996instantons}.

\subsection{Eulerian Spontaneous Stochasticity}

Our RG approach can be applied also to study Eulerian spontaneous stochasticity. Very attractive 
cases for numerical RG analysis are the self-similar ``equilibrium'' states in turbulent shear-flows and 
gravitationally-unstable layers, where evidence has already emerged for robust spontaneous stochasticity
\cite{biferale2018rayleigh,mailybaev2017toward,thalabard2020butterfly}. In these flows the H\"older
exponent $h$ that governs the self-similar scaling is set by the singular initial data, so that no
self-consistent determination of $h$ is required. Numerical RG in this application will allow the 
long-time self-similar regime to be more accurately probed, alleviating the restriction on direct 
numerical simulations imposed by finite domain size. These canonical flows are simple enough that analytical 
RG analysis may also be possible. An interesting feature here is that both the singular initial 
data and also the limiting ideal Euler dynamics are $PT$-invariant, e.g. under the transformation
\be u(x,y,t)\mapsto -u(x,-y,-t), \quad v(x,y,t)\mapsto v(x,-y,-t) \lb{PT} \ee 
for the 2D singular vortex sheet in \cite{thalabard2020butterfly}. However, the linear fluid instability breaks 
this symmetry \cite{qin2019kelvin} \footnote{The ``PT-symmetry'' considered by those authors, as they note, 
is physically just time-reversal $T$-symmetry, which interchanges the growing and decaying linear eigenmodes. 
It is easy to check that these linear eigenmodes are eigenfunctions of the reflection $P$-symmetry.}
and the spontaneous statistics associated to a growing turbulent layer 
are obviously not $PT$-invariant. Here the viscosity $\nu$ of the fluid plays the role of a symmetry-breaking 
field in a phase-transition, whose effect does not disappear in the inviscid, zero-noise limit.   

An ultimate aim of the RG theory of spontaneous stochasticity must be to treat singular Euler solutions 
which describe turbulent inertial-ranges and which, for generic initial data, should experience the 
``inverse error-cascade'' that was predicted by Lorenz \cite{lorenz1969predictability} and verified by following 
work \cite{leith1972predictability,boffetta2017chaos,berera2018chaotic,palmer2014real,mailybaev2016spontaneously}.  
This inverse cascade is characterized by an error field that grows self-similarly in scale 
\cite{lorenz1969predictability,leith1972predictability} or, in logarithmic wavenumber and time variables,
by a ``stochastic front'' that leaves universal spontaneous statistics in its wake \cite{mailybaev2016spontaneously}.
RG analysis of this type of multiscale spontaneous stochasticity presumably requires not only windowing 
out early times but also eliminating high wave-number modes. The effective dynamics for long times and 
low wavenumbers at the RG fixed point should contain not only ``eddy-viscosity'' effects that account for dissipative
anomaly but also ``eddy noise'' associated to the stochastic anomaly \cite{rose1977eddy,eyink1996turbulence}. 
One suspects that some type of turbulent fluctuation-dissipation relation will connect these 
two effects, which would be great practical interest in numerical modelling \cite{palmer2019stochastic}.  
A very remarkable numerical observation in \cite{mailybaev2016spontaneously} is that the spontaneous 
statistics achieved in a finite time for individual turbulent velocity realizations is identical to that observed 
for infinite-time attractors in forced steady-states. This result underlines the fundamental role of spontaneous 
stochasticity in achieving universal statistics for general turbulent flows. It presents also an imposing 
challenge, because the anomalous scaling and multifractality that characterize the turbulence state 
have so far defied theoretical analysis. Functional renormalization group methods that are non-perturbative
and formally exact \cite{delamotte2012introduction,canet2011general} have already yielded some novel 
predictions for high-$Re$ steady-states of Navier-Stokes turbulence \cite{canet2016fully,tarpin2018breaking}
and these methods should be capable also to describe spontaneously stochastic solutions. 

\vspace{10pt} 
Work on many of these directions is currently in progress and will be reported in following publications.

\acknowledgements{We acknowledge the Simons Foundation for support of this work through Targeted Grant 
in MPS-663054 at JHU and MPS-662985 at UIUC, ``Revisiting the Turbulence Problem Using Statistical Mechanics''. 
We wish to thank also many colleagues in that collaboration for fruitful discussions, especially 
Freddy Bouchet, L\'eonie Canet, Theodore Drivas, Gregory Falkovich, Nigel Goldenfeld, and Alexei Mailybaev.}

\appendix

\section{Stabilization of the Origin}\lb{stable0} 

We have shown in section \ref{approach} that 
\be p_0:=\lim_{Re\to\infty}\bar{P}_{Re}(|\bar{x}(\bar{t})|<\delta_0)=0 \ee 
for all $\bar{t}>0$ and for any sufficiently small $\delta_0>0,$ whenever $v'(x)\geq 0$
and the velocity regularization keeps the origin unstable. We shall show here that the probability 
to remain near zero may instead be positive, if the regularization makes the origin sufficiently stable. 
In the dissipation-range formulation of the model the requirement is that the particle must remain 
in a small neighborhood of zero with positive probability in the joint limit $Re\to\infty$ and 
$t= Re^{\frac{1-h}{1+h}} \bar{t}\to\infty$. Because this must hold even at times going to infinity, 
a very strong stabilizing drift back to the origin is required. Without attempting to identify 
necessary conditions, we study the specific example of a linear drift and show that by making it 
suitably diverge with $Re$ in the dissipation-range description, the particle can be prevented 
from escaping a neighborhood of the origin even at very long times. 

Our sufficient condition for $p_0>0$ follows from classical results in probability theory \cite{darling1953first}
on the first-passage time for the Ornstein-Uhlenbeck process
\be dx/dt = -x + \sqrt{2}\, \eta(t), \lb{OU-1} \ee 
that is, the time $T_{a,x}$ to first exit the interval $(-a,a)$ when $x(0)=x\in (-a,a).$ Paper \cite{darling1953first} showed that 
the Laplace transform of the probability density of $T_{a,x}$ is given by the formula
\begin{eqnarray}
{\mathbb E}[e^{-\lambda T_{a,x}}] &=& \frac{D_{-\lambda}(x)+D_{-\lambda}(-x)}{D_{-\lambda}(a)+D_{-\lambda}(-a)}
\exp\left(\frac{1}{4}(x^2-a^2)\right), \cr
&& \hspace{100pt} \lambda>0 
\end{eqnarray} 
in terms of the parabolic cylinder functions $D_\nu(x).$ This result can be put into a more useful form 
in terms of Kummer's confluent hypergeometric functions $M(a,b;z)$ using the identity:
\be e^{\frac{1}{4}x^2} [D_{-\lambda}(x)+D_{-\lambda}(-x)] = \frac{\sqrt{\pi} 2^{1-\lambda/2}}{\Gamma\left(\frac{1+\lambda}{2}\right)}
M\left(\frac{\lambda}{2},\frac{1}{2};\frac{x^2}{2}\right). \ee 
See \cite{abramowitz2012handbook}, \S 19.2.3. Thus
\be {\mathbb E}[e^{-\lambda T_{a,x}}] = \frac{M\left(\frac{\lambda}{2},\frac{1}{2};\frac{x^2}{2}\right)}{M\left(\frac{\lambda}{2},\frac{1}{2};\frac{a^2}{2}\right)}
, \quad \lambda>0. \ee

We can now study the effect of a drift velocity multiplied by $\gamma,$ or the solution $x^{(\gamma)}(t)$ of the 
modified Langevin equation 
\be dx/dt = -\gamma x + \sqrt{2}\, \eta(t), \lb{OU} \ee 
by rescaling time and space as $t'=\gamma t,$ $x'=\sqrt{\gamma} x$ to bring the equation to the form 
(\ref{OU-1}) with $\gamma=1.$ This implies that the exit time $T^{(\gamma)}_{a,x}$ for the general 
equation (\ref{OU}) can be related to that studied in \cite{darling1953first} as
\be T^{(\gamma)}_{a,x}=\frac{1}{\gamma} T_{\sqrt{\gamma}a,\sqrt{\gamma}x} \ee  
and thus
\be {\mathbb E}[e^{-\lambda T^{(\gamma)}_{a,x}}] = \frac{M\left(\frac{\lambda}{2\gamma},\frac{1}{2};\frac{\gamma x^2}{2}\right)}
{M\left(\frac{\lambda}{2\gamma},\frac{1}{2};\frac{\gamma a^2}{2}\right)}
, \quad \lambda>0. \ee
We can then estimate the probability that the particle exits the interval $(-a,a)$ before time $t$ by the 
Chebyshebv-Bienaym\'e inequality for any $\lambda>0$
\be P(T^{(\gamma)}_{a,x}<t) \leq e^{\lambda t} {\mathbb E}[e^{-\lambda T^{(\gamma)}_{a,x}}]. \lb{CBineq} \ee 

To exploit this inequality for the strong-drift limit, we can use a standard asymptotic estimate 
of the Kummer functions (\cite{abramowitz2012handbook}, \S 13.1.4) to infer that 
\be M\left(\frac{\lambda}{2\gamma},\frac{1}{2};\frac{\gamma x^2}{2}\right)\sim 
\frac{\lambda}{\gamma}\left(\frac{\pi}{2\gamma x^2}\right)^{1/2}\exp(\gamma x^2/2), \quad \gamma\to\infty \ee
so that
\be {\mathbb E}[e^{-\lambda T^{(\gamma)}_{a,x}}] \sim \frac{a}{|x|} \exp(-\gamma (a^2-x^2)/2), \quad 
\gamma\to\infty. \ee
If we take in the inequality (\ref{CBineq}) the fixed value $\lambda=(a^2-x^2)/2,$ then we infer that 
\be P(T^{(\gamma)}_{a,x}<t) \leq ({\rm const.)} \frac{a}{|x|} \exp(-(\gamma-t) (a^2-x^2)/2) \ee
for some sufficiently large $\gamma,$ and thus
\be \lim_{\gamma\to\infty,\gamma-t\to\infty} P(T^{(\gamma)}_{a,x}<t) = 0. \ee 
Because
\begin{eqnarray}
 P(T^{(\gamma)}_{a,x}>t) &= &P(\max_{s\in [0,t]}|x^{(\gamma)}(s)|<a\,|\,x^{(\gamma)}(0)=x) \cr
  &\leq& P(|x^{(\gamma)}(t)|<a\,|\,x^{(\gamma)}(0)=x) 
 \end{eqnarray} 
we therefore get for all $x_0\in(-a,a)$ that 
\be  \lim_{\gamma\to\infty,\gamma-t\to\infty} \int_{-a}^a dx\, P^{(\gamma)}(x,t|x_0,0)=1, \ee
that is, the stochastic particle in this strong-drift limit $\gamma\to \infty$ never escapes from the interval 
$(-a,a)$ up to time $t,$ even if $t\to\infty,$ as long as $\gamma-t\to \infty.$

We can apply this result to the 1D model of SS in its dissipation-range description by taking $a<1,$ 
so that the interval $(-a,a)$ lies in the regularization region $|x|<1.$ If we regularize the problem by taking 
\be v_{Re}(x)= -\gamma(Re) x, \quad |x|<a, \qquad \mbox{$\gamma(Re)=Re^{\frac{1-h}{1+h}}\gamma_0$} \ee
with $\gamma_0>\bar{t},$ then for any initial distribution $P_0(x_0)$ with $p_a=\int_{-a}^a dx_0\ P_0(x_0)>0$  
\begin{eqnarray}
 \lim_{Re\to\infty} p_0(t) &:=&\lim_{Re\to\infty} \int_{-a}^a dx\ P_{Re}(x,t)\geq p_a,  \cr
 && \hspace{60pt} t=Re^{\frac{1-h}{1+h}} \bar{t},
\end{eqnarray}  
with equality if no particles from outside the interval $(-a,a)$ enter it with any positive probability. 
Expressed in inertial-range units, this means that for any $\bar{t}<\gamma_0$
\be p_0(\bar{t}):=\lim_{Re\to\infty} \int_{-a/Re^{1+h}}^{a/Re^{1+h}} d\bar{x}\ P_{Re}(\bar{x},\bar{t})\geq p_a,  \ee 
so that the probability is positive to remain in a vanishingly small neighborhood of the origin
as $Re\to\infty.$

The velocity regularization required to stabilize the particle at $x=0$ is admittedly ``unnatural'',
corresponding to a potential well which is increasingly deep as $Re\to\infty.$ However, from the 
macroscopic inertial-range point of view the regularization region $|\bar{x}|\leq Re^{-1/(1+h)}$ 
disappears entirely from view as $Re\to\infty$ and only the unstable singular drift field 
$v_*(\bar{x})$ is observed. In such ``unnatural'' models the exact solution $\bar{x}^\infty(\bar{t})\equiv 0$
with infinite waiting-time may occur with positive probability in addition to the two extremal solutions
$\bar{x}^\pm(\bar{t})$ with zero waiting-time. Note that these three solutions of the singular initial-value 
problem are exactly those that are invariant under the basic scaling symmetry \eqref{scal-sym}. 

\vspace{10pt} 

\section{Domain of Attraction of the Symmetric RG Fixed-Point} \lb{symdomain} 

We show here that the symmetric RG fixed point $Q_*$ in \eqref{fxpt} attracts some non-symmetric 
initial probability distributions $P_0.$ For simplicity we assume that the velocity $v(x)$ is symmetric,
$v(-x)=-v(x),$ although it should not be too difficult to remove that restriction. 

We begin by defining the function 
\be \delta p(x):=p_+(x)-p_-(x)= (1/N) \int_{-x}^x dy \ \exp(-Sc\  V(y)), \ee 
which is non-negative, even in $x$, and strictly increasing to 1 as $|x|\to \infty.$ 
For any $a>0,$ there are infinitely-many square-integrable functions $\Delta\in {\mathcal L}^2[0,\infty]$
which are both orthogonal to $\delta p$ and supported inside the interval $[0,a]$:
\be \int_0^\infty dx\ \delta p(x) \Delta(x) =\int_0^a dx\ \delta p(x) \Delta(x)=0.  \ee 
It is also easy to construct many examples that are bounded in magnitude by the constant 1, e.g.
\be \Delta(x) = \left\{\begin{array}{ll}
                        A\frac{\sin(2\pi x/a)}{\delta p(x)} & 0\leq x\leq a \cr
                        0 & x>a
                        \end{array} \right. \ee
for a suitable choice of $A.$                        
Such functions can be extended anti-symmetrically to the entire real line 
by setting $\Delta(-x)=-\Delta(x),$ so that by \eqref{p-sym} 
\be \int_{-\infty}^\infty dx\ p_+(x) \Delta(x) =\int_{-\infty}^\infty dx\ p_-(x) \Delta(x)=0. \ee

In that case, the function 
\be P_0(x_0):=\left\{\begin{array}{ll}
                             \frac{1}{2a} (1+\theta\,\Delta(x_0)) & |x_0|\leq a \cr
                             0 & \mbox{o.w.} 
                             \end{array} \right. \ee 
for any $0<\theta <1$ is a non-symmetric probability density that nevertheless has $p_+=p_-=1/2.$ 

\section{Numerical Methods}\lb{num-meth}  

The 1D Langevin equation was numerically integrated with initial condition $x(0) = 0$ at a sequence 
of times $t_i=i(\Delta t),$ using the Euler-Maruyama method \cite{kloeden2013numerical}:
\be x(t_{i+1}) = x(t_i) + v(x_i) \Delta t + \sqrt{2\hat{D}\Delta t}\, N_i, \quad i=0,1,2,...\ee
where $\hat{D}=1/Sc$ for \eqref{langevin-dis} and $\hat{D}=1/Pe$ for \eqref{langevin-int}. 
Here $N_i\sim \mathcal{N}(0,1)$ is a pseudo-random sequence of standard normal random variables,
which we obtained by generating uniform random variables with the Mersenne twister algorithm 
\cite{matsumoto1998mersenne} and applying a 256-step Ziggurat method
to obtain normal variables \cite{marsaglia2000ziggurat}.  
Statistics presented are empirical averages over $N$ independent samples and empirical probability density functions $P(x,t)$
of particle positions are further smoothed using Gaussian kernel-density estimators \cite{silverman2018density}. 

For each average calculated, weak convergence with respect to step size $\Delta t$ was verified 
by comparing the average for two step sizes that differ by a factor of 2. The result 
was deemed to be converged with respect to step size if the error was dominated by the number of 
samples used. For Figures \ref{4.png}, \ref{1.png}, \ref{6.png}, \ref{5.png}, \ref{3.png}, 
\ref{2.png} we chose $\Delta t=0.001$, $0.0156$, $0.005$, $0.00333$, $0.00333$, $0.00333,$ respectively. 
The number of samples $N$ was chosen to be large enough for the goal at hand. For Figures \ref{1.png}, \ref{3.png}, 
\ref{2.png} of position PDF's $N=10^7$ was found to be sufficient 
so that the error is smaller than the line width of the plot. 

\begin{figure}[h!]
 \begin{center}
  \includegraphics[width=240pt]{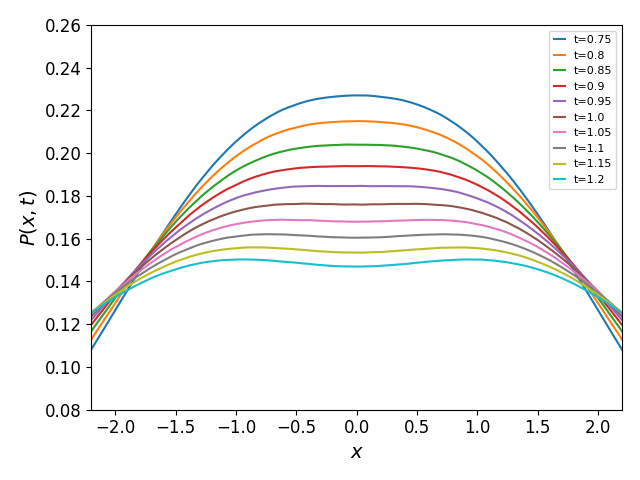}
 \end{center}
 \caption{Probability density function $P(x,t)$ for particle positions at ten different times, 
 all of the quantities in dissipation-range scaling units, near the moment of bifurcation at 
 $t \doteq 0.95$.}\label{7.png}
 \end{figure}

\begin{figure*}[t]
  \centering
  \begin{subfigure}[b]{0.24\linewidth}
    \includegraphics[width=\linewidth]{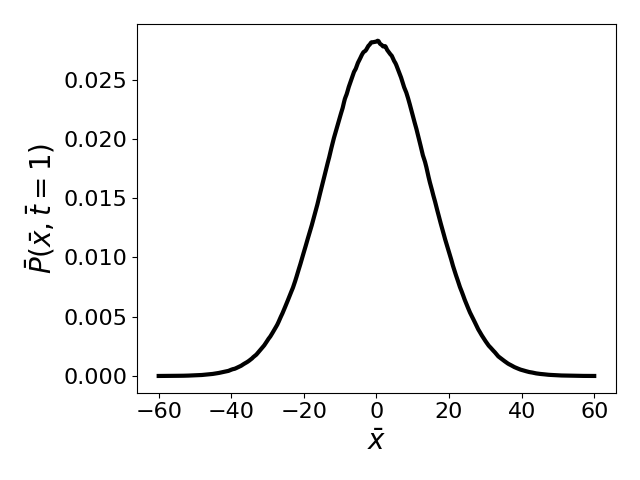}
    \caption{$Re = Pe = 10^2$}
  \end{subfigure}
  \begin{subfigure}[b]{0.24\linewidth}
    \includegraphics[width=\linewidth]{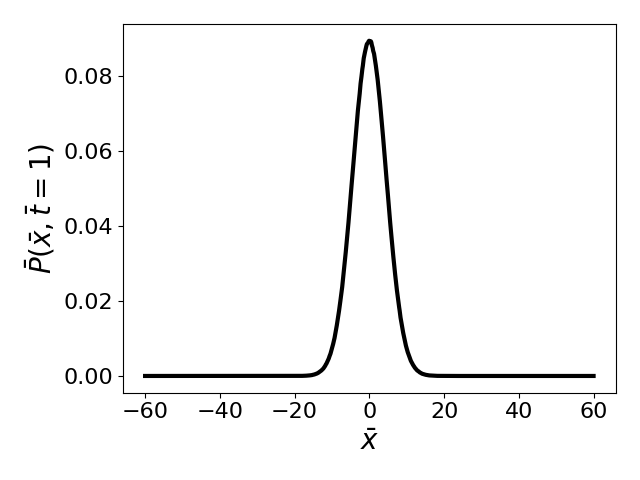}
    \caption{$Re = Pe = 10^3$}
  \end{subfigure}
    \begin{subfigure}[b]{0.24\linewidth}
    \includegraphics[width=\linewidth]{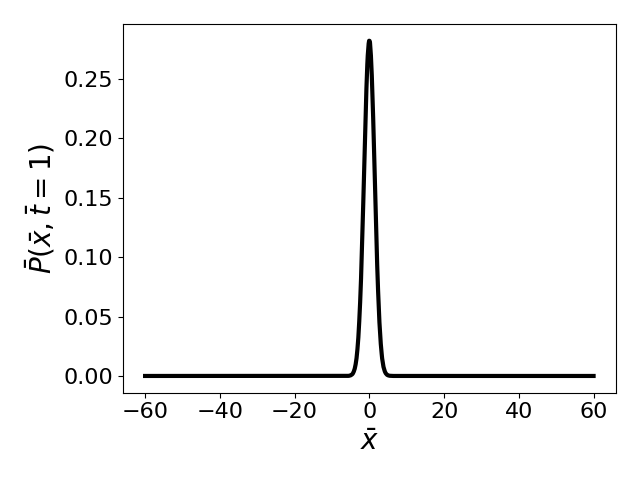}
    \caption{$Re = Pe =10^4 $}
  \end{subfigure}
    \begin{subfigure}[b]{0.24\linewidth}
    \includegraphics[width=\linewidth]{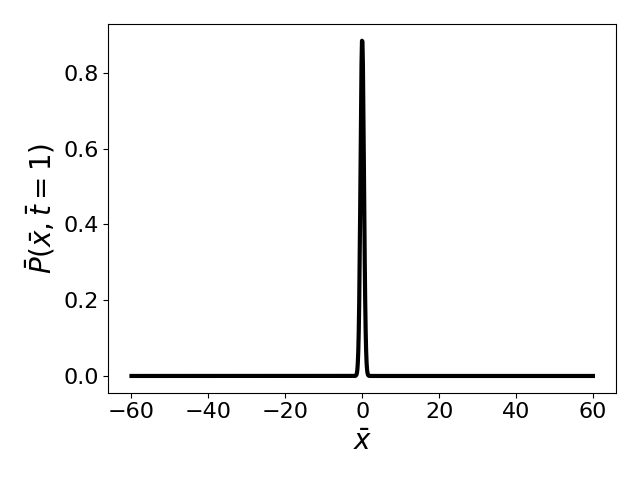}
    \caption{$Re = Pe = 10^5$}
  \end{subfigure}
   \caption{Probability density functions $\bar{P}(\bar{x},\bar{t})$ for particle positions at time $\bar{t} = 1$ in the 
   model with $h=1,$ for  four different values of $Re=Pe,$ all quantities in inertial-range scaling units.}\label{2.png} 
\end{figure*}

This number had to increased to $N=10^9$ in order to evaluate the bifurcation diagram in Figure \ref{6.png}, 
since the PDF becomes nearly flat near the bifurcation and this greatly increases the error in estimation 
of the location of local extrema. See Figure \ref{7.png}. To identify those extrema, we 
split the $x-t$ plane into parts such that each must have only one local extremum and then found the global 
maximum/minimum on each part. The error was determined by finding the intervals where the values of the 
PDF differ by less than $0.1\%$ from the extremal value.

To arrive at the phase diagram in Figure \ref{4.png} we integrated the Langevin model
in the inertial-range form \eqref{langevin-int} up to time 1 for the set of $Re$ and $Pe$ 
values shown and the variance for each was evaluated by Monte Carlo average with $N=2680$ samples. 
Convergence was checked by increasing $N$ and assessing by eye the difference in the phase diagram. 


Figure \ref{5.png} corresponds conceptually to slices of the phase diagram in Figure \ref{1.png}
along lines of constant $Sc,$ each with slope 1, and rescaled from inertial to dissipative units. 
However, the range of $Re$ and $Pe$ was modified to better 
illustrate all of the scaling regimes. Furthermore, the number of samples was increased to 
$N=5 \times 10^5$ in order to ensure that the error in the estimated variance is smaller than the width 
of the plotted curves. 

To generate Figures \ref{1.png}, \ref{6.png}, \ref{3.png}, \ref{2.png}, \ref{7.png} we approximated the PDF's 
using kernel-density estimators \cite{silverman2018density}. In this scheme the PDF is estimated 
by the $N$-sample average:

\vspace{-15pt}
\be P(x) = \frac{1}{N} \sum_{n=1}^N K_{{\hit}}(x_n), \ee
where $K_{{\hit}}(x) = \frac{1}{{\hit}}K(\frac{x}{{\hit}})$ is a filter kernel with 
bandwidth ${\hit}$. We chose a Gaussian filter kernel and for each PDF the bandwidth was chosen 
as in \cite{drivas2017lagrangian} via the least sensitivity principle \cite{stevenson1981optimized}. 
According to this principle one selects the bandwidth that has the least influence on the estimated PDF. 
For each PDF the bandwidth was selected individually by evaluating the $L_1$ norm of the increments of the PDF 
for varying ${\hit}$. Then the bandwidth that minimizes the increments was chosen. 

Finally, the constant $\epsilon_0$ in the theoretical curves in 
Figure \ref{3.png} was calculated with the MATSLISE 2.0 software,  
which uses a high-order piecewise constant perturbation method to calculate both eigenvalues and eigenfunctions \cite{ledoux2016matslise}. The value obtained was $\epsilon_0\doteq 0.55447399023$
with estimated error $1.3529\times 10^{-11}.$  We also verified 
that the numerically obtained eigenfunction $\varPsi_0(x)$ (not shown here) satisfies the asymptotics \eqref{Psi-asympt} for $|x|\gg 1.$

\section{Control Experiment with $h=1$}\lb{inverted}

To confirm that the persistent stochasticity observed 
in Figure \ref{3.png} is a consequence of the H\"older 
regularity exponent $h<1$, we have carried out an identical  
analysis for the same model with only the exponent changed 
to $h=1$ and with all other parameters the same. Note that 
the regularized velocity field \eqref{vreg} for $h=1$ reduces 
to a linear drift $v(x)=Ax,$
corresponding to diffusion in an inverted parabolic potential. 
In Figure \ref{2.png} we plot for this $h=1$ case 
the inertial-range PDF's of particle positions  $\bar{P}(\bar{x},\bar{t})$ for the same increasing sequence 
of values $Re=Pe$ as in Figure \ref{3.png}. Unlike there, however, 
we see now that the PDF's converge to a single delta-function 
at $\bar{x}=0,$ corresponding to a deterministic limit. 
This null result verifies that the stochasticity in the 
infinite-$Re,$ $Pe$ limit is not a trivial consequence of the instability of the fixed point $x=0$ 
but requires the quasi-singular velocity with exponent $h<1.$ The results obtained in our model
with $h=1$ are typical of smooth dynamical systems, which may exhibit deterministic chaos with 
exponential error growth but for which determinism is restored at any finite time $\bar{t}$ in the zero-noise limit. 

\bibliography{bibliography.bib}

\providecommand{\noopsort}[1]{}\providecommand{\singleletter}[1]{#1}%
\begin{thebibliography}{108}%
\makeatletter
\providecommand \@ifxundefined [1]{%
 \@ifx{#1\undefined}
}%
\providecommand \@ifnum [1]{%
 \ifnum #1\expandafter \@firstoftwo
 \else \expandafter \@secondoftwo
 \fi
}%
\providecommand \@ifx [1]{%
 \ifx #1\expandafter \@firstoftwo
 \else \expandafter \@secondoftwo
 \fi
}%
\providecommand \natexlab [1]{#1}%
\providecommand \enquote  [1]{``#1''}%
\providecommand \bibnamefont  [1]{#1}%
\providecommand \bibfnamefont [1]{#1}%
\providecommand \citenamefont [1]{#1}%
\providecommand \href@noop [0]{\@secondoftwo}%
\providecommand \href [0]{\begingroup \@sanitize@url \@href}%
\providecommand \@href[1]{\@@startlink{#1}\@@href}%
\providecommand \@@href[1]{\endgroup#1\@@endlink}%
\providecommand \@sanitize@url [0]{\catcode `\\12\catcode `\$12\catcode
  `\&12\catcode `\#12\catcode `\^12\catcode `\_12\catcode `\%12\relax}%
\providecommand \@@startlink[1]{}%
\providecommand \@@endlink[0]{}%
\providecommand \url  [0]{\begingroup\@sanitize@url \@url }%
\providecommand \@url [1]{\endgroup\@href {#1}{\urlprefix }}%
\providecommand \urlprefix  [0]{URL }%
\providecommand \Eprint [0]{\href }%
\providecommand \doibase [0]{http://dx.doi.org/}%
\providecommand \selectlanguage [0]{\@gobble}%
\providecommand \bibinfo  [0]{\@secondoftwo}%
\providecommand \bibfield  [0]{\@secondoftwo}%
\providecommand \translation [1]{[#1]}%
\providecommand \BibitemOpen [0]{}%
\providecommand \bibitemStop [0]{}%
\providecommand \bibitemNoStop [0]{.\EOS\space}%
\providecommand \EOS [0]{\spacefactor3000\relax}%
\providecommand \BibitemShut  [1]{\csname bibitem#1\endcsname}%
\let\auto@bib@innerbib\@empty
\bibitem [{\citenamefont {Gaw\c{e}dzki}(2001)}]{gawedzki2001turbulent}%
  \BibitemOpen
  \bibfield  {author} {\bibinfo {author} {\bibfnamefont {K.}~\bibnamefont
  {Gaw\c{e}dzki}},\ }\bibfield  {title} {\enquote {\bibinfo {title} {Turbulent
  advection and breakdown of the {L}agrangian flow},}\ }in\ \href@noop {}
  {\emph {\bibinfo {booktitle} {Intermittency in Turbulent Flows}}},\ \bibinfo
  {editor} {edited by\ \bibinfo {editor} {\bibfnamefont {J.~C}\ \bibnamefont
  {Vassilicos}}}\ (\bibinfo  {publisher} {Cambridge University Press},\
  \bibinfo {address} {Cambridge, UK},\ \bibinfo {year} {2001})\ pp.\ \bibinfo
  {pages} {86--104}\BibitemShut {NoStop}%
\bibitem [{\citenamefont {Kupiainen}(2003)}]{kupiainen2003nondeterministic}%
  \BibitemOpen
  \bibfield  {author} {\bibinfo {author} {\bibfnamefont {A.}~\bibnamefont
  {Kupiainen}},\ }\bibfield  {title} {\enquote {\bibinfo {title}
  {Nondeterministic dynamics and turbulent transport},}\ }\href@noop {}
  {\bibfield  {journal} {\bibinfo  {journal} {Annales Henri Poincar{\'e}}\
  }\textbf {\bibinfo {volume} {4}},\ \bibinfo {pages} {713--726} (\bibinfo
  {year} {2003})}\BibitemShut {NoStop}%
\bibitem [{\citenamefont {Chaves}\ \emph {et~al.}(2003)\citenamefont {Chaves},
  \citenamefont {Gaw\c{e}dzki}, \citenamefont {Horvai}, \citenamefont
  {Kupiainen},\ and\ \citenamefont {Vergassola}}]{chaves2003lagrangian}%
  \BibitemOpen
  \bibfield  {author} {\bibinfo {author} {\bibfnamefont {Marta}\ \bibnamefont
  {Chaves}}, \bibinfo {author} {\bibfnamefont {Krzysztof}\ \bibnamefont
  {Gaw\c{e}dzki}}, \bibinfo {author} {\bibfnamefont {Peter}\ \bibnamefont
  {Horvai}}, \bibinfo {author} {\bibfnamefont {Antti}\ \bibnamefont
  {Kupiainen}}, \ and\ \bibinfo {author} {\bibfnamefont {Massimo}\ \bibnamefont
  {Vergassola}},\ }\bibfield  {title} {\enquote {\bibinfo {title} {Lagrangian
  dispersion in {G}aussian self-similar velocity ensembles},}\ }\href@noop {}
  {\bibfield  {journal} {\bibinfo  {journal} {Journal of statistical physics}\
  }\textbf {\bibinfo {volume} {113}},\ \bibinfo {pages} {643--692} (\bibinfo
  {year} {2003})}\BibitemShut {NoStop}%
\bibitem [{\citenamefont {Richardson}(1926)}]{richardson1926atmospheric}%
  \BibitemOpen
  \bibfield  {author} {\bibinfo {author} {\bibfnamefont {L.~F.}\ \bibnamefont
  {Richardson}},\ }\bibfield  {title} {\enquote {\bibinfo {title} {Atmospheric
  diffusion on a distance-neighbor graph},}\ }\href@noop {} {\bibfield
  {journal} {\bibinfo  {journal} {Proc. R. Soc. A}\ }\textbf {\bibinfo {volume}
  {110}},\ \bibinfo {pages} {709--737} (\bibinfo {year} {1926})}\BibitemShut
  {NoStop}%
\bibitem [{\citenamefont {Bernard}\ \emph {et~al.}(1998)\citenamefont
  {Bernard}, \citenamefont {Gawedzki},\ and\ \citenamefont
  {Kupiainen}}]{bernard1998slow}%
  \BibitemOpen
  \bibfield  {author} {\bibinfo {author} {\bibfnamefont {Denis}\ \bibnamefont
  {Bernard}}, \bibinfo {author} {\bibfnamefont {Krzysztof}\ \bibnamefont
  {Gawedzki}}, \ and\ \bibinfo {author} {\bibfnamefont {Antti}\ \bibnamefont
  {Kupiainen}},\ }\bibfield  {title} {\enquote {\bibinfo {title} {Slow modes in
  passive advection},}\ }\href@noop {} {\bibfield  {journal} {\bibinfo
  {journal} {Journal of Statistical Physics}\ }\textbf {\bibinfo {volume}
  {90}},\ \bibinfo {pages} {519--569} (\bibinfo {year} {1998})}\BibitemShut
  {NoStop}%
\bibitem [{\citenamefont {E}\ and\ \citenamefont
  {Vanden-Eijnden}(2000)}]{eijnden2000generalized}%
  \BibitemOpen
  \bibfield  {author} {\bibinfo {author} {\bibfnamefont {W.}~\bibnamefont {E}}\
  and\ \bibinfo {author} {\bibfnamefont {E.}~\bibnamefont {Vanden-Eijnden}},\
  }\bibfield  {title} {\enquote {\bibinfo {title} {Generalized flows, intrinsic
  stochasticity, and turbulent transport},}\ }\href@noop {} {\bibfield
  {journal} {\bibinfo  {journal} {Proceedings of the National Academy of
  Sciences}\ }\textbf {\bibinfo {volume} {97}},\ \bibinfo {pages} {8200--8205}
  (\bibinfo {year} {2000})}\BibitemShut {NoStop}%
\bibitem [{\citenamefont {E}\ and\ \citenamefont
  {Vanden-Eijnden}(2003)}]{e2003note}%
  \BibitemOpen
  \bibfield  {author} {\bibinfo {author} {\bibfnamefont {W.}~\bibnamefont {E}}\
  and\ \bibinfo {author} {\bibfnamefont {Eric}\ \bibnamefont
  {Vanden-Eijnden}},\ }\bibfield  {title} {\enquote {\bibinfo {title} {A note
  on generalized flows},}\ }\href@noop {} {\bibfield  {journal} {\bibinfo
  {journal} {Physica D: Nonlinear Phenomena}\ }\textbf {\bibinfo {volume}
  {183}},\ \bibinfo {pages} {159--174} (\bibinfo {year} {2003})}\BibitemShut
  {NoStop}%
\bibitem [{\citenamefont {Drivas}\ and\ \citenamefont
  {Eyink}(2017)}]{drivas2017lagrangian}%
  \BibitemOpen
  \bibfield  {author} {\bibinfo {author} {\bibfnamefont {Theodore~D}\
  \bibnamefont {Drivas}}\ and\ \bibinfo {author} {\bibfnamefont {Gregory~L}\
  \bibnamefont {Eyink}},\ }\bibfield  {title} {\enquote {\bibinfo {title} {A
  {L}agrangian fluctuation--dissipation relation for scalar turbulence. {P}art
  {I}. flows with no bounding walls},}\ }\href@noop {} {\bibfield  {journal}
  {\bibinfo  {journal} {Journal of Fluid Mechanics}\ }\textbf {\bibinfo
  {volume} {829}},\ \bibinfo {pages} {153--189} (\bibinfo {year}
  {2017})}\BibitemShut {NoStop}%
\bibitem [{\citenamefont {Lazarian}\ and\ \citenamefont
  {Vishniac}(1999)}]{lazarian1999reconnection}%
  \BibitemOpen
  \bibfield  {author} {\bibinfo {author} {\bibfnamefont {A}~\bibnamefont
  {Lazarian}}\ and\ \bibinfo {author} {\bibfnamefont {Ethan~T}\ \bibnamefont
  {Vishniac}},\ }\bibfield  {title} {\enquote {\bibinfo {title} {Reconnection
  in a weakly stochastic field},}\ }\href@noop {} {\bibfield  {journal}
  {\bibinfo  {journal} {The Astrophysical Journal}\ }\textbf {\bibinfo {volume}
  {517}},\ \bibinfo {pages} {700} (\bibinfo {year} {1999})}\BibitemShut
  {NoStop}%
\bibitem [{\citenamefont {Eyink}\ \emph {et~al.}(2011)\citenamefont {Eyink},
  \citenamefont {Lazarian},\ and\ \citenamefont {Vishniac}}]{eyink2011fast}%
  \BibitemOpen
  \bibfield  {author} {\bibinfo {author} {\bibfnamefont {Gregory~L}\
  \bibnamefont {Eyink}}, \bibinfo {author} {\bibfnamefont {Alex}\ \bibnamefont
  {Lazarian}}, \ and\ \bibinfo {author} {\bibfnamefont {Ethan~T}\ \bibnamefont
  {Vishniac}},\ }\bibfield  {title} {\enquote {\bibinfo {title} {Fast magnetic
  reconnection and spontaneous stochasticity},}\ }\href@noop {} {\bibfield
  {journal} {\bibinfo  {journal} {The Astrophysical Journal}\ }\textbf
  {\bibinfo {volume} {743}},\ \bibinfo {pages} {51} (\bibinfo {year}
  {2011})}\BibitemShut {NoStop}%
\bibitem [{\citenamefont {Eyink}\ \emph {et~al.}(2013)\citenamefont {Eyink},
  \citenamefont {Vishniac}, \citenamefont {Lalescu}, \citenamefont {Aluie},
  \citenamefont {Kanov}, \citenamefont {B{\"u}rger}, \citenamefont {Burns},
  \citenamefont {Meneveau},\ and\ \citenamefont {Szalay}}]{eyink2013flux}%
  \BibitemOpen
  \bibfield  {author} {\bibinfo {author} {\bibfnamefont {Gregory}\ \bibnamefont
  {Eyink}}, \bibinfo {author} {\bibfnamefont {Ethan}\ \bibnamefont {Vishniac}},
  \bibinfo {author} {\bibfnamefont {Cristian}\ \bibnamefont {Lalescu}},
  \bibinfo {author} {\bibfnamefont {Hussein}\ \bibnamefont {Aluie}}, \bibinfo
  {author} {\bibfnamefont {Kalin}\ \bibnamefont {Kanov}}, \bibinfo {author}
  {\bibfnamefont {Kai}\ \bibnamefont {B{\"u}rger}}, \bibinfo {author}
  {\bibfnamefont {Randal}\ \bibnamefont {Burns}}, \bibinfo {author}
  {\bibfnamefont {Charles}\ \bibnamefont {Meneveau}}, \ and\ \bibinfo {author}
  {\bibfnamefont {Alexander}\ \bibnamefont {Szalay}},\ }\bibfield  {title}
  {\enquote {\bibinfo {title} {Flux-freezing breakdown in high-conductivity
  magnetohydrodynamic turbulence},}\ }\href@noop {} {\bibfield  {journal}
  {\bibinfo  {journal} {Nature}\ }\textbf {\bibinfo {volume} {497}},\ \bibinfo
  {pages} {466--469} (\bibinfo {year} {2013})}\BibitemShut {NoStop}%
\bibitem [{\citenamefont {Eyink}(2018)}]{eyink2018cascades}%
  \BibitemOpen
  \bibfield  {author} {\bibinfo {author} {\bibfnamefont {Gregory~L}\
  \bibnamefont {Eyink}},\ }\bibfield  {title} {\enquote {\bibinfo {title}
  {Cascades and dissipative anomalies in nearly collisionless plasma
  turbulence},}\ }\href@noop {} {\bibfield  {journal} {\bibinfo  {journal}
  {Physical Review X}\ }\textbf {\bibinfo {volume} {8}},\ \bibinfo {pages}
  {041020} (\bibinfo {year} {2018})}\BibitemShut {NoStop}%
\bibitem [{\citenamefont {Bardos}\ \emph {et~al.}(2019)\citenamefont {Bardos},
  \citenamefont {Besse},\ and\ \citenamefont {Nguyen}}]{bardos2019onsager}%
  \BibitemOpen
  \bibfield  {author} {\bibinfo {author} {\bibfnamefont {Claude}\ \bibnamefont
  {Bardos}}, \bibinfo {author} {\bibfnamefont {Nicolas}\ \bibnamefont {Besse}},
  \ and\ \bibinfo {author} {\bibfnamefont {Toan~T}\ \bibnamefont {Nguyen}},\
  }\bibfield  {title} {\enquote {\bibinfo {title} {Onsager's type conjecture
  and renormalized solutions for the relativistic {V}lasov {M}axwell system},}\
  }\href@noop {} {\bibfield  {journal} {\bibinfo  {journal} {arXiv preprint
  arXiv:1903.04878}\ } (\bibinfo {year} {2019})}\BibitemShut {NoStop}%
\bibitem [{\citenamefont {Lorenz}(1969)}]{lorenz1969predictability}%
  \BibitemOpen
  \bibfield  {author} {\bibinfo {author} {\bibfnamefont {Edward~N}\
  \bibnamefont {Lorenz}},\ }\bibfield  {title} {\enquote {\bibinfo {title} {The
  predictability of a flow which possesses many scales of motion},}\
  }\href@noop {} {\bibfield  {journal} {\bibinfo  {journal} {Tellus}\ }\textbf
  {\bibinfo {volume} {21}},\ \bibinfo {pages} {289--307} (\bibinfo {year}
  {1969})}\BibitemShut {NoStop}%
\bibitem [{\citenamefont {De~Lellis}\ and\ \citenamefont
  {Sz{\'e}kelyhidi}(2010)}]{delellis2010admissibility}%
  \BibitemOpen
  \bibfield  {author} {\bibinfo {author} {\bibfnamefont {Camillo}\ \bibnamefont
  {De~Lellis}}\ and\ \bibinfo {author} {\bibfnamefont {L{\'a}szl{\'o}}\
  \bibnamefont {Sz{\'e}kelyhidi}},\ }\bibfield  {title} {\enquote {\bibinfo
  {title} {On admissibility criteria for weak solutions of the {E}uler
  equations},}\ }\href@noop {} {\bibfield  {journal} {\bibinfo  {journal}
  {Archive for rational mechanics and analysis}\ }\textbf {\bibinfo {volume}
  {195}},\ \bibinfo {pages} {225--260} (\bibinfo {year} {2010})}\BibitemShut
  {NoStop}%
\bibitem [{\citenamefont {De~Lellis}\ and\ \citenamefont
  {Sz{\'e}kelyhidi~Jr}(2017)}]{delellis2017high}%
  \BibitemOpen
  \bibfield  {author} {\bibinfo {author} {\bibfnamefont {Camillo}\ \bibnamefont
  {De~Lellis}}\ and\ \bibinfo {author} {\bibfnamefont {L{\'a}szl{\'o}}\
  \bibnamefont {Sz{\'e}kelyhidi~Jr}},\ }\bibfield  {title} {\enquote {\bibinfo
  {title} {High dimensionality and $h$-principle in {PDE}},}\ }\href@noop {}
  {\bibfield  {journal} {\bibinfo  {journal} {Bulletin of the American
  Mathematical Society}\ }\textbf {\bibinfo {volume} {54}},\ \bibinfo {pages}
  {247--282} (\bibinfo {year} {2017})}\BibitemShut {NoStop}%
\bibitem [{\citenamefont {Daneri}\ \emph {et~al.}(2020)\citenamefont {Daneri},
  \citenamefont {Runa},\ and\ \citenamefont {Szekelyhidi~Jr}}]{daneri2020non}%
  \BibitemOpen
  \bibfield  {author} {\bibinfo {author} {\bibfnamefont {Sara}\ \bibnamefont
  {Daneri}}, \bibinfo {author} {\bibfnamefont {Eris}\ \bibnamefont {Runa}}, \
  and\ \bibinfo {author} {\bibfnamefont {Laszlo}\ \bibnamefont
  {Szekelyhidi~Jr}},\ }\bibfield  {title} {\enquote {\bibinfo {title}
  {Non-uniqueness for the {E}uler equations up to {O}nsager's critical
  exponent},}\ }\href@noop {} {\bibfield  {journal} {\bibinfo  {journal} {arXiv
  preprint arXiv:2004.00391}\ } (\bibinfo {year} {2020})}\BibitemShut {NoStop}%
\bibitem [{\citenamefont {Palmer}\ \emph {et~al.}(2014)\citenamefont {Palmer},
  \citenamefont {D{\"o}ring},\ and\ \citenamefont {Seregin}}]{palmer2014real}%
  \BibitemOpen
  \bibfield  {author} {\bibinfo {author} {\bibfnamefont {TN}~\bibnamefont
  {Palmer}}, \bibinfo {author} {\bibfnamefont {A}~\bibnamefont {D{\"o}ring}}, \
  and\ \bibinfo {author} {\bibfnamefont {G}~\bibnamefont {Seregin}},\
  }\bibfield  {title} {\enquote {\bibinfo {title} {The real butterfly
  effect},}\ }\href@noop {} {\bibfield  {journal} {\bibinfo  {journal}
  {Nonlinearity}\ }\textbf {\bibinfo {volume} {27}},\ \bibinfo {pages} {R123}
  (\bibinfo {year} {2014})}\BibitemShut {NoStop}%
\bibitem [{\citenamefont {Palmer}(2019)}]{palmer2019stochastic}%
  \BibitemOpen
  \bibfield  {author} {\bibinfo {author} {\bibfnamefont {TN}~\bibnamefont
  {Palmer}},\ }\bibfield  {title} {\enquote {\bibinfo {title} {Stochastic
  weather and climate models},}\ }\href@noop {} {\bibfield  {journal} {\bibinfo
   {journal} {Nature Reviews Physics}\ }\textbf {\bibinfo {volume} {1}},\
  \bibinfo {pages} {463--471} (\bibinfo {year} {2019})}\BibitemShut {NoStop}%
\bibitem [{\citenamefont {Mailybaev}(2016)}]{mailybaev2016spontaneously}%
  \BibitemOpen
  \bibfield  {author} {\bibinfo {author} {\bibfnamefont {Alexei~A}\
  \bibnamefont {Mailybaev}},\ }\bibfield  {title} {\enquote {\bibinfo {title}
  {Spontaneously stochastic solutions in one-dimensional inviscid systems},}\
  }\href@noop {} {\bibfield  {journal} {\bibinfo  {journal} {Nonlinearity}\
  }\textbf {\bibinfo {volume} {29}},\ \bibinfo {pages} {2238} (\bibinfo {year}
  {2016})}\BibitemShut {NoStop}%
\bibitem [{\citenamefont {Biferale}\ \emph {et~al.}(2018)\citenamefont
  {Biferale}, \citenamefont {Boffetta}, \citenamefont {Mailybaev},\ and\
  \citenamefont {Scagliarini}}]{biferale2018rayleigh}%
  \BibitemOpen
  \bibfield  {author} {\bibinfo {author} {\bibfnamefont {Luca}\ \bibnamefont
  {Biferale}}, \bibinfo {author} {\bibfnamefont {Guido}\ \bibnamefont
  {Boffetta}}, \bibinfo {author} {\bibfnamefont {Alexei~A}\ \bibnamefont
  {Mailybaev}}, \ and\ \bibinfo {author} {\bibfnamefont {Andrea}\ \bibnamefont
  {Scagliarini}},\ }\bibfield  {title} {\enquote {\bibinfo {title}
  {Rayleigh-{T}aylor turbulence with singular nonuniform initial conditions},}\
  }\href@noop {} {\bibfield  {journal} {\bibinfo  {journal} {Physical Review
  Fluids}\ }\textbf {\bibinfo {volume} {3}},\ \bibinfo {pages} {092601}
  (\bibinfo {year} {2018})}\BibitemShut {NoStop}%
\bibitem [{\citenamefont {Thalabard}\ \emph {et~al.}(2020)\citenamefont
  {Thalabard}, \citenamefont {Bec},\ and\ \citenamefont
  {Mailybaev}}]{thalabard2020butterfly}%
  \BibitemOpen
  \bibfield  {author} {\bibinfo {author} {\bibfnamefont {Simon}\ \bibnamefont
  {Thalabard}}, \bibinfo {author} {\bibfnamefont {J{\'e}r{\'e}mie}\
  \bibnamefont {Bec}}, \ and\ \bibinfo {author} {\bibfnamefont {Alexei~A}\
  \bibnamefont {Mailybaev}},\ }\bibfield  {title} {\enquote {\bibinfo {title}
  {From the butterfly effect to spontaneous stochasticity in singular shear
  flows},}\ }\href@noop {} {\bibfield  {journal} {\bibinfo  {journal}
  {Communications Physics}\ }\textbf {\bibinfo {volume} {3}},\ \bibinfo {pages}
  {1--8} (\bibinfo {year} {2020})}\BibitemShut {NoStop}%
\bibitem [{\citenamefont {Wilson}\ and\ \citenamefont
  {Kogut}(1974)}]{wilson1974renormalization}%
  \BibitemOpen
  \bibfield  {author} {\bibinfo {author} {\bibfnamefont {Kenneth~G}\
  \bibnamefont {Wilson}}\ and\ \bibinfo {author} {\bibfnamefont {John}\
  \bibnamefont {Kogut}},\ }\bibfield  {title} {\enquote {\bibinfo {title} {The
  renormalization group and the $\epsilon$ expansion},}\ }\href@noop {}
  {\bibfield  {journal} {\bibinfo  {journal} {Physics reports}\ }\textbf
  {\bibinfo {volume} {12}},\ \bibinfo {pages} {75--199} (\bibinfo {year}
  {1974})}\BibitemShut {NoStop}%
\bibitem [{\citenamefont {Wilson}(1975)}]{wilson1975renormalization}%
  \BibitemOpen
  \bibfield  {author} {\bibinfo {author} {\bibfnamefont {Kenneth~G}\
  \bibnamefont {Wilson}},\ }\bibfield  {title} {\enquote {\bibinfo {title} {The
  renormalization group: {C}ritical phenomena and the {K}ondo problem},}\
  }\href@noop {} {\bibfield  {journal} {\bibinfo  {journal} {Reviews of modern
  physics}\ }\textbf {\bibinfo {volume} {47}},\ \bibinfo {pages} {773}
  (\bibinfo {year} {1975})}\BibitemShut {NoStop}%
\bibitem [{\citenamefont {Goldenfeld}(2018)}]{goldenfeld2018lectures}%
  \BibitemOpen
  \bibfield  {author} {\bibinfo {author} {\bibfnamefont {N.}~\bibnamefont
  {Goldenfeld}},\ }\href {https://books.google.com/books?id=HQpQDwAAQBAJ}
  {\emph {\bibinfo {title} {Lectures On Phase Transitions And The
  Renormalization Group}}}\ (\bibinfo  {publisher} {CRC Press},\ \bibinfo
  {year} {2018})\BibitemShut {NoStop}%
\bibitem [{\citenamefont {Onsager}(1949)}]{onsager1949statistical}%
  \BibitemOpen
  \bibfield  {author} {\bibinfo {author} {\bibfnamefont {Lars}\ \bibnamefont
  {Onsager}},\ }\bibfield  {title} {\enquote {\bibinfo {title} {Statistical
  hydrodynamics},}\ }\href@noop {} {\bibfield  {journal} {\bibinfo  {journal}
  {Il Nuovo Cimento (1943-1954)}\ }\textbf {\bibinfo {volume} {6}},\ \bibinfo
  {pages} {279--287} (\bibinfo {year} {1949})}\BibitemShut {NoStop}%
\bibitem [{\citenamefont {Eyink}\ and\ \citenamefont
  {Sreenivasan}(2006)}]{eyink2006onsager}%
  \BibitemOpen
  \bibfield  {author} {\bibinfo {author} {\bibfnamefont {Gregory~L}\
  \bibnamefont {Eyink}}\ and\ \bibinfo {author} {\bibfnamefont {Katepalli~R}\
  \bibnamefont {Sreenivasan}},\ }\bibfield  {title} {\enquote {\bibinfo {title}
  {Onsager and the theory of hydrodynamic turbulence},}\ }\href@noop {}
  {\bibfield  {journal} {\bibinfo  {journal} {Reviews of modern physics}\
  }\textbf {\bibinfo {volume} {78}},\ \bibinfo {pages} {87} (\bibinfo {year}
  {2006})}\BibitemShut {NoStop}%
\bibitem [{\citenamefont {Hartman}(1982)}]{hartman1982ordinary}%
  \BibitemOpen
  \bibfield  {author} {\bibinfo {author} {\bibfnamefont {P.}~\bibnamefont
  {Hartman}},\ }\href {https://books.google.com/books?id=NEkkJ93O9okC} {\emph
  {\bibinfo {title} {Ordinary Differential Equations}}},\ \bibinfo {edition}
  {2nd}\ ed.,\ Classics in Applied Mathematics\ (\bibinfo  {publisher} {Society
  for Industrial and Applied Mathematics},\ \bibinfo {address} {SIAM,
  Philadelphia, USA},\ \bibinfo {year} {1982})\BibitemShut {NoStop}%
\bibitem [{\citenamefont {Bafico}\ and\ \citenamefont
  {Baldi}(1982)}]{bafico1982small}%
  \BibitemOpen
  \bibfield  {author} {\bibinfo {author} {\bibfnamefont {R}~\bibnamefont
  {Bafico}}\ and\ \bibinfo {author} {\bibfnamefont {P}~\bibnamefont {Baldi}},\
  }\bibfield  {title} {\enquote {\bibinfo {title} {Small random perturbations
  of {P}eano phenomena},}\ }\href@noop {} {\bibfield  {journal} {\bibinfo
  {journal} {Stochastics}\ }\textbf {\bibinfo {volume} {6}},\ \bibinfo {pages}
  {279--292} (\bibinfo {year} {1982})}\BibitemShut {NoStop}%
\bibitem [{\citenamefont {Gradinaru}\ \emph {et~al.}(2001)\citenamefont
  {Gradinaru}, \citenamefont {Herrmann},\ and\ \citenamefont
  {Roynette}}]{gradinaru2001singular}%
  \BibitemOpen
  \bibfield  {author} {\bibinfo {author} {\bibfnamefont {Mihai}\ \bibnamefont
  {Gradinaru}}, \bibinfo {author} {\bibfnamefont {Samuel}\ \bibnamefont
  {Herrmann}}, \ and\ \bibinfo {author} {\bibfnamefont {Bernard}\ \bibnamefont
  {Roynette}},\ }\bibfield  {title} {\enquote {\bibinfo {title} {A singular
  large deviations phenomenon},}\ }\href@noop {} {\bibfield  {journal}
  {\bibinfo  {journal} {Annales de l'Institut Henri Poincar{\'e} -
  Probabilit{\'e}s et statistiques}\ }\textbf {\bibinfo {volume} {37}},\
  \bibinfo {pages} {555--580} (\bibinfo {year} {2001})}\BibitemShut {NoStop}%
\bibitem [{\citenamefont {Attanasio}\ and\ \citenamefont
  {Flandoli}(2009)}]{attanasio2009zero}%
  \BibitemOpen
  \bibfield  {author} {\bibinfo {author} {\bibfnamefont {Stefano}\ \bibnamefont
  {Attanasio}}\ and\ \bibinfo {author} {\bibfnamefont {Franco}\ \bibnamefont
  {Flandoli}},\ }\bibfield  {title} {\enquote {\bibinfo {title} {Zero-noise
  solutions of linear transport equations without uniqueness: an example},}\
  }\href@noop {} {\bibfield  {journal} {\bibinfo  {journal} {Comptes Rendus
  Mathematique}\ }\textbf {\bibinfo {volume} {347}},\ \bibinfo {pages}
  {753--756} (\bibinfo {year} {2009})}\BibitemShut {NoStop}%
\bibitem [{\citenamefont {Flandoli}(2013)}]{flandoli2013topics}%
  \BibitemOpen
  \bibfield  {author} {\bibinfo {author} {\bibfnamefont {Franco}\ \bibnamefont
  {Flandoli}},\ }\href@noop {} {\enquote {\bibinfo {title} {Topics on
  regularization by noise. {L}ecture {N}otes, {U}niversity of {P}isa},}\
  }\bibinfo {howpublished}
  {\url{http://users.dma.unipi.it/~flandoli/Berlino_Lectures_Flandoli.pdf}}
  (\bibinfo {year} {2013})\BibitemShut {NoStop}%
\bibitem [{\citenamefont {Freidlin}\ and\ \citenamefont
  {Wentzell}(2012)}]{freidlin2012random}%
  \BibitemOpen
  \bibfield  {author} {\bibinfo {author} {\bibfnamefont {M.I.}\ \bibnamefont
  {Freidlin}}\ and\ \bibinfo {author} {\bibfnamefont {A.D.}\ \bibnamefont
  {Wentzell}},\ }\href {https://books.google.com/books?id=3H\_gBwAAQBAJ} {\emph
  {\bibinfo {title} {Random Perturbations of Dynamical Systems}}},\ Grundlehren
  der mathematischen Wissenschaften\ (\bibinfo  {publisher} {Springer New
  York},\ \bibinfo {year} {2012})\BibitemShut {NoStop}%
\bibitem [{\citenamefont {Onsager}\ and\ \citenamefont
  {Machlup}(1953)}]{onsager1953fluctuations}%
  \BibitemOpen
  \bibfield  {author} {\bibinfo {author} {\bibfnamefont {Lars}\ \bibnamefont
  {Onsager}}\ and\ \bibinfo {author} {\bibfnamefont {Stefan}\ \bibnamefont
  {Machlup}},\ }\bibfield  {title} {\enquote {\bibinfo {title} {Fluctuations
  and irreversible processes},}\ }\href@noop {} {\bibfield  {journal} {\bibinfo
   {journal} {Physical Review}\ }\textbf {\bibinfo {volume} {91}},\ \bibinfo
  {pages} {1505} (\bibinfo {year} {1953})}\BibitemShut {NoStop}%
\bibitem [{\citenamefont {Wio}(2013)}]{wio2013path}%
  \BibitemOpen
  \bibfield  {author} {\bibinfo {author} {\bibfnamefont {H.S.}\ \bibnamefont
  {Wio}},\ }\href {https://books.google.com/books?id=jpG6CgAAQBAJ} {\emph
  {\bibinfo {title} {Path Integrals for Stochastic Processes: An
  Introduction}}}\ (\bibinfo  {publisher} {World Scientific},\ \bibinfo {year}
  {2013})\BibitemShut {NoStop}%
\bibitem [{\citenamefont {Nelson}\ and\ \citenamefont
  {Fisher}(1975)}]{nelson1975soluble}%
  \BibitemOpen
  \bibfield  {author} {\bibinfo {author} {\bibfnamefont {David~R}\ \bibnamefont
  {Nelson}}\ and\ \bibinfo {author} {\bibfnamefont {Michael~E}\ \bibnamefont
  {Fisher}},\ }\bibfield  {title} {\enquote {\bibinfo {title} {Soluble
  renormalization groups and scaling fields for low-dimensional {I}sing
  systems},}\ }\href@noop {} {\bibfield  {journal} {\bibinfo  {journal} {Annals
  of Physics}\ }\textbf {\bibinfo {volume} {91}},\ \bibinfo {pages} {226--274}
  (\bibinfo {year} {1975})}\BibitemShut {NoStop}%
\bibitem [{\citenamefont {Frisch}\ and\ \citenamefont
  {Kolmogorov}(1995)}]{frisch1995turbulence}%
  \BibitemOpen
  \bibfield  {author} {\bibinfo {author} {\bibfnamefont {U.}~\bibnamefont
  {Frisch}}\ and\ \bibinfo {author} {\bibfnamefont {A.N.}\ \bibnamefont
  {Kolmogorov}},\ }\href {https://books.google.com/books?id=K-Pf7RuYkf0C}
  {\emph {\bibinfo {title} {Turbulence: The Legacy of A. N. Kolmogorov}}}\
  (\bibinfo  {publisher} {Cambridge University Press},\ \bibinfo {year}
  {1995})\BibitemShut {NoStop}%
\bibitem [{\citenamefont {Sachdev}(2000)}]{sachdev2000quantum}%
  \BibitemOpen
  \bibfield  {author} {\bibinfo {author} {\bibfnamefont {Subir}\ \bibnamefont
  {Sachdev}},\ }\bibfield  {title} {\enquote {\bibinfo {title} {Quantum
  criticality: competing ground states in low dimensions},}\ }\href@noop {}
  {\bibfield  {journal} {\bibinfo  {journal} {Science}\ }\textbf {\bibinfo
  {volume} {288}},\ \bibinfo {pages} {475--480} (\bibinfo {year}
  {2000})}\BibitemShut {NoStop}%
\bibitem [{\citenamefont {Coleman}\ and\ \citenamefont
  {Schofield}(2005)}]{coleman2005quantum}%
  \BibitemOpen
  \bibfield  {author} {\bibinfo {author} {\bibfnamefont {Piers}\ \bibnamefont
  {Coleman}}\ and\ \bibinfo {author} {\bibfnamefont {Andrew~J}\ \bibnamefont
  {Schofield}},\ }\bibfield  {title} {\enquote {\bibinfo {title} {Quantum
  criticality},}\ }\href@noop {} {\bibfield  {journal} {\bibinfo  {journal}
  {Nature}\ }\textbf {\bibinfo {volume} {433}},\ \bibinfo {pages} {226--229}
  (\bibinfo {year} {2005})}\BibitemShut {NoStop}%
\bibitem [{\citenamefont {Lorenz}(1963)}]{lorenz1963deterministic}%
  \BibitemOpen
  \bibfield  {author} {\bibinfo {author} {\bibfnamefont {Edward~N}\
  \bibnamefont {Lorenz}},\ }\bibfield  {title} {\enquote {\bibinfo {title}
  {Deterministic nonperiodic flow},}\ }\href@noop {} {\bibfield  {journal}
  {\bibinfo  {journal} {Journal of the atmospheric sciences}\ }\textbf
  {\bibinfo {volume} {20}},\ \bibinfo {pages} {130--141} (\bibinfo {year}
  {1963})}\BibitemShut {NoStop}%
\bibitem [{\citenamefont {Cvitanovi{\'c}}\ \emph {et~al.}(2016)\citenamefont
  {Cvitanovi{\'c}}, \citenamefont {Artuso}, \citenamefont {Mainieri},
  \citenamefont {Tanner},\ and\ \citenamefont {Vattay}}]{ChaosBook}%
  \BibitemOpen
  \bibfield  {author} {\bibinfo {author} {\bibfnamefont {P.}~\bibnamefont
  {Cvitanovi{\'c}}}, \bibinfo {author} {\bibfnamefont {R.}~\bibnamefont
  {Artuso}}, \bibinfo {author} {\bibfnamefont {R.}~\bibnamefont {Mainieri}},
  \bibinfo {author} {\bibfnamefont {G.}~\bibnamefont {Tanner}}, \ and\ \bibinfo
  {author} {\bibfnamefont {G.}~\bibnamefont {Vattay}},\ }\href@noop {}
  {\enquote {\bibinfo {title} {{C}haos: {C}lassical and {Q}uantum},}\ }\bibinfo
  {howpublished} {Niels Bohr Inst., \url{http://ChaosBook.org}} (\bibinfo
  {year} {2016})\BibitemShut {NoStop}%
\bibitem [{\citenamefont {Monin}\ and\ \citenamefont
  {Yaglom}(2013)}]{monin2013statistical}%
  \BibitemOpen
  \bibfield  {author} {\bibinfo {author} {\bibfnamefont {A.S.}\ \bibnamefont
  {Monin}}\ and\ \bibinfo {author} {\bibfnamefont {A.M.}\ \bibnamefont
  {Yaglom}},\ }\href {https://books.google.com/books?id=6xPEAgAAQBAJ} {\emph
  {\bibinfo {title} {Statistical Fluid Mechanics, Volume II: Mechanics of
  Turbulence}}},\ Dover Books on Physics\ (\bibinfo  {publisher} {Dover
  Publications},\ \bibinfo {year} {2013})\BibitemShut {NoStop}%
\bibitem [{\citenamefont {Brown}\ and\ \citenamefont
  {Roshko}(2012)}]{brown2012turbulent}%
  \BibitemOpen
  \bibfield  {author} {\bibinfo {author} {\bibfnamefont {Garry~L}\ \bibnamefont
  {Brown}}\ and\ \bibinfo {author} {\bibfnamefont {Anatol}\ \bibnamefont
  {Roshko}},\ }\bibfield  {title} {\enquote {\bibinfo {title} {Turbulent shear
  layers and wakes},}\ }\href@noop {} {\bibfield  {journal} {\bibinfo
  {journal} {Journal of Turbulence}\ ,\ \bibinfo {pages} {N51}} (\bibinfo
  {year} {2012})}\BibitemShut {NoStop}%
\bibitem [{\citenamefont {Suryanarayanan}\ \emph {et~al.}(2014)\citenamefont
  {Suryanarayanan}, \citenamefont {Narasimha},\ and\ \citenamefont
  {Dass}}]{suryanarayanan2014free}%
  \BibitemOpen
  \bibfield  {author} {\bibinfo {author} {\bibfnamefont {Saikishan}\
  \bibnamefont {Suryanarayanan}}, \bibinfo {author} {\bibfnamefont {Roddam}\
  \bibnamefont {Narasimha}}, \ and\ \bibinfo {author} {\bibfnamefont {ND~Hari}\
  \bibnamefont {Dass}},\ }\bibfield  {title} {\enquote {\bibinfo {title} {Free
  turbulent shear layer in a point vortex gas as a problem in nonequilibrium
  statistical mechanics},}\ }\href@noop {} {\bibfield  {journal} {\bibinfo
  {journal} {Physical Review E}\ }\textbf {\bibinfo {volume} {89}},\ \bibinfo
  {pages} {013009} (\bibinfo {year} {2014})}\BibitemShut {NoStop}%
\bibitem [{\citenamefont {Donzis}\ \emph {et~al.}(2005)\citenamefont {Donzis},
  \citenamefont {Sreenivasan},\ and\ \citenamefont {Yeung}}]{donzis2005scalar}%
  \BibitemOpen
  \bibfield  {author} {\bibinfo {author} {\bibfnamefont {DA}~\bibnamefont
  {Donzis}}, \bibinfo {author} {\bibfnamefont {KR}~\bibnamefont {Sreenivasan}},
  \ and\ \bibinfo {author} {\bibfnamefont {PKc}\ \bibnamefont {Yeung}},\
  }\bibfield  {title} {\enquote {\bibinfo {title} {Scalar dissipation rate and
  dissipative anomaly in isotropic turbulence},}\ }\href@noop {} {\bibfield
  {journal} {\bibinfo  {journal} {Journal of Fluid Mechanics}\ }\textbf
  {\bibinfo {volume} {532}},\ \bibinfo {pages} {199--216} (\bibinfo {year}
  {2005})}\BibitemShut {NoStop}%
\bibitem [{\citenamefont {Bourgoin}\ \emph {et~al.}(2006)\citenamefont
  {Bourgoin}, \citenamefont {Ouellette}, \citenamefont {Xu}, \citenamefont
  {Berg},\ and\ \citenamefont {Bodenschatz}}]{bourgoin2006role}%
  \BibitemOpen
  \bibfield  {author} {\bibinfo {author} {\bibfnamefont {Micka{\"e}l}\
  \bibnamefont {Bourgoin}}, \bibinfo {author} {\bibfnamefont {Nicholas~T}\
  \bibnamefont {Ouellette}}, \bibinfo {author} {\bibfnamefont {Haitao}\
  \bibnamefont {Xu}}, \bibinfo {author} {\bibfnamefont {Jacob}\ \bibnamefont
  {Berg}}, \ and\ \bibinfo {author} {\bibfnamefont {Eberhard}\ \bibnamefont
  {Bodenschatz}},\ }\bibfield  {title} {\enquote {\bibinfo {title} {The role of
  pair dispersion in turbulent flow},}\ }\href@noop {} {\bibfield  {journal}
  {\bibinfo  {journal} {Science}\ }\textbf {\bibinfo {volume} {311}},\ \bibinfo
  {pages} {835--838} (\bibinfo {year} {2006})}\BibitemShut {NoStop}%
\bibitem [{\citenamefont {Ouellette}\ \emph {et~al.}(2006)\citenamefont
  {Ouellette}, \citenamefont {Xu}, \citenamefont {Bourgoin},\ and\
  \citenamefont {Bodenschatz}}]{ouellette2006experimental}%
  \BibitemOpen
  \bibfield  {author} {\bibinfo {author} {\bibfnamefont {Nicholas~T}\
  \bibnamefont {Ouellette}}, \bibinfo {author} {\bibfnamefont {Haitao}\
  \bibnamefont {Xu}}, \bibinfo {author} {\bibfnamefont {Micka{\"e}l}\
  \bibnamefont {Bourgoin}}, \ and\ \bibinfo {author} {\bibfnamefont {Eberhard}\
  \bibnamefont {Bodenschatz}},\ }\bibfield  {title} {\enquote {\bibinfo {title}
  {An experimental study of turbulent relative dispersion models},}\
  }\href@noop {} {\bibfield  {journal} {\bibinfo  {journal} {New Journal of
  Physics}\ }\textbf {\bibinfo {volume} {8}},\ \bibinfo {pages} {109} (\bibinfo
  {year} {2006})}\BibitemShut {NoStop}%
\bibitem [{\citenamefont {Kolmogorov}(1941)}]{kolmogorov1941local}%
  \BibitemOpen
  \bibfield  {author} {\bibinfo {author} {\bibfnamefont {Andrei~Nikolaevitch}\
  \bibnamefont {Kolmogorov}},\ }\bibfield  {title} {\enquote {\bibinfo {title}
  {Local turbulence structure in incompressible fluids at very high reynolds
  numbers},}\ }\href@noop {} {\bibfield  {journal} {\bibinfo  {journal} {Dokl.
  Akad. Nauk SSSR}\ }\textbf {\bibinfo {volume} {30}},\ \bibinfo {pages}
  {299--303} (\bibinfo {year} {1941})}\BibitemShut {NoStop}%
\bibitem [{\citenamefont {Bitane}\ \emph {et~al.}(2013)\citenamefont {Bitane},
  \citenamefont {Homann},\ and\ \citenamefont {Bec}}]{bitane2013geometry}%
  \BibitemOpen
  \bibfield  {author} {\bibinfo {author} {\bibfnamefont {Rehab}\ \bibnamefont
  {Bitane}}, \bibinfo {author} {\bibfnamefont {Holger}\ \bibnamefont {Homann}},
  \ and\ \bibinfo {author} {\bibfnamefont {J{\'e}r{\'e}mie}\ \bibnamefont
  {Bec}},\ }\bibfield  {title} {\enquote {\bibinfo {title} {Geometry and
  violent events in turbulent pair dispersion},}\ }\href@noop {} {\bibfield
  {journal} {\bibinfo  {journal} {Journal of Turbulence}\ }\textbf {\bibinfo
  {volume} {14}},\ \bibinfo {pages} {23--45} (\bibinfo {year}
  {2013})}\BibitemShut {NoStop}%
\bibitem [{Note1()}]{Note1}%
  \BibitemOpen
  \bibinfo {note} {Paper \cite {eyink2011fast} used the terminology
  ``(magnetic) Prandtl number'' rather than ``Schmidt number'', because the
  application considered there was the turbulent kinematic dynamo}\BibitemShut
  {NoStop}%
\bibitem [{\citenamefont {Boffetta}\ and\ \citenamefont
  {Musacchio}(2017)}]{boffetta2017chaos}%
  \BibitemOpen
  \bibfield  {author} {\bibinfo {author} {\bibfnamefont {Guido}\ \bibnamefont
  {Boffetta}}\ and\ \bibinfo {author} {\bibfnamefont {Stefano}\ \bibnamefont
  {Musacchio}},\ }\bibfield  {title} {\enquote {\bibinfo {title} {Chaos and
  predictability of homogeneous-isotropic turbulence},}\ }\href@noop {}
  {\bibfield  {journal} {\bibinfo  {journal} {Physical review letters}\
  }\textbf {\bibinfo {volume} {119}},\ \bibinfo {pages} {054102} (\bibinfo
  {year} {2017})}\BibitemShut {NoStop}%
\bibitem [{\citenamefont {Berera}\ and\ \citenamefont
  {Ho}(2018)}]{berera2018chaotic}%
  \BibitemOpen
  \bibfield  {author} {\bibinfo {author} {\bibfnamefont {Arjun}\ \bibnamefont
  {Berera}}\ and\ \bibinfo {author} {\bibfnamefont {Richard~DJG}\ \bibnamefont
  {Ho}},\ }\bibfield  {title} {\enquote {\bibinfo {title} {Chaotic properties
  of a turbulent isotropic fluid},}\ }\href@noop {} {\bibfield  {journal}
  {\bibinfo  {journal} {Physical review letters}\ }\textbf {\bibinfo {volume}
  {120}},\ \bibinfo {pages} {024101} (\bibinfo {year} {2018})}\BibitemShut
  {NoStop}%
\bibitem [{\citenamefont {Leith}\ and\ \citenamefont
  {Kraichnan}(1972)}]{leith1972predictability}%
  \BibitemOpen
  \bibfield  {author} {\bibinfo {author} {\bibfnamefont {CE}~\bibnamefont
  {Leith}}\ and\ \bibinfo {author} {\bibfnamefont {RH}~\bibnamefont
  {Kraichnan}},\ }\bibfield  {title} {\enquote {\bibinfo {title}
  {Predictability of turbulent flows},}\ }\href@noop {} {\bibfield  {journal}
  {\bibinfo  {journal} {Journal of the Atmospheric Sciences}\ }\textbf
  {\bibinfo {volume} {29}},\ \bibinfo {pages} {1041--1058} (\bibinfo {year}
  {1972})}\BibitemShut {NoStop}%
\bibitem [{\citenamefont {Buckmaster}\ \emph {et~al.}(2019)\citenamefont
  {Buckmaster}, \citenamefont {Shkoller},\ and\ \citenamefont
  {Vicol}}]{buckmaster2019nonuniqueness}%
  \BibitemOpen
  \bibfield  {author} {\bibinfo {author} {\bibfnamefont {Tristan}\ \bibnamefont
  {Buckmaster}}, \bibinfo {author} {\bibfnamefont {Steve}\ \bibnamefont
  {Shkoller}}, \ and\ \bibinfo {author} {\bibfnamefont {Vlad}\ \bibnamefont
  {Vicol}},\ }\bibfield  {title} {\enquote {\bibinfo {title} {Nonuniqueness of
  weak solutions to the {SQG} equation},}\ }\href@noop {} {\bibfield  {journal}
  {\bibinfo  {journal} {Communications on Pure and Applied Mathematics}\
  }\textbf {\bibinfo {volume} {72}},\ \bibinfo {pages} {1809--1874} (\bibinfo
  {year} {2019})}\BibitemShut {NoStop}%
\bibitem [{\citenamefont {Sz{\'e}kelyhidi~Jr}(2011)}]{szekelyhidi2011weak}%
  \BibitemOpen
  \bibfield  {author} {\bibinfo {author} {\bibfnamefont {L{\'a}szl{\'o}}\
  \bibnamefont {Sz{\'e}kelyhidi~Jr}},\ }\bibfield  {title} {\enquote {\bibinfo
  {title} {Weak solutions to the incompressible {E}uler equations with vortex
  sheet initial data},}\ }\href@noop {} {\bibfield  {journal} {\bibinfo
  {journal} {Comptes Rendus Mathematique}\ }\textbf {\bibinfo {volume} {349}},\
  \bibinfo {pages} {1063--1066} (\bibinfo {year} {2011})}\BibitemShut {NoStop}%
\bibitem [{\citenamefont {Mengual}\ and\ \citenamefont
  {Sz{\'e}kelyhidi~Jr}(2020)}]{mengual2020dissipative}%
  \BibitemOpen
  \bibfield  {author} {\bibinfo {author} {\bibfnamefont {Francisco}\
  \bibnamefont {Mengual}}\ and\ \bibinfo {author} {\bibfnamefont
  {L{\'a}szl{\'o}}\ \bibnamefont {Sz{\'e}kelyhidi~Jr}},\ }\bibfield  {title}
  {\enquote {\bibinfo {title} {Dissipative {E}uler flows for vortex sheet
  initial data without distinguished sign},}\ }\href@noop {} {\bibfield
  {journal} {\bibinfo  {journal} {arXiv preprint arXiv:2005.08333}\ } (\bibinfo
  {year} {2020})}\BibitemShut {NoStop}%
\bibitem [{\citenamefont {Athanassoulis}\ and\ \citenamefont
  {Paul}(2012)}]{athanassoulis2012strong}%
  \BibitemOpen
  \bibfield  {author} {\bibinfo {author} {\bibfnamefont {Agissilaos}\
  \bibnamefont {Athanassoulis}}\ and\ \bibinfo {author} {\bibfnamefont
  {Thierry}\ \bibnamefont {Paul}},\ }\bibfield  {title} {\enquote {\bibinfo
  {title} {Strong and weak semiclassical limit for some rough
  {H}amiltonians},}\ }\href@noop {} {\bibfield  {journal} {\bibinfo  {journal}
  {Mathematical Models and Methods in Applied Sciences}\ }\textbf {\bibinfo
  {volume} {22}},\ \bibinfo {pages} {1250038} (\bibinfo {year}
  {2012})}\BibitemShut {NoStop}%
\bibitem [{\citenamefont {Eyink}\ and\ \citenamefont
  {Drivas}(2015{\natexlab{a}})}]{eyink2015quantum}%
  \BibitemOpen
  \bibfield  {author} {\bibinfo {author} {\bibfnamefont {Gregory~L}\
  \bibnamefont {Eyink}}\ and\ \bibinfo {author} {\bibfnamefont {Theodore~D}\
  \bibnamefont {Drivas}},\ }\href@noop {} {\enquote {\bibinfo {title} {Quantum
  spontaneous stochasticity},}\ }\bibinfo {howpublished} {arXiv preprint
  arXiv:1509.04941, \url{https://arxiv.org/abs/1509.04941}} (\bibinfo {year}
  {2015}{\natexlab{a}})\BibitemShut {NoStop}%
\bibitem [{\citenamefont {Le~Jan}\ \emph {et~al.}(2002)\citenamefont {Le~Jan},
  \citenamefont {Raimond} \emph {et~al.}}]{lejan2002integration}%
  \BibitemOpen
  \bibfield  {author} {\bibinfo {author} {\bibfnamefont {Yves}\ \bibnamefont
  {Le~Jan}}, \bibinfo {author} {\bibfnamefont {Olivier}\ \bibnamefont
  {Raimond}},  \emph {et~al.},\ }\bibfield  {title} {\enquote {\bibinfo {title}
  {Integration of {B}rownian vector fields},}\ }\href@noop {} {\bibfield
  {journal} {\bibinfo  {journal} {the Annals of Probability}\ }\textbf
  {\bibinfo {volume} {30}},\ \bibinfo {pages} {826--873} (\bibinfo {year}
  {2002})}\BibitemShut {NoStop}%
\bibitem [{\citenamefont {Le~Jan}\ \emph {et~al.}(2004)\citenamefont {Le~Jan},
  \citenamefont {Raimond} \emph {et~al.}}]{lejan2004flows}%
  \BibitemOpen
  \bibfield  {author} {\bibinfo {author} {\bibfnamefont {Yves}\ \bibnamefont
  {Le~Jan}}, \bibinfo {author} {\bibfnamefont {Olivier}\ \bibnamefont
  {Raimond}},  \emph {et~al.},\ }\bibfield  {title} {\enquote {\bibinfo {title}
  {Flows, coalescence and noise},}\ }\href@noop {} {\bibfield  {journal}
  {\bibinfo  {journal} {The Annals of Probability}\ }\textbf {\bibinfo {volume}
  {32}},\ \bibinfo {pages} {1247--1315} (\bibinfo {year} {2004})}\BibitemShut
  {NoStop}%
\bibitem [{\citenamefont {Eyink}\ and\ \citenamefont
  {Drivas}(2015{\natexlab{b}})}]{eyink2015spontaneous}%
  \BibitemOpen
  \bibfield  {author} {\bibinfo {author} {\bibfnamefont {Gregory~L}\
  \bibnamefont {Eyink}}\ and\ \bibinfo {author} {\bibfnamefont {Theodore~D}\
  \bibnamefont {Drivas}},\ }\bibfield  {title} {\enquote {\bibinfo {title}
  {Spontaneous stochasticity and anomalous dissipation for {B}urgers
  equation},}\ }\href@noop {} {\bibfield  {journal} {\bibinfo  {journal}
  {Journal of Statistical Physics}\ }\textbf {\bibinfo {volume} {158}},\
  \bibinfo {pages} {386--432} (\bibinfo {year}
  {2015}{\natexlab{b}})}\BibitemShut {NoStop}%
\bibitem [{\citenamefont {Delamotte}(2012)}]{delamotte2012introduction}%
  \BibitemOpen
  \bibfield  {author} {\bibinfo {author} {\bibfnamefont {Bertrand}\
  \bibnamefont {Delamotte}},\ }\bibfield  {title} {\enquote {\bibinfo {title}
  {An introduction to the nonperturbative renormalization group},}\ }in\
  \href@noop {} {\emph {\bibinfo {booktitle} {Renormalization Group and
  Effective Field Theory Approaches to Many-Body Systems}}}\ (\bibinfo
  {publisher} {Springer},\ \bibinfo {year} {2012})\ pp.\ \bibinfo {pages}
  {49--132}\BibitemShut {NoStop}%
\bibitem [{\citenamefont {Canet}\ \emph {et~al.}(2011)\citenamefont {Canet},
  \citenamefont {Chat{\'e}},\ and\ \citenamefont
  {Delamotte}}]{canet2011general}%
  \BibitemOpen
  \bibfield  {author} {\bibinfo {author} {\bibfnamefont {L{\'e}onie}\
  \bibnamefont {Canet}}, \bibinfo {author} {\bibfnamefont {Hugues}\
  \bibnamefont {Chat{\'e}}}, \ and\ \bibinfo {author} {\bibfnamefont
  {Bertrand}\ \bibnamefont {Delamotte}},\ }\bibfield  {title} {\enquote
  {\bibinfo {title} {General framework of the non-perturbative renormalization
  group for non-equilibrium steady states},}\ }\href@noop {} {\bibfield
  {journal} {\bibinfo  {journal} {Journal of Physics A: Mathematical and
  Theoretical}\ }\textbf {\bibinfo {volume} {44}},\ \bibinfo {pages} {495001}
  (\bibinfo {year} {2011})}\BibitemShut {NoStop}%
\bibitem [{\citenamefont {Risken}(2012)}]{risken2012fokker}%
  \BibitemOpen
  \bibfield  {author} {\bibinfo {author} {\bibfnamefont {H.}~\bibnamefont
  {Risken}},\ }\href {https://books.google.com/books?id=dXvpCAAAQBAJ} {\emph
  {\bibinfo {title} {The Fokker-Planck Equation: Methods of Solution and
  Applications}}},\ Springer Series in Synergetics\ (\bibinfo  {publisher}
  {Springer Berlin Heidelberg},\ \bibinfo {year} {2012})\BibitemShut {NoStop}%
\bibitem [{\citenamefont {Goldenfeld}\ \emph {et~al.}(1990)\citenamefont
  {Goldenfeld}, \citenamefont {Martin}, \citenamefont {Oono},\ and\
  \citenamefont {Liu}}]{goldenfeld1990anomalous}%
  \BibitemOpen
  \bibfield  {author} {\bibinfo {author} {\bibfnamefont {Nigel}\ \bibnamefont
  {Goldenfeld}}, \bibinfo {author} {\bibfnamefont {Olivier}\ \bibnamefont
  {Martin}}, \bibinfo {author} {\bibfnamefont {Yoshitsugu}\ \bibnamefont
  {Oono}}, \ and\ \bibinfo {author} {\bibfnamefont {Fong}\ \bibnamefont
  {Liu}},\ }\bibfield  {title} {\enquote {\bibinfo {title} {Anomalous
  dimensions and the renormalization group in a nonlinear diffusion process},}\
  }\href@noop {} {\bibfield  {journal} {\bibinfo  {journal} {Physical review
  letters}\ }\textbf {\bibinfo {volume} {64}},\ \bibinfo {pages} {1361}
  (\bibinfo {year} {1990})}\BibitemShut {NoStop}%
\bibitem [{\citenamefont {Goldenfeld}\ \emph {et~al.}(1991)\citenamefont
  {Goldenfeld}, \citenamefont {Martin},\ and\ \citenamefont
  {Oono}}]{goldenfeld1991asymptotics}%
  \BibitemOpen
  \bibfield  {author} {\bibinfo {author} {\bibfnamefont {Nigel}\ \bibnamefont
  {Goldenfeld}}, \bibinfo {author} {\bibfnamefont {Olivier}\ \bibnamefont
  {Martin}}, \ and\ \bibinfo {author} {\bibfnamefont {Y}~\bibnamefont {Oono}},\
  }\bibfield  {title} {\enquote {\bibinfo {title} {Asymptotics of partial
  differential equations and the renormalisation group},}\ }in\ \href@noop {}
  {\emph {\bibinfo {booktitle} {Asymptotics Beyond All Orders}}}\ (\bibinfo
  {publisher} {Springer},\ \bibinfo {year} {1991})\ pp.\ \bibinfo {pages}
  {375--383}\BibitemShut {NoStop}%
\bibitem [{\citenamefont {Bricmont}\ and\ \citenamefont
  {Kupiainen}(1995)}]{bricmont1995renormalizing}%
  \BibitemOpen
  \bibfield  {author} {\bibinfo {author} {\bibfnamefont {Jean}\ \bibnamefont
  {Bricmont}}\ and\ \bibinfo {author} {\bibfnamefont {Antti}\ \bibnamefont
  {Kupiainen}},\ }\bibfield  {title} {\enquote {\bibinfo {title} {Renormalizing
  partial differential equations},}\ }in\ \href@noop {} {\emph {\bibinfo
  {booktitle} {Constructive Physics Results in Field Theory, Statistical
  Mechanics and Condensed Matter Physics}}}\ (\bibinfo  {publisher}
  {Springer},\ \bibinfo {year} {1995})\ pp.\ \bibinfo {pages}
  {83--115}\BibitemShut {NoStop}%
\bibitem [{\citenamefont {Kadanoff}\ and\ \citenamefont
  {Wegner}(1971)}]{kadanoff1971some}%
  \BibitemOpen
  \bibfield  {author} {\bibinfo {author} {\bibfnamefont {Leo~P}\ \bibnamefont
  {Kadanoff}}\ and\ \bibinfo {author} {\bibfnamefont {Franz~J}\ \bibnamefont
  {Wegner}},\ }\bibfield  {title} {\enquote {\bibinfo {title} {Some critical
  properties of the eight-vertex model},}\ }\href@noop {} {\bibfield  {journal}
  {\bibinfo  {journal} {Physical Review B}\ }\textbf {\bibinfo {volume} {4}},\
  \bibinfo {pages} {3989} (\bibinfo {year} {1971})}\BibitemShut {NoStop}%
\bibitem [{\citenamefont {Van~Leeuwen}(1975)}]{van1975singularities}%
  \BibitemOpen
  \bibfield  {author} {\bibinfo {author} {\bibfnamefont {JMJ}\ \bibnamefont
  {Van~Leeuwen}},\ }\bibfield  {title} {\enquote {\bibinfo {title}
  {Singularities in the critical surface and universality for {I}sing-like spin
  systems},}\ }\href@noop {} {\bibfield  {journal} {\bibinfo  {journal}
  {Physical Review Letters}\ }\textbf {\bibinfo {volume} {34}},\ \bibinfo
  {pages} {1056} (\bibinfo {year} {1975})}\BibitemShut {NoStop}%
\bibitem [{Note2()}]{Note2}%
  \BibitemOpen
  \bibinfo {note} {The exponent $\protect \frac {1-h}{1+h}\not =1$ is not,
  however, an ``anomalous dimension'' in the sense of critical phenomenon and
  quantum-field theory. As discussed in section \ref {dimension}, the factor
  $Re^{\protect \frac {1-h}{1+h}}$ appears in the ``bridging relation''
  \protect \textup {\hbox {\mathsurround \z@ \protect \normalfont
  (\ignorespaces \ref {large-Re-b}\unskip \@@italiccorr )}} because of simple
  dimensional analysis.}\BibitemShut {Stop}%
\bibitem [{Note3()}]{Note3}%
  \BibitemOpen
  \bibinfo {note} {The equation (2) of \cite {gradinaru2001singular} differs
  from our inertial-range Langevin equation \protect \textup {\hbox
  {\mathsurround \z@ \protect \normalfont (\ignorespaces \ref
  {langevin-int}\unskip \@@italiccorr )}} by some simple factors of 2. Setting
  $x\to x/2^{1+h},$ $t\to t/2^{\protect \frac {1-h}{1+h}}$ in (2) of \cite
  {gradinaru2001singular} gives our \protect \textup {\hbox {\mathsurround \z@
  \protect \normalfont (\ignorespaces \ref {langevin-int}\unskip \@@italiccorr
  )}}, with their $\varepsilon $ related to our P\'eclet number by $\varepsilon
  ^2=1/Pe.$ By this change of variables, all results of \cite
  {gradinaru2001singular} are transformed into the $Sc\to 0$ limit of
  ours.}\BibitemShut {Stop}%
\bibitem [{\citenamefont {Gulisashvili}\ and\ \citenamefont
  {Kon}(1996)}]{gulisashvili1996exact}%
  \BibitemOpen
  \bibfield  {author} {\bibinfo {author} {\bibfnamefont {Archil}\ \bibnamefont
  {Gulisashvili}}\ and\ \bibinfo {author} {\bibfnamefont {Mark~A}\ \bibnamefont
  {Kon}},\ }\bibfield  {title} {\enquote {\bibinfo {title} {Exact smoothing
  properties of {S}chr{\"o}dinger semigroups},}\ }\href@noop {} {\bibfield
  {journal} {\bibinfo  {journal} {American Journal of Mathematics}\ }\textbf
  {\bibinfo {volume} {118}},\ \bibinfo {pages} {1215--1248} (\bibinfo {year}
  {1996})}\BibitemShut {NoStop}%
\bibitem [{Note4()}]{Note4}%
  \BibitemOpen
  \bibinfo {note} {An attentive reader will have noticed that this same
  argument derives directly the singular large-deviations estimate \protect
  \textup {\hbox {\mathsurround \z@ \protect \normalfont (\ignorespaces \ref
  {Grad-int}\unskip \@@italiccorr )}}, without any use of the RG flow
  equations. In fact, the relation \protect \textup {\hbox {\mathsurround \z@
  \protect \normalfont (\ignorespaces \ref {int-dis}\unskip \@@italiccorr )}}
  between inertial and dissipation range variables implies the relation of
  their PDF's $\protect \mathaccentV {bar}016{P}_{Re}(\protect \mathaccentV
  {bar}016{x},\protect \mathaccentV {bar}016{t}) = Re^{\protect \frac {1}{1+h}}
  P( Re^{\protect \frac {1}{1+h}}\protect \mathaccentV {bar}016{x},Re^{\protect
  \frac {1-h}{1+h}}\protect \mathaccentV {bar}016{t})$ and this result may be
  use to estimate $\protect \mathaccentV {bar}016{P}_{Re}(\protect \mathaccentV
  {bar}016{x},\protect \mathaccentV {bar}016{t})$ as $Re\to \infty $ in the
  same manner as \protect \textup {\hbox {\mathsurround \z@ \protect
  \normalfont (\ignorespaces \ref {Q-def}\unskip \@@italiccorr )}} was used to
  derive an estimate for $Q_b(x_w,t_w)$ as $b\to \infty .$}\BibitemShut
  {NoStop}%
\bibitem [{Note5()}]{Note5}%
  \BibitemOpen
  \bibinfo {note} {For example, the Tanaka formula that was used in \cite
  {gradinaru2001singular} for the It$\protect \mathaccentV {bar}016{\protect
  \rm o}$-differential $d|B(t)|$ of the absolute value of a Wiener process
  $B(t)$ is derived by a regularization such as $|x|\to \protect \frac
  {x^2}{2\ell }+\protect \frac {\ell }{2}$ for $|x|<\ell $ with $\ell \to
  0$}\BibitemShut {NoStop}%
\bibitem [{Note6()}]{Note6}%
  \BibitemOpen
  \bibinfo {note} {There is nothing special about the origin for our argument
  and we could consider equally well the probabilities of the two complementary
  events $x(t)>a$ and $x(t)<a,$ for any fixed real constant $a$. However, it is
  natural to choose $a=0.$}\BibitemShut {NoStop}%
\bibitem [{Note7()}]{Note7}%
  \BibitemOpen
  \bibinfo {note} {These zero modes belong to the Banach space ${\protect
  \mathcal L}^\infty $ of bounded, measurable functions, where the Markov
  semigroup $e^{t\protect \mathaccentV {hat}05E{L}^*}$ naturally acts. None of
  these modes appears in the eigenfunction expansion (\ref {expand}), because
  the corresponding wavefunctions $\varPsi $ via the relation (\ref {Phi-Psi})
  are not square-integrable.}\BibitemShut {Stop}%
\bibitem [{\citenamefont {Lin}\ \emph {et~al.}(2011)\citenamefont {Lin},
  \citenamefont {Guo-sheng},\ and\ \citenamefont {Wen-biao}}]{lin2011gaussian}%
  \BibitemOpen
  \bibfield  {author} {\bibinfo {author} {\bibfnamefont {Wang}\ \bibnamefont
  {Lin}}, \bibinfo {author} {\bibfnamefont {Rui}\ \bibnamefont {Guo-sheng}}, \
  and\ \bibinfo {author} {\bibfnamefont {Tian}\ \bibnamefont {Wen-biao}},\
  }\bibfield  {title} {\enquote {\bibinfo {title} {Gaussian white noise
  generating based on {FPGA [J]}},}\ }\href@noop {} {\bibfield  {journal}
  {\bibinfo  {journal} {Modern Electronics Technique}\ }\textbf {\bibinfo
  {volume} {3}},\ \bibinfo {pages} {104--106} (\bibinfo {year}
  {2011})}\BibitemShut {NoStop}%
\bibitem [{\citenamefont {Shultz}\ and\ \citenamefont
  {Haak}(2018)}]{schultz2018pocket}%
  \BibitemOpen
  \bibfield  {author} {\bibinfo {author} {\bibfnamefont {Aaron}\ \bibnamefont
  {Shultz}}\ and\ \bibinfo {author} {\bibfnamefont {Peter}\ \bibnamefont
  {Haak}},\ }\bibfield  {title} {\enquote {\bibinfo {title} {Pocket-size white
  noise generator for quickly testing circuit signal response},}\ }\href@noop
  {} {\bibfield  {journal} {\bibinfo  {journal} {Analog Dialogue}\ }\textbf
  {\bibinfo {volume} {52}},\ \bibinfo {pages} {1--4} (\bibinfo {year}
  {2018})}\BibitemShut {NoStop}%
\bibitem [{\citenamefont {Strogatz}(2018)}]{strogatz2018nonlinear}%
  \BibitemOpen
  \bibfield  {author} {\bibinfo {author} {\bibfnamefont {S.H.}\ \bibnamefont
  {Strogatz}},\ }\href {https://books.google.com/books?id=1kpnDwAAQBAJ} {\emph
  {\bibinfo {title} {Nonlinear Dynamics and Chaos: With Applications to
  Physics, Biology, Chemistry, and Engineering}}}\ (\bibinfo  {publisher} {CRC
  Press},\ \bibinfo {year} {2018})\BibitemShut {NoStop}%
\bibitem [{\citenamefont {Norton}(2003)}]{norton2003causation}%
  \BibitemOpen
  \bibfield  {author} {\bibinfo {author} {\bibfnamefont {John~D}\ \bibnamefont
  {Norton}},\ }\bibfield  {title} {\enquote {\bibinfo {title} {Causation as
  folk science},}\ }\href {http://www.philosophersimprint.org/003004}
  {\bibfield  {journal} {\bibinfo  {journal} {Philosophers' Imprint}\ }\textbf
  {\bibinfo {volume} {3}} (\bibinfo {year} {2003})}\BibitemShut {NoStop}%
\bibitem [{\citenamefont {Vanpoucke}\ and\ \citenamefont
  {Wenmackers}(2020)}]{vanpoucke2020assigning}%
  \BibitemOpen
  \bibfield  {author} {\bibinfo {author} {\bibfnamefont {Danny~EP}\
  \bibnamefont {Vanpoucke}}\ and\ \bibinfo {author} {\bibfnamefont {Sylvia}\
  \bibnamefont {Wenmackers}},\ }\bibfield  {title} {\enquote {\bibinfo {title}
  {Assigning probabilities to non-{L}ipschitz mechanical systems},}\
  }\href@noop {} {\bibfield  {journal} {\bibinfo  {journal} {arXiv preprint
  arXiv:2001.10375}\ } (\bibinfo {year} {2020})}\BibitemShut {NoStop}%
\bibitem [{\citenamefont {Kloeden}\ and\ \citenamefont
  {Platen}(2013)}]{kloeden2013numerical}%
  \BibitemOpen
  \bibfield  {author} {\bibinfo {author} {\bibfnamefont {P.E.}\ \bibnamefont
  {Kloeden}}\ and\ \bibinfo {author} {\bibfnamefont {E.}~\bibnamefont
  {Platen}},\ }\href {https://books.google.com/books?id=r9r6CAAAQBAJ} {\emph
  {\bibinfo {title} {Numerical Solution of Stochastic Differential
  Equations}}},\ Stochastic Modelling and Applied Probability\ (\bibinfo
  {publisher} {Springer Berlin Heidelberg},\ \bibinfo {year}
  {2013})\BibitemShut {NoStop}%
\bibitem [{\citenamefont {Ledoux}\ and\ \citenamefont
  {Van~Daele}(2016)}]{ledoux2016matslise}%
  \BibitemOpen
  \bibfield  {author} {\bibinfo {author} {\bibfnamefont {Veerle}\ \bibnamefont
  {Ledoux}}\ and\ \bibinfo {author} {\bibfnamefont {Marnix}\ \bibnamefont
  {Van~Daele}},\ }\bibfield  {title} {\enquote {\bibinfo {title} {Matslise 2.0:
  A {M}atlab toolbox for {S}turm-{L}iouville computations},}\ }\href@noop {}
  {\bibfield  {journal} {\bibinfo  {journal} {ACM Transactions on Mathematical
  Software (TOMS)}\ }\textbf {\bibinfo {volume} {42}},\ \bibinfo {pages}
  {1--18} (\bibinfo {year} {2016})}\BibitemShut {NoStop}%
\bibitem [{\citenamefont {Diacu}(1992)}]{diacu1992singularities}%
  \BibitemOpen
  \bibfield  {author} {\bibinfo {author} {\bibfnamefont {F.}~\bibnamefont
  {Diacu}},\ }\href {https://books.google.com/books?id=rnDvAAAAMAAJ} {\emph
  {\bibinfo {title} {Singularities of the {N}-body problem: an introduction to
  celestial mechanics}}}\ (\bibinfo  {publisher} {Publications CRM},\ \bibinfo
  {year} {1992})\BibitemShut {NoStop}%
\bibitem [{\citenamefont {Sreenivasan}\ \emph {et~al.}(1996)\citenamefont
  {Sreenivasan}, \citenamefont {Vainshtein}, \citenamefont {Bhiladvala},
  \citenamefont {San~Gil}, \citenamefont {Chen},\ and\ \citenamefont
  {Cao}}]{sreenivasan1996asymmetry}%
  \BibitemOpen
  \bibfield  {author} {\bibinfo {author} {\bibfnamefont {KR}~\bibnamefont
  {Sreenivasan}}, \bibinfo {author} {\bibfnamefont {SI}~\bibnamefont
  {Vainshtein}}, \bibinfo {author} {\bibfnamefont {R}~\bibnamefont
  {Bhiladvala}}, \bibinfo {author} {\bibfnamefont {I}~\bibnamefont {San~Gil}},
  \bibinfo {author} {\bibfnamefont {S}~\bibnamefont {Chen}}, \ and\ \bibinfo
  {author} {\bibfnamefont {N}~\bibnamefont {Cao}},\ }\bibfield  {title}
  {\enquote {\bibinfo {title} {Asymmetry of velocity increments in fully
  developed turbulence and the scaling of low-order moments},}\ }\href@noop {}
  {\bibfield  {journal} {\bibinfo  {journal} {Physical review letters}\
  }\textbf {\bibinfo {volume} {77}},\ \bibinfo {pages} {1488} (\bibinfo {year}
  {1996})}\BibitemShut {NoStop}%
\bibitem [{\citenamefont {Iyer}\ \emph {et~al.}(2020)\citenamefont {Iyer},
  \citenamefont {Sreenivasan},\ and\ \citenamefont {Yeung}}]{iyer2020scaling}%
  \BibitemOpen
  \bibfield  {author} {\bibinfo {author} {\bibfnamefont {Kartik~P}\
  \bibnamefont {Iyer}}, \bibinfo {author} {\bibfnamefont {Katepalli~R}\
  \bibnamefont {Sreenivasan}}, \ and\ \bibinfo {author} {\bibfnamefont
  {PK}~\bibnamefont {Yeung}},\ }\bibfield  {title} {\enquote {\bibinfo {title}
  {Scaling exponents saturate in three-dimensional isotropic turbulence},}\
  }\href@noop {} {\bibfield  {journal} {\bibinfo  {journal} {Physical Review
  Fluids}\ }\textbf {\bibinfo {volume} {5}},\ \bibinfo {pages} {054605}
  (\bibinfo {year} {2020})}\BibitemShut {NoStop}%
\bibitem [{\citenamefont {Bohr}\ \emph {et~al.}(2005)\citenamefont {Bohr},
  \citenamefont {Jensen}, \citenamefont {Paladin},\ and\ \citenamefont
  {Vulpiani}}]{bohr2005dynamical}%
  \BibitemOpen
  \bibfield  {author} {\bibinfo {author} {\bibfnamefont {T.}~\bibnamefont
  {Bohr}}, \bibinfo {author} {\bibfnamefont {M.H.}\ \bibnamefont {Jensen}},
  \bibinfo {author} {\bibfnamefont {G.}~\bibnamefont {Paladin}}, \ and\
  \bibinfo {author} {\bibfnamefont {A.}~\bibnamefont {Vulpiani}},\ }\href
  {https://books.google.com/books?id=9kmdql4kcA8C} {\emph {\bibinfo {title}
  {Dynamical Systems Approach to Turbulence}}},\ Cambridge Nonlinear Science
  Series\ (\bibinfo  {publisher} {Cambridge University Press},\ \bibinfo {year}
  {2005})\BibitemShut {NoStop}%
\bibitem [{\citenamefont {Eckmann}(1981)}]{eckmann1981roads}%
  \BibitemOpen
  \bibfield  {author} {\bibinfo {author} {\bibfnamefont {J-P}\ \bibnamefont
  {Eckmann}},\ }\bibfield  {title} {\enquote {\bibinfo {title} {Roads to
  turbulence in dissipative dynamical systems},}\ }\href@noop {} {\bibfield
  {journal} {\bibinfo  {journal} {Reviews of Modern Physics}\ }\textbf
  {\bibinfo {volume} {53}},\ \bibinfo {pages} {643} (\bibinfo {year}
  {1981})}\BibitemShut {NoStop}%
\bibitem [{\citenamefont {Drivas}\ and\ \citenamefont
  {Mailybaev}(2018)}]{drivas2018life}%
  \BibitemOpen
  \bibfield  {author} {\bibinfo {author} {\bibfnamefont {Theodore~D}\
  \bibnamefont {Drivas}}\ and\ \bibinfo {author} {\bibfnamefont {Alexei~A}\
  \bibnamefont {Mailybaev}},\ }\bibfield  {title} {\enquote {\bibinfo {title}
  {$\,${`L}ife after death{'} in ordinary differential equations with a
  non-{L}ipschitz singularity},}\ }\href@noop {} {\bibfield  {journal}
  {\bibinfo  {journal} {arXiv preprint arXiv:1806.09001}\ } (\bibinfo {year}
  {2018})}\BibitemShut {NoStop}%
\bibitem [{\citenamefont {Drivas}\ \emph {et~al.}(2020)\citenamefont {Drivas},
  \citenamefont {Mailybaev},\ and\ \citenamefont
  {Raibekas}}]{drivas2020statistical}%
  \BibitemOpen
  \bibfield  {author} {\bibinfo {author} {\bibfnamefont {Theodore~D}\
  \bibnamefont {Drivas}}, \bibinfo {author} {\bibfnamefont {Alexei~A}\
  \bibnamefont {Mailybaev}}, \ and\ \bibinfo {author} {\bibfnamefont {Artem}\
  \bibnamefont {Raibekas}},\ }\bibfield  {title} {\enquote {\bibinfo {title}
  {Statistical determinism in non-{L}ipschitz dynamical systems},}\ }\href@noop
  {} {\bibfield  {journal} {\bibinfo  {journal} {arXiv preprint
  arXiv:2004.03075}\ } (\bibinfo {year} {2020})}\BibitemShut {NoStop}%
\bibitem [{Note8()}]{Note8}%
  \BibitemOpen
  \bibinfo {note} {It may be worth remarking in this context that there is no
  Eulerian spontaneous stochasticity of the scalar advection equations for the
  Kraichnan model. At least for incompressible (divergence-free) velocities, it
  has been shown that the scalar advection equations have strong stochastic
  solutions for fixed velocity realizations \cite {lototskii2004passive}. This
  provides another example in addition to the Burgers equation \cite
  {eyink2015spontaneous} where Lagrangian spontaneous stochasticity occurs
  without its Eulerian counterpart.}\BibitemShut {Stop}%
\bibitem [{\citenamefont {Chen}\ and\ \citenamefont
  {Goldenfeld}(1995)}]{chen1995numerical}%
  \BibitemOpen
  \bibfield  {author} {\bibinfo {author} {\bibfnamefont {Lin-Yuan}\
  \bibnamefont {Chen}}\ and\ \bibinfo {author} {\bibfnamefont {Nigel}\
  \bibnamefont {Goldenfeld}},\ }\bibfield  {title} {\enquote {\bibinfo {title}
  {Numerical renormalization-group calculations for similarity solutions and
  traveling waves},}\ }\href@noop {} {\bibfield  {journal} {\bibinfo  {journal}
  {Physical Review E}\ }\textbf {\bibinfo {volume} {51}},\ \bibinfo {pages}
  {5577} (\bibinfo {year} {1995})}\BibitemShut {NoStop}%
\bibitem [{\citenamefont {Frisch}\ and\ \citenamefont
  {Parisi}(1985)}]{frisch1985singularity}%
  \BibitemOpen
  \bibfield  {author} {\bibinfo {author} {\bibfnamefont {U.}~\bibnamefont
  {Frisch}}\ and\ \bibinfo {author} {\bibfnamefont {G.}~\bibnamefont
  {Parisi}},\ }\bibfield  {title} {\enquote {\bibinfo {title} {On the
  singularity structure of fully developed turbulence},}\ }in\ \href@noop {}
  {\emph {\bibinfo {booktitle} {Turbulence and Predictability in Geophysical
  Fluid Dynamics and Climate Dynamics}}}\ (\bibinfo  {publisher} {Amsterdam,
  North-Holland, Elsevier},\ \bibinfo {year} {1985})\ pp.\ \bibinfo {pages}
  {84--88}\BibitemShut {NoStop}%
\bibitem [{\citenamefont {Falkovich}\ \emph {et~al.}(1996)\citenamefont
  {Falkovich}, \citenamefont {Kolokolov}, \citenamefont {Lebedev},\ and\
  \citenamefont {Migdal}}]{falkovich1996instantons}%
  \BibitemOpen
  \bibfield  {author} {\bibinfo {author} {\bibfnamefont {G}~\bibnamefont
  {Falkovich}}, \bibinfo {author} {\bibfnamefont {I}~\bibnamefont {Kolokolov}},
  \bibinfo {author} {\bibfnamefont {V}~\bibnamefont {Lebedev}}, \ and\ \bibinfo
  {author} {\bibfnamefont {A}~\bibnamefont {Migdal}},\ }\bibfield  {title}
  {\enquote {\bibinfo {title} {Instantons and intermittency},}\ }\href@noop {}
  {\bibfield  {journal} {\bibinfo  {journal} {Physical Review E}\ }\textbf
  {\bibinfo {volume} {54}},\ \bibinfo {pages} {4896} (\bibinfo {year}
  {1996})}\BibitemShut {NoStop}%
\bibitem [{\citenamefont {Mailybaev}(2017)}]{mailybaev2017toward}%
  \BibitemOpen
  \bibfield  {author} {\bibinfo {author} {\bibfnamefont {Alexei~A}\
  \bibnamefont {Mailybaev}},\ }\bibfield  {title} {\enquote {\bibinfo {title}
  {Toward analytic theory of the {R}ayleigh--{T}aylor instability: lessons from
  a toy model},}\ }\href@noop {} {\bibfield  {journal} {\bibinfo  {journal}
  {Nonlinearity}\ }\textbf {\bibinfo {volume} {30}},\ \bibinfo {pages} {2466}
  (\bibinfo {year} {2017})}\BibitemShut {NoStop}%
\bibitem [{\citenamefont {Qin}\ \emph {et~al.}(2019)\citenamefont {Qin},
  \citenamefont {Zhang}, \citenamefont {Glasser},\ and\ \citenamefont
  {Xiao}}]{qin2019kelvin}%
  \BibitemOpen
  \bibfield  {author} {\bibinfo {author} {\bibfnamefont {Hong}\ \bibnamefont
  {Qin}}, \bibinfo {author} {\bibfnamefont {Ruili}\ \bibnamefont {Zhang}},
  \bibinfo {author} {\bibfnamefont {Alexander~S}\ \bibnamefont {Glasser}}, \
  and\ \bibinfo {author} {\bibfnamefont {Jianyuan}\ \bibnamefont {Xiao}},\
  }\bibfield  {title} {\enquote {\bibinfo {title} {Kelvin-{H}elmholtz
  instability is the result of parity-time symmetry breaking},}\ }\href@noop {}
  {\bibfield  {journal} {\bibinfo  {journal} {Physics of Plasmas}\ }\textbf
  {\bibinfo {volume} {26}},\ \bibinfo {pages} {032102} (\bibinfo {year}
  {2019})}\BibitemShut {NoStop}%
\bibitem [{Note9()}]{Note9}%
  \BibitemOpen
  \bibinfo {note} {The ``PT-symmetry'' considered by those authors, as they
  note, is physically just time-reversal $T$-symmetry, which interchanges the
  growing and decaying linear eigenmodes. It is easy to check that these linear
  eigenmodes are eigenfunctions of the reflection $P$-symmetry.}\BibitemShut
  {Stop}%
\bibitem [{\citenamefont {Rose}(1977)}]{rose1977eddy}%
  \BibitemOpen
  \bibfield  {author} {\bibinfo {author} {\bibfnamefont {Harvey~A}\
  \bibnamefont {Rose}},\ }\bibfield  {title} {\enquote {\bibinfo {title} {Eddy
  diffusivity, eddy noise and subgrid-scale modelling},}\ }\href@noop {}
  {\bibfield  {journal} {\bibinfo  {journal} {Journal of Fluid Mechanics}\
  }\textbf {\bibinfo {volume} {81}},\ \bibinfo {pages} {719--734} (\bibinfo
  {year} {1977})}\BibitemShut {NoStop}%
\bibitem [{\citenamefont {Eyink}(1996)}]{eyink1996turbulence}%
  \BibitemOpen
  \bibfield  {author} {\bibinfo {author} {\bibfnamefont {Gregory~L}\
  \bibnamefont {Eyink}},\ }\bibfield  {title} {\enquote {\bibinfo {title}
  {Turbulence noise},}\ }\href@noop {} {\bibfield  {journal} {\bibinfo
  {journal} {Journal of statistical physics}\ }\textbf {\bibinfo {volume}
  {83}},\ \bibinfo {pages} {955--1019} (\bibinfo {year} {1996})}\BibitemShut
  {NoStop}%
\bibitem [{\citenamefont {Canet}\ \emph {et~al.}(2016)\citenamefont {Canet},
  \citenamefont {Delamotte},\ and\ \citenamefont {Wschebor}}]{canet2016fully}%
  \BibitemOpen
  \bibfield  {author} {\bibinfo {author} {\bibfnamefont {L{\'e}onie}\
  \bibnamefont {Canet}}, \bibinfo {author} {\bibfnamefont {Bertrand}\
  \bibnamefont {Delamotte}}, \ and\ \bibinfo {author} {\bibfnamefont
  {Nicol{\'a}s}\ \bibnamefont {Wschebor}},\ }\bibfield  {title} {\enquote
  {\bibinfo {title} {Fully developed isotropic turbulence: Nonperturbative
  renormalization group formalism and fixed-point solution},}\ }\href@noop {}
  {\bibfield  {journal} {\bibinfo  {journal} {Physical Review E}\ }\textbf
  {\bibinfo {volume} {93}},\ \bibinfo {pages} {063101} (\bibinfo {year}
  {2016})}\BibitemShut {NoStop}%
\bibitem [{\citenamefont {Tarpin}\ \emph {et~al.}(2018)\citenamefont {Tarpin},
  \citenamefont {Canet},\ and\ \citenamefont {Wschebor}}]{tarpin2018breaking}%
  \BibitemOpen
  \bibfield  {author} {\bibinfo {author} {\bibfnamefont {Malo}\ \bibnamefont
  {Tarpin}}, \bibinfo {author} {\bibfnamefont {L{\'e}onie}\ \bibnamefont
  {Canet}}, \ and\ \bibinfo {author} {\bibfnamefont {Nicol{\'a}s}\ \bibnamefont
  {Wschebor}},\ }\bibfield  {title} {\enquote {\bibinfo {title} {Breaking of
  scale invariance in the time dependence of correlation functions in isotropic
  and homogeneous turbulence},}\ }\href@noop {} {\bibfield  {journal} {\bibinfo
   {journal} {Physics of Fluids}\ }\textbf {\bibinfo {volume} {30}},\ \bibinfo
  {pages} {055102} (\bibinfo {year} {2018})}\BibitemShut {NoStop}%
\bibitem [{\citenamefont {Darling}\ and\ \citenamefont
  {Siegert}(1953)}]{darling1953first}%
  \BibitemOpen
  \bibfield  {author} {\bibinfo {author} {\bibfnamefont {Donald~A}\
  \bibnamefont {Darling}}\ and\ \bibinfo {author} {\bibfnamefont {A.~J.~F.}\
  \bibnamefont {Siegert}},\ }\bibfield  {title} {\enquote {\bibinfo {title}
  {The first passage problem for a continuous {M}arkov process},}\ }\href@noop
  {} {\bibfield  {journal} {\bibinfo  {journal} {The Annals of Mathematical
  Statistics}\ }\textbf {\bibinfo {volume} {24}},\ \bibinfo {pages} {624--639}
  (\bibinfo {year} {1953})}\BibitemShut {NoStop}%
\bibitem [{\citenamefont {Abramowitz}\ and\ \citenamefont
  {Stegun}(2012)}]{abramowitz2012handbook}%
  \BibitemOpen
  \bibfield  {author} {\bibinfo {author} {\bibfnamefont {M.}~\bibnamefont
  {Abramowitz}}\ and\ \bibinfo {author} {\bibfnamefont {I.A.}\ \bibnamefont
  {Stegun}},\ }\href {https://books.google.com/books?id=KiPCAgAAQBAJ} {\emph
  {\bibinfo {title} {Handbook of Mathematical Functions: with Formulas, Graphs,
  and Mathematical Tables}}},\ Dover Books on Mathematics\ (\bibinfo
  {publisher} {Dover Publications},\ \bibinfo {year} {2012})\BibitemShut
  {NoStop}%
\bibitem [{\citenamefont {Matsumoto}\ and\ \citenamefont
  {Nishimura}(1998)}]{matsumoto1998mersenne}%
  \BibitemOpen
  \bibfield  {author} {\bibinfo {author} {\bibfnamefont {M.}~\bibnamefont
  {Matsumoto}}\ and\ \bibinfo {author} {\bibfnamefont {T.}~\bibnamefont
  {Nishimura}},\ }\bibfield  {title} {\enquote {\bibinfo {title} {Mersenne
  twister: a 623-dimensionally equidistributed uniform pseudo-random number
  generator},}\ }\href@noop {} {\bibfield  {journal} {\bibinfo  {journal} {ACM
  Transactions on Modeling and Computer Simulation (TOMACS)}\ }\textbf
  {\bibinfo {volume} {8}},\ \bibinfo {pages} {3--30} (\bibinfo {year}
  {1998})}\BibitemShut {NoStop}%
\bibitem [{\citenamefont {Marsaglia}\ and\ \citenamefont
  {Tsang}(2000)}]{marsaglia2000ziggurat}%
  \BibitemOpen
  \bibfield  {author} {\bibinfo {author} {\bibfnamefont {George}\ \bibnamefont
  {Marsaglia}}\ and\ \bibinfo {author} {\bibfnamefont {Wai~Wan}\ \bibnamefont
  {Tsang}},\ }\bibfield  {title} {\enquote {\bibinfo {title} {The ziggurat
  method for generating random variables},}\ }\href@noop {} {\bibfield
  {journal} {\bibinfo  {journal} {Journal of Statistical Software}\ }\textbf
  {\bibinfo {volume} {5}},\ \bibinfo {pages} {1--7} (\bibinfo {year}
  {2000})}\BibitemShut {NoStop}%
\bibitem [{\citenamefont {Silverman}(2018)}]{silverman2018density}%
  \BibitemOpen
  \bibfield  {author} {\bibinfo {author} {\bibfnamefont {B.W.}\ \bibnamefont
  {Silverman}},\ }\href {https://books.google.com/books?id=fExnDwAAQBAJ} {\emph
  {\bibinfo {title} {Density Estimation for Statistics and Data Analysis}}}\
  (\bibinfo  {publisher} {CRC Press},\ \bibinfo {year} {2018})\BibitemShut
  {NoStop}%
\bibitem [{\citenamefont {Stevenson}(1981)}]{stevenson1981optimized}%
  \BibitemOpen
  \bibfield  {author} {\bibinfo {author} {\bibfnamefont {Paul~M}\ \bibnamefont
  {Stevenson}},\ }\bibfield  {title} {\enquote {\bibinfo {title} {Optimized
  perturbation theory},}\ }\href@noop {} {\bibfield  {journal} {\bibinfo
  {journal} {Physical Review D}\ }\textbf {\bibinfo {volume} {23}},\ \bibinfo
  {pages} {2916} (\bibinfo {year} {1981})}\BibitemShut {NoStop}%
\bibitem [{\citenamefont {Lototskii}\ and\ \citenamefont
  {Rozovskii}(2004)}]{lototskii2004passive}%
  \BibitemOpen
  \bibfield  {author} {\bibinfo {author} {\bibfnamefont {SV}~\bibnamefont
  {Lototskii}}\ and\ \bibinfo {author} {\bibfnamefont {Boris~L}\ \bibnamefont
  {Rozovskii}},\ }\bibfield  {title} {\enquote {\bibinfo {title} {Passive
  scalar equation in a turbulent incompressible {G}aussian velocity field},}\
  }\href@noop {} {\bibfield  {journal} {\bibinfo  {journal} {Russian
  Mathematical Surveys}\ }\textbf {\bibinfo {volume} {59}},\ \bibinfo {pages}
  {297} (\bibinfo {year} {2004})}\BibitemShut {NoStop}%
\end{thebibliography}%

\end{document}